\documentclass[stslayout, noinfoline]{imsart}
\usepackage{geometry} 
\usepackage{graphicx}	
\usepackage{amssymb, amsmath}
\usepackage{epstopdf}
\usepackage{natbib}
\usepackage{subfigure}
\usepackage{algcompatible}
\usepackage[floatrow]{trivfloat}
\usepackage{comment}
\usepackage{tikz}
\usepackage[colorlinks,citecolor=blue,urlcolor=blue]{hyperref}

\usetikzlibrary{intersections, backgrounds, calc}

\definecolor{light}{RGB}{220, 188, 188}
\definecolor{mid}{RGB}{185, 124, 124}
\definecolor{dark}{RGB}{143, 39, 39}
\definecolor{highlight}{RGB}{0, 255, 0}
\definecolor{gray10}{gray}{0.1}
\definecolor{gray20}{gray}{0.2}
\definecolor{gray30}{gray}{0.3}
\definecolor{gray40}{gray}{0.4}
\definecolor{gray60}{gray}{0.6}
\definecolor{gray70}{gray}{0.7}
\definecolor{gray80}{gray}{0.8}
\definecolor{gray90}{gray}{0.9}
\definecolor{gray95}{gray}{0.95}
\definecolor{comment}{gray}{0.50}

\trivfloat{algorithm}

\newcommand{\mcpoint}[2]{  
  \fill[color=dark] (#1, #2) circle (7pt); 
  \fill[color=light] (#1, #2) circle (5pt);
}

\graphicspath{{external_figures/}}

\begin{document}

\begin{frontmatter}

\title{A Conceptual Introduction to Hamiltonian Monte Carlo}
\runtitle{A Conceptual Introduction to Hamiltonian Monte Carlo}

\begin{aug}
  \author{Michael Betancourt%
  \ead[label=e1]{betanalpha@gmail.com}}

  \runauthor{Betancourt}

  \address{Michael Betancourt is a research scientist in the Applied Statistics
           Center at Columbia University.  Much of this review was completed as 
           a Research Fellow at the Centre for Research in Statistical Methodology, 
           University of Warwick, Coventry CV4 7AL, UK \printead{e1}.}

\end{aug}

\begin{abstract}
Hamiltonian Monte Carlo has proven a remarkable empirical success, but
only recently have we begun to develop a rigorous understanding of why
it performs so well on difficult problems and how it is best applied in
practice.  Unfortunately, that understanding is confined within the 
mathematics of differential geometry which has limited its dissemination, 
especially to the applied communities for which it is particularly 
important.  

In this review I provide a comprehensive conceptual account of these 
theoretical foundations, focusing on developing a principled intuition 
behind the method and its optimal implementations rather of any exhaustive 
rigor.  Whether a practitioner or a statistician, the dedicated reader 
will acquire a solid grasp of how Hamiltonian Monte Carlo works, when 
it succeeds, and, perhaps most importantly, when it fails.
\end{abstract}

\end{frontmatter}

\pagebreak

\setcounter{tocdepth}{2}
\tableofcontents

\pagebreak

Hamiltonian Monte Carlo has followed a long and winding path into modern
statistical computing.  The method was originally developed in the late 1980s 
as \emph{Hybrid Monte Carlo} to tackle calculations in Lattice Quantum 
Chromodynamics~\citep{DuaneEtAl:1987}, a field focused on understanding the 
structure of the protons and neutrons that comprise nuclei, atoms, and 
ultimately the world around us.  Within a few years Radford Neal recognized 
the potential of the method for problems in applied statistics in his pioneering 
work on Bayesian neural networks \citep{Neal:1995}.  Over the next decade 
the method began to make appearances in textbooks, notably \cite{MacKay:2003},
who first used the term Hamiltonian Monte Carlo instead of Hybrid Monte Carlo, 
and \cite{Bishop:2006}.  Neal's influential review~\citep{Neal:2011}, however, 
really introduced the approach into the mainstream of statistical computing.  
With the rise of high-performance software implementations such as 
Stan~\citep{Stan:2017}, the method has now become a pervasive tool across many 
scientific, medical, and industrial applications.

Only recently, however, have we begun to understand why the success of
Hamiltonian Monte Carlo has been so extensive.  Instead of relying on 
fragile heuristics, the method is built upon a rich theoretical foundation 
that makes it uniquely suited to the high-dimensional problems of applied 
interest~\citep{BetancourtEtAl:2014a}.  Unfortunately, this theoretical 
foundation is formulated in terms of differential geometry, an advanced 
field of mathematics that is rarely included in statistical pedagogy.  
Consequently this formal construction is often out of reach of theoretical 
and applied statisticians alike.

The aim of this paper is to introduce the intuition behind the success of 
Hamiltonian Monte Carlo while avoiding the mathematical details; in particular, 
I assume only a basic familiarity with probability and calculus.  Such an 
approach necessarily sacrifices rigor, but I hope the concepts will be 
sufficiently intuitive to satisfy readers working in applied fields.  I 
highly encourage those interested in learning more about Hamiltonian Monte Carlo, 
or even contributing to its development, to follow up on the references 
discussed in \citealp[Section 2]{BetancourtEtAl:2014a}.

Our story will begin with an introduction to the geometry of high-dimensional
probability distributions and how that geometry frustrates efficient statistical
computing.  We will then consider Markov chain Monte Carlo from this
geometric perspective, motiving the features necessary to scale the approach
to such high-dimensional problems.  By developing a method that inherently 
satisfies these criteria we will very naturally be led to Hamiltonian Monte 
Carlo.  Finally I will discuss how this understanding can be extended to motivate 
not just the method itself but also its efficient practical implementation, 
including optimized tuning as well as inherent diagnostics of pathological 
behavior.

\section{Computing Expectations By Exploring Probability Distributions}
\label{sec:exploring}

The ultimate undertaking in statistical computing is evaluating expectations 
with respect to some distinguished \emph{target} probability distribution.  For 
example, we might be interested in extracting information from a posterior 
distribution over model configuration space in Bayesian inference, or computing 
coverage of an estimator with respect to the likelihood over data space in 
frequentist statistics.  Here we will be agnostic, considering only a target 
distribution, $\pi$, on a $D$-dimensional sample space, $Q$, and the 
corresponding expectations of functions, $\mathbb{E}_{\pi} \! \left[ f \right]$.

Probability distributions, and the corresponding expectations, are rather
abstract objects, however, and if we want to use them in practice then we
need a more explicit means of specifying them.  Here we will assume that 
the sample space is smooth, in which case that we can represent the target 
distribution with a \emph{probability density function} and expectations as 
integrals.  Care must be taken, however, as this representation hides some 
of the more subtle behaviors of high-dimensional spaces that are critical
towards understanding how to compute these integrals, and hence the desired
expectations, efficiently.

\subsection{Computing Expectations in Practice}

We begin by assuming that the target sample space, $Q$, can be parameterized 
by the real numbers such that every point $q \in Q$ can be specified with 
$D$ real numbers.  Given a parameter space, $\mathcal{Q}$, we can then specify 
the target distribution as a smooth probability density function, 
$\pi \! \left( q \right)$, while expectations reduce to integrals over 
parameter space,
\begin{equation*}
\mathbb{E}_{\pi} \! \left[ f \right] 
=
\int_{\mathcal{Q}} \mathrm{d} q \, 
\pi \! \left( q \right) f \! \left( q \right).
\end{equation*}

Parameterizations are not unique: we can always take another parameterization, 
$\mathcal{Q}'$, over which we specify the target distribution with a different 
probability density function, $\pi' \! \left( q' \right)$, while expectations 
reduce to the new integral,
\begin{equation*}
\mathbb{E}_{\pi} \! \left[ f \right] 
=
\int_{\mathcal{Q}'} \mathrm{d} q' \, 
\pi ( q' ) f ( q' ).
\end{equation*}
Critically, however, the expectations values themselves are invariant to any 
particular choice of parameterization, so the integrals must be equal,
\begin{equation*}
\mathbb{E}_{\pi} \! \left[ f \right] 
=
\int_{\mathcal{Q}} \mathrm{d} q \, 
\pi ( q ) f ( q )
=
\int_{\mathcal{Q}'} \mathrm{d} q' \, 
\pi' ( q' ) f ( q' ).
\end{equation*}
In this review we will consider only a single parameterization for computing
expectations, but we must be careful to ensure that any such computation does 
not depend on the irrelevant details of that parameterization, such as the 
particular shape of the probability density function.

Once we have chosen a parameterization, the abstract expectation becomes
the concrete integral.  Unfortunately, for any nontrivial target distribution 
we will not be able to evaluate these integrals analytically, and we must 
instead resort to numerical methods which only \emph{approximate} them.  
The accuracy of these approximations, and hence the utility of any given 
algorithm, however, is limited by our finite computational power.

For a method to scale to the complex problems at the frontiers of applied
statistics, it has to make effective use of each and every evaluation of the 
target density, $\pi \! \left( q \right)$, and relevant functions, 
$f \! \left( q \right)$.  Optimizing these evaluations is a subtle problem 
frustrated by the natural geometry of probability distributions, especially 
over high-dimensional parameter spaces.

\subsection{Parsimonious Expectation Computation}

One way to ensure computational \emph{inefficiency} is to waste 
computational resources evaluating the target density and relevant 
functions in regions of parameter space that have negligible contribution 
to the desired expectation.  In order to avoid these regions, and focus 
our computation only on significant regions of parameter space, we first 
need to identify how the target density and target function contribute 
to the overall expectation. 

Because integration is a linear operation, scaling the integrand 
proportionately scales the integral.  Consequently, a common intuition 
is to focus on regions where the integrand is largest.  This intuition 
suggests that we consider regions where the target density and target 
function take on their largest values.

In practice we often are interested in computing expectations with 
respect to many target functions, for example in Bayesian inference
we typically summarize our uncertainty with both means and variances,
or multiple quantiles.  Any method that depends on the specific details 
of any one function will then have to be repeatedly adjusted for each 
new function we encounter, expanding a single computational problem into 
many.  Consequently, from here on in we will assume that any relevant 
function is sufficiently uniform in parameter space that its variation 
does not strongly effect the integrand.  Keep in mind, however, that 
if only a single expectation is in fact of interest then exploiting 
the structure of that function can provide significant 
improvements~\citep{MiraEtAl:2013, OatesEtAl:2016}.

This assumption implies that the variation in the integrand is dominated
by the target density, and hence we should consider the neighborhood 
around the mode where the density is maximized.  This intuition is 
consistent with the many statistical methods that utilize the mode, such 
as maximum likelihood estimators and Laplace approximations, although
conflicts with our desire to avoid the specific details of the target
density.  Indeed, this intuition is fatally naive as it misses a critical 
detail.  

Expectation values are given by accumulating the integrand over a 
\emph{volume} of parameter space and, while the density is largest 
around the mode, there is not much volume there.  To identify the 
regions of parameter space that dominate expectations we need to 
consider the behavior of both the density and the volume.  In 
high-dimensional spaces the volume behaves very differently from the 
density, resulting in a tension that concentrates the significant
regions of parameter space away from either extreme.

\subsection{The Geometry of High-Dimensional Spaces}

One of the characteristic properties of high-dimensional spaces is
that there is much more volume outside any given neighborhood than 
inside of it.  Although this may at first appear strange, we can build 
intuition to demystify this behavior by considering a few simple cases.  

\begin{figure*}
\centering
% One-dimensional
\subfigure[]{
\begin{tikzpicture}[scale=0.29, thick]
 \draw[white] (-8, -8) rectangle (8, 8);
  \draw[|-|, dashed, gray60] (-6, 0) -- (6, 0);
  \draw[|-|] (-2, 0) -- (2, 0);
\end{tikzpicture}
}
% Two-dimensional
\subfigure[]{
\begin{tikzpicture}[scale=0.29, thick]
 \draw[white] (-8, -8) rectangle (8, 8);
  \foreach \i in {-6, -2, 2, 6} {
    \draw[dashed, gray60] (\i, -6) -- (\i, 6);
    \draw[dashed, gray60] (-6, \i) -- (6, \i);
  }
  \draw[] (-2, -2) -- (-2, 2) -- (2, 2) -- (2, -2) -- (-2, -2);
\end{tikzpicture}
}
% Three-dimensional
\subfigure[]{
\begin{tikzpicture}[scale=0.29 / 1.375, thick]
 \draw[white] (-11, -11) rectangle (11, 11);
 \pgfmathsetmacro{\xy}{2}
 \pgfmathsetmacro{\z}{1}
  \foreach \i in {-3, -1, 1, 3} {
    \foreach \j in {1, 3} {
      \draw[dashed, gray60] (\xy * \i + \z * \j, -\xy * 3 + \z * \j) -- 
                              (\xy * \i + \z * \j, \xy * 3 + \z * \j);
      \draw[dashed, gray60] (-\xy * 3 + \z * \j, \xy * \i + \z * \j) -- 
                              (\xy * 3 + \z * \j, \xy * \i + \z * \j);
    }
  }
  \foreach \i in {0, 1, 2, 3} {
    \foreach \j in {0, 1, 2, 3} {
      \draw[dashed, gray60] 
        (-3 * \xy - 3 *\z + 2 * \xy * \i, -3 * \xy - 3 * \z + 2 * \xy * \j) -- 
        +(6 * \z, 6 * \z);
    }
  }
  \draw[black,fill=white] (-\xy + \z, -\xy + \z) -- (-\xy + \z, \xy + \z) -- 
                (\xy + \z, \xy + \z) -- (\xy + \z, -\xy + \z) -- 
                (-\xy + \z, -\xy + \z);
  \draw[black,fill=white] (\xy - \z, \xy - \z) -- (\xy + \z, \xy + \z) -- 
                (\xy + \z, -\xy + \z) -- (\xy - \z, -\xy - \z) -- 
                (\xy - \z, \xy - \z);
  \draw[black,fill=white] (-\xy + \z, \xy + \z) -- (\xy + \z, \xy + \z) -- 
                (\xy - \z, \xy - \z) -- (-\xy - \z, \xy - \z) -- 
                (-\xy + \z, \xy + \z);
  \foreach \i in {-3, -1, 1, 3} {
    \foreach \j in {-1} {
      \draw[dashed, gray60] (\xy * \i + \z * \j, -\xy * 3 + \z * \j) -- 
                              (\xy * \i + \z * \j, \xy * 3 + \z * \j);
      \draw[dashed, gray60] (-\xy * 3 + \z * \j, \xy * \i + \z * \j) -- 
                             (\xy * 3 + \z * \j, \xy * \i + \z * \j);
    }
  } 
  \draw[black,fill=white] (-\xy - \z, -\xy - \z) -- (-\xy - \z, \xy - \z) -- 
                (\xy - \z, \xy - \z) -- (\xy - \z, -\xy - \z) -- 
                (-\xy - \z, -\xy - \z);  
  \foreach \i in {-3, -1, 1, 3} {
    \foreach \j in {-3} {
      \draw[dashed, gray60] (\xy * \i + \z * \j, -\xy * 3 + \z * \j) -- 
                              (\xy * \i + \z * \j, \xy * 3 + \z * \j);
      \draw[dashed, gray60] (-\xy * 3 + \z * \j, \xy * \i + \z * \j) -- 
                             (\xy * 3 + \z * \j, \xy * \i + \z * \j);
    }
  }        
\end{tikzpicture}
}
\caption{To understand how the distribution of volume behaves with 
increasing dimension we can consider a rectangular partitioning centered 
around a distinguished point, such as the mode.  (a) In one dimension the 
relative weight of the center partition is $1/3$, (b) in two dimensions it 
is $1/9$, (c) and in three dimensions it is only $1/27$.  Very quickly the 
volume in the center partition becomes negligible compared to the neighboring 
volume.
}
\label{fig:rectangular_volumes}
\end{figure*}
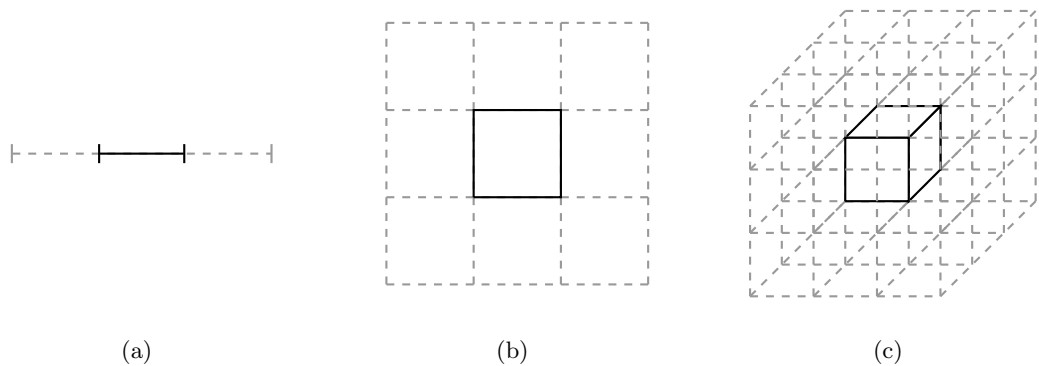

For example, consider partitioning parameter space into rectangular 
boxes centered around the mode (Figure \ref{fig:rectangular_volumes}).
In one dimension there are only two partitions neighboring the center
partition, leaving a significant volume around the mode.  Adding one
more dimension, however, introduces eight neighboring partitions, and 
in three dimensions there are already 26.  In general there are 
$3^{D} - 1$ neighboring partitions in a $D$-dimensional space, and for 
even small $D$ the volume neighboring the mode dominates the volume 
immediately around the mode.  This pattern only amplifies if we consider 
more partitions further away from the mode that multiply even faster.

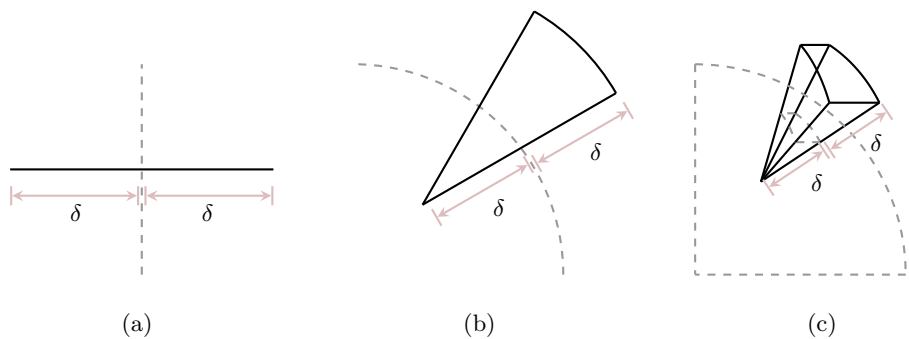
\begin{figure*}
\centering
% One-dimensional
\subfigure[]{
\begin{tikzpicture}[scale=0.35, thick]
  \draw[white] (-6, -6) rectangle (6, 6);
  \draw[dashed, color=gray60] (0, -5) -- (0, 3);
  \draw (-5, -1) -- (5, -1);
  \draw[|<->|, >=stealth, light] (-5, -2) -- (-0.1, -2);
  \node[below] at (-2.5, -2) { $\delta$};
  \draw[|<->|, >=stealth, light] (5, -2) -- (0.1, -2);
  \node[below] at (2.5, -2) { $\delta$};
\end{tikzpicture}
}
% Two-dimensional
\subfigure[]{
\begin{tikzpicture}[scale=0.35, thick]
  \draw[white] (-1, -1) rectangle (11, 11);
  \draw[dashed, color=gray60] (8, 0) arc (0:90:8);
  \draw ({8  / sqrt(2) - 3}, {8  / sqrt(2) - 3}) -- + ({sqrt(2) * cos(30) * 6}, {sqrt(2) * sin(30) * 6});
  \draw ({8  / sqrt(2) - 3}, {8  / sqrt(2) - 3}) -- + ({sqrt(2) * cos(60) * 6}, {sqrt(2) * sin(60) * 6});
  \draw ({8  / sqrt(2) - 3 + sqrt(2) * cos(30) * 6}, 
             {8  / sqrt(2) - 3 + sqrt(2) * sin(30) * 6}) arc (30:60:{sqrt(2) * 6});
  \draw[|<->|, >=stealth, light] ({8  / sqrt(2) - 3 + 0.5}, {8  / sqrt(2) - 3 - 0.5}) 
                      -- +({sqrt(2) * cos(30) * 2.9}, {sqrt(2) * sin(30) * 2.9});
  \draw[|<->|, >=stealth, light] ({8  / sqrt(2) - 3 + 0.5 + sqrt(2) * cos(30) * 3.1}, 
                      {8  / sqrt(2) - 3 - 0.5 + sqrt(2) * sin(30) * 2.9}) 
                      -- +({sqrt(2) * cos(30) * 2.9}, {sqrt(2) * sin(30) * 2.9});
  \node[below right] at ({8  / sqrt(2) - 3 + 0.5 + sqrt(2) * cos(30) * 1.5}, 
                                     {8  / sqrt(2) - 3 - 0.5 + sqrt(2) * sin(30) * 1.5}) { $\delta$ };
  \node[below right] at ({8  / sqrt(2) - 3 + 0.5 + sqrt(2) * cos(30) * 4.5}, 
                                     {8  / sqrt(2) - 3 - 0.5 + sqrt(2) * sin(30) * 4.5}) { $\delta$ };
\end{tikzpicture}
}
% Three-dimensional
\subfigure[]{
\begin{tikzpicture}[scale=0.35, thick]
  \draw[white] (-1, -1) rectangle (11, 11);
  \draw        ({8 * sin(45) * cos(45) - 3 * sin(45) * cos(45)}, {8 * cos(45) - 3 * cos(45)}) 
            -- + ({3 * sin(30) * cos(30)}, {3 * cos(30) });
  \draw        ({8 * sin(45) * cos(45) - 3 * sin(45) * cos(45)}, {8 * cos(45) - 3 * cos(45)}) 
            -- + ({3 * sin(30) * cos(60)}, {3 * cos(30)});
  \draw        ({8 * sin(45) * cos(45) - 3 * sin(45) * cos(45)}, {8 * cos(45) - 3 * cos(45)}) 
            -- + ({3 * sin(60) * cos(30)}, {3 * cos(60)});
  \draw        ({8 * sin(45) * cos(45) - 3 * sin(45) * cos(45)}, {8 * cos(45) - 3 * cos(45)}) 
            -- + ({3 * sin(60) * cos(60)}, {3 * cos(60)});
  \draw [dashed, color=gray60, domain=0:1, samples=25] 
    plot ({8 * sin(45) * cos(45) - 3 * sin(45) * cos(45) + 3 * sin(30 * \x + 30) * cos(30)}, 
            {8 * cos(45) - 3 * cos(45) + 3 * cos(30 * \x + 30)} );
  \draw [dashed, color=gray60, domain=0:1, samples=25] 
    plot ({8 * sin(45) * cos(45) - 3 * sin(45) * cos(45) + 3 * sin(30 * \x + 30) * cos(60)}, 
            {8 * cos(45) - 3 * cos(45) + 3 * cos(30 * \x + 30)} );
  \draw [dashed, color=gray60, domain=0:1, samples=25] 
    plot ({8 * sin(45) * cos(45) - 3 * sin(45) * cos(45) + 3 * sin(30) * cos(30 * \x + 30)}, 
            {8 * cos(45) - 3 * cos(45) + 3 * cos(30)} ); 
  \draw [dashed, color=gray60, domain=0:1, samples=25] 
    plot ({8 * sin(45) * cos(45) - 3 * sin(45) * cos(45) + 3 * sin(60) * cos(30 * \x + 30)}, 
            {8 * cos(45) - 3 * cos(45) + 3 * cos(60)} ); 
  \draw        ({8 * sin(45) * cos(45) - 3 * sin(45) * cos(45) + 3 * sin(30) * cos(30)}, 
                    {8 * cos(45) - 3 * cos(45) + 3 * cos(30)}) 
            -- + ({3 * sin(30) * cos(30)}, {3 * cos(30)});
  \draw       ({8 * sin(45) * cos(45) - 3 * sin(45) * cos(45) + 3 * sin(30) * cos(60)}, 
                    {8 * cos(45) - 3 * cos(45) + 3 * cos(30)}) 
            -- + ({3 * sin(30) * cos(60)}, {3 * cos(30)});
  \draw        ({8 * sin(45) * cos(45) - 3 * sin(45) * cos(45) + 3 * sin(60) * cos(30)}, 
                    {8 * cos(45) - 3 * cos(45) + 3 *cos(60)}) 
            -- + ({3 * sin(60) * cos(30)}, {3 * cos(60)});
  \draw        ({8 * sin(45) * cos(45) - 3 * sin(45) * cos(45) + 3 * sin(60) * cos(60)}, 
                    {8 * cos(45) - 3 * cos(45) + 3 * cos(60)}) 
            -- + ({3 * sin(60) * cos(60)}, {3 * cos(60)});
  \draw [domain=0:1, samples=25] 
    plot ({8 * sin(45) * cos(45) - 3 * sin(45) * cos(45) + 6 * sin(30 * \x + 30) * cos(30)}, 
            {8 * cos(45) - 3 * cos(45) + 6 * cos(30 * \x + 30)} );
  \draw [domain=0:1, samples=25] 
    plot ({8 * sin(45) * cos(45) - 3 * sin(45) * cos(45) + 6 * sin(30 * \x + 30) * cos(60)}, 
            {8 * cos(45) - 3 * cos(45) + 6 * cos(30 * \x + 30)} );
  \draw [domain=0:1, samples=25] 
    plot ({8 * sin(45) * cos(45) - 3 * sin(45) * cos(45) + 6 * sin(30) * cos(30 * \x + 30)}, 
            {8 * cos(45) - 3 * cos(45) + 6 * cos(30)} ); 
  \draw [domain=0:1, samples=25] 
    plot ({8 * sin(45) * cos(45) - 3 * sin(45) * cos(45) + 6 * sin(60) * cos(30 * \x + 30)}, 
            {8 * cos(45) - 3 * cos(45) + 6 * cos(60)} ); 
  \draw [dashed, color=gray60, domain=0:1, samples=25] 
    plot ({8 * sin(90 * \x) * cos(0)}, 
            {8 * cos(90 * \x)} );
  \draw [dashed, color=gray60, domain=0:1, samples=25] 
    plot ({8 * sin(90 * \x) * cos(90)}, 
            {8 * cos(90 * \x)} );
  \draw [dashed, color=gray60, domain=0:1, samples=25] 
    plot ({8 * sin(90) * cos(90 * \x)}, 
            {8 * cos(90)} );
  \draw[|<->|, >=stealth, light] 
    ({8 * sin(45) * cos(45) - 3 * sin(45) * cos(45) + 0.25} , 
     {8 * cos(45) - 3 * cos(45) - 0.25}) 
     -- + ({2.9 * sin(60) * cos(30)}, {2.9 * cos(60)});
  \node[below right] at ({8 * sin(45) * cos(45) - 3 * sin(45) * cos(45) + 1.75 * sin(60) * cos(30) + 0.25}, 
                                     {8 * cos(45) - 3 * cos(45) + 1.75 * cos(60) - 0.25}) { $\delta$ };
  \draw[|<->|, >=stealth, light] 
    ({8 * sin(45) * cos(45) - 3 * sin(45) * cos(45) + 3.1 * sin(60) * cos(30) + 0.25}, 
     {8 * cos(45) - 3.1 * cos(45) + 3 * cos(60) - 0.25}) 
     -- + ({3 * sin(60) * cos(30)}, {3 * cos(60)});
  \node[below right] at ({8 * sin(45) * cos(45) - 3 * sin(45) * cos(45) + 4.75 * sin(60) * cos(30) + 0.25}, 
                                     {8 * cos(45) - 3 * cos(45) + 4.75 * cos(60) - 0.25}) { $\delta$ };
\end{tikzpicture}
}
\caption{The dominance of volume away from any point in parameter 
space can also be seen from a spherical perspective, where we consider
the volume contained radial distance $\delta$ both interior to and exterior 
to a $D$-dimensional spherical shell, shown here with dashed lines.
(a) In one dimension the spherical shell is a line and volumes interior 
and exterior are equivalent.  (b) In two dimensions the spherical shell 
becomes circle and there is more volume immediately outside the shell than 
immediately inside.  (c) The exterior volume grows even larger relative 
to the interior volume in three dimensions, where the spherical shell is 
now a the surface of a sphere.  In fact, with increasing dimension the 
exterior volume grows exponentially large relative to the interior volume, 
and very quickly the volume around the mode is dwarfed by the volume away 
from the mode.
}
\label{fig:spherical_volumes}
\end{figure*}

Alternatively, we can take a spherical view of parameter space and 
consider the relative volume a distance $\delta$ inside and outside 
of a spherical shell (Figure \ref{fig:spherical_volumes}).  In one 
dimension the interior and exterior volumes are equal, but in two 
and three dimensions more and more volume concentrates on the outside 
of the shell.  Centering the shell at the mode we can see once again 
that the volume in any neighborhood containing the mode becomes more 
and more negligible as the dimension of the parameter space increases.

Generically, then, volume is largest out in the \emph{tails} of the target
distribution away from the mode, and this disparity grows exponentially
with the dimension of parameter space.  Consequently, the massive volume 
over which we integrate can compensate to give a significant contribution 
to the target expectation despite the smaller density.  In order to identify 
the neighborhoods that most contribute to expectations, we need to carefully 
balance the behavior of both the density and the volume.

\subsection{The Geometry of High-Dimensional Probability Distributions}

\begin{figure*}
\centering
\includegraphics[width=3in]{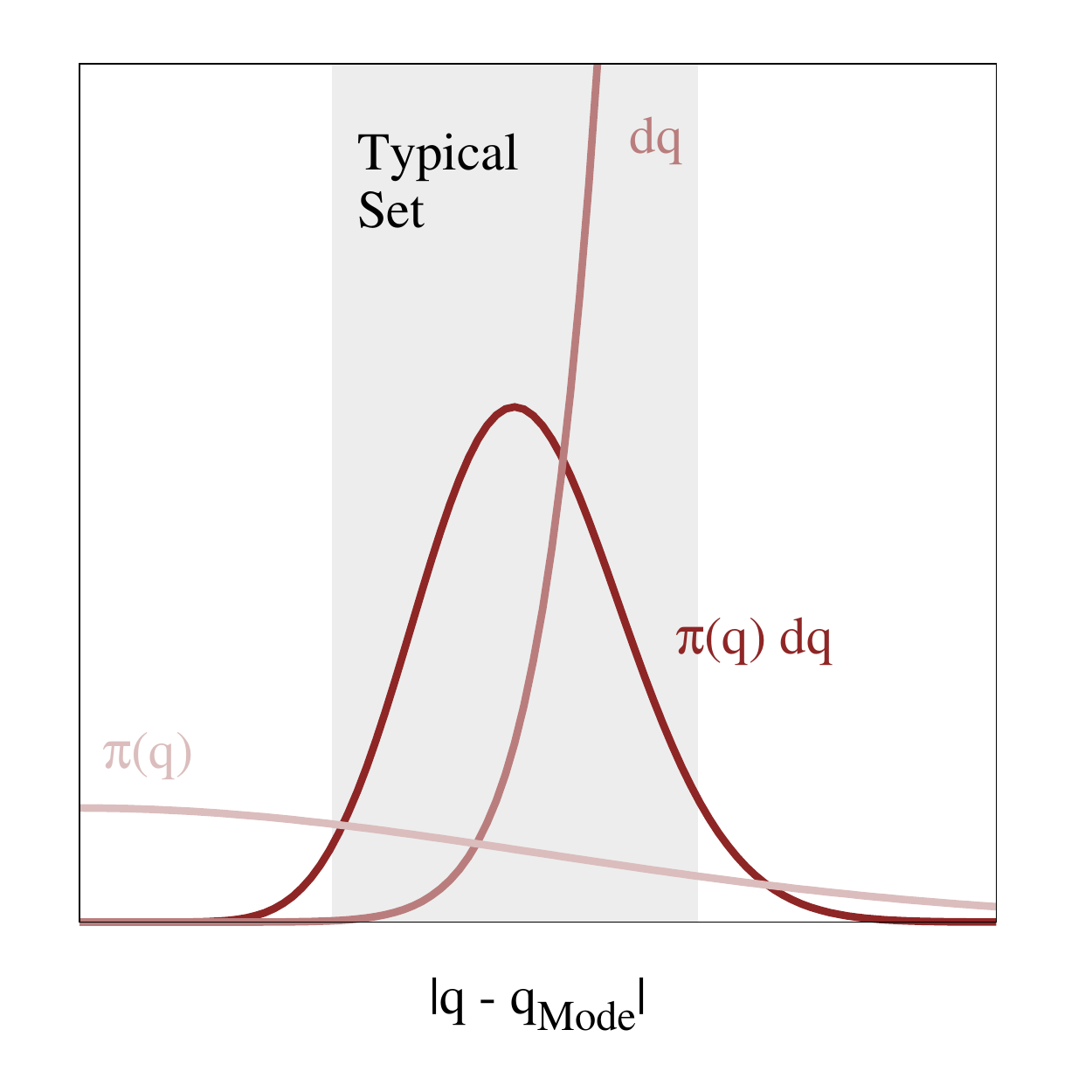}
\caption{In high dimensions a probability density, $\pi \! \left( q \right)$, will 
concentrate around its mode, but the volume over which we integrate that density, 
$\mathrm{d}q$, is much larger away from the mode.  Contributions to any expectation 
are determined by the product of density and volume, $\pi \! \left( q \right) 
\mathrm{d} q$, which then concentrates in a nearly-singular neighborhood called the 
typical set (grey).}
\label{fig:conc_of_meas_anal}
\end{figure*}

The neighborhood immediately around the mode features large densities,
but in more than a few dimensions the small volume of that neighborhood
prevents it from having much contribution to any expectation.  On the 
other hand, the complimentary neighborhood far away from the mode features 
a much larger volume, but the vanishing densities lead to similarly negligible 
contributions expectations.  The only significant contributions come from the 
neighborhood between these two extremes known as the \emph{typical set} (Figure \ref{fig:conc_of_meas_anal}).  Importantly, because probability densities and
volumes transform oppositely under any reparameterization, the typical set
is an invariant object that does not depend on the irrelevant details of any
particular choice of parameters.

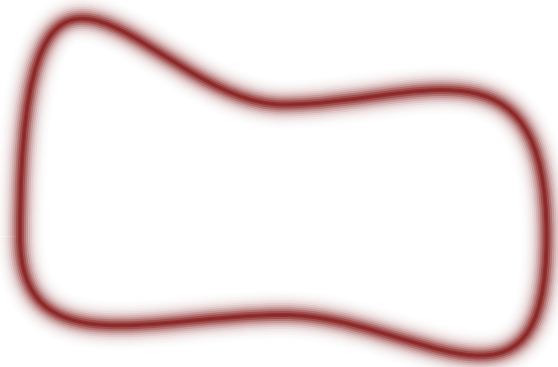
\begin{figure*}
\centering
\begin{tikzpicture}[scale=0.35, thick]
  \begin{scope}
    \clip (-12, -6) rectangle (12, 10);
    \foreach \i in {0, 0.05,..., 1} {
      \draw[line width={30 * \i}, opacity={exp(-8 * \i)}, dark] 
        (-10, 0) .. controls (-10, 15) and (-5, 5) .. (0, 5)
        .. controls (5, 5) and (10, 8) .. (10, 0)
        .. controls (10, -8) and (5, -3) .. (0, -3)
        .. controls (-5, -3) and (-10, -5) .. (-10, 0);
    }  
  \end{scope}
\end{tikzpicture}
\caption{In high-dimensional parameter spaces probability mass, 
$\pi \! \left( q \right) \mathrm{d} q$, and hence the dominant 
contributions to expectations, concentrates in a neighborhood 
called the \emph{typical set}.  In order to accurately estimate 
expectations we have to be able to identify where the typical set 
lies in parameter space so that we can focus our computational 
resources where they are most effective.
}
\label{fig:typical_set}
\end{figure*}

As the dimension of parameter space increases, the tension between the density 
and the volume grows and the regions where the density and volume are both large 
enough to yield a significant contribution becomes more and more narrow.  
Consequently the typical set becomes more singular with increasing dimension, 
a manifestation of \emph{concentration of measure}.  The immediate consequence of 
concentration of measure is that the only significant contributions to any expectation 
come from the typical set; evaluating the integrand outside of the typical set has 
negligible effect on expectations and hence is a waste of precious computational 
resources. In other words, we can accurately estimate expectations by averaging 
over the typical set instead of the entirety of parameter space.  Consequently, 
in order to compute expectations efficiently, we have to be able to identify, 
and then focus our computational resources into, the typical set 
(Figure \ref{fig:typical_set}).

This helps to explain, for example, why brute force methods like naive
quadrature scale so poorly with dimension.  A grid of length $N$ distributed 
uniformly in a $D$-dimensional parameter space requires $N^{D}$ points and 
hence $N^{D}$ evaluations of the integrand.  Unless $N$ is incredibly large, 
however, it is unlikely that any of these points will intersect the narrow 
typical set, and the exponentially-growing cost of averaging over the grid
yields worse and worse approximations to expectations.  In general, framing 
algorithms by how they quantify the typical set is a powerful way to quickly 
intuit how an algorithm will perform in practice.

Of course, understanding \emph{why} we want to focus on the typical set in only 
the first step.  \emph{How} to construct an algorithm that can quantify the 
typical set of an arbitrary target distribution is another problem altogether.  
There are many strategies for this task, but one of the most generic, and hence 
most useful in applied practice, is \emph{Markov chain Monte 
Carlo}~\citep{RobertEtAl:1999, BrooksEtAl:2011}.

\section{Markov Chain Monte Carlo}

Markov chain Monte Carlo uses a Markov chain to stochastically explore
the typical set, generating a random grid across the region of high
probability from which we can construct accurate expectation estimates.
Given sufficient computational resources a properly designed Markov
chain will \emph{eventually} explore the typical set of any distribution.
The more practical, and much more challenging question, however, is
whether a given Markov chain will explore a typical set in the finite
time available in a real analysis.

In this section I'll introduce a more substantial definition of Markov 
chain Monte Carlo and discuss both its ideal and pathological behaviors.  
Finally we'll consider how to implement Markov chain Monte Carlo in 
practice and see how fragile of an endeavor it can be. 

\subsection{Estimating Expectations with Markov Chains}

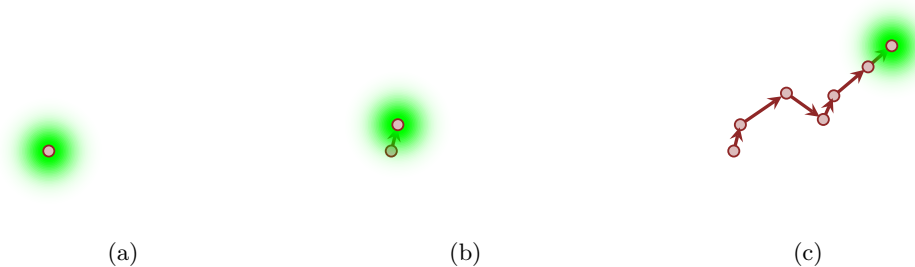
\begin{figure*}
\centering
\subfigure[]{
\begin{tikzpicture}[scale=0.35, thick]
  \draw[white] (-4, -4) rectangle (8, 8);
 
  \foreach \i in {0, 0.02,..., 1} {
    \fill[opacity={exp(-8 * \i)}, green] (-1, -1) circle ({3 * \i});
  }
  \mcpoint{-1}{-1}

\end{tikzpicture}
}
\subfigure[]{
\begin{tikzpicture}[scale=0.35, thick]
   \draw[white] (-4, -4) rectangle (8, 8);
    
  \draw[color=dark, ->, >=stealth, line width=1.25] (-1, -1) -- +(0.9 * 0.25, 0.9 * 1);
  \mcpoint{-1}{-1}

  \foreach \i in {0, 0.02,..., 1} {
    \fill[opacity={exp(-8 * \i)}, green] (-0.75, 0) circle ({3 * \i});
  }
  \mcpoint{-0.75}{0}
  
\end{tikzpicture}
}
\subfigure[]{
\begin{tikzpicture}[scale=0.35, thick]
  \draw[white] (-4, -4) rectangle (8, 8);
   
  \draw[color=dark, ->, >=stealth, line width=1.25] (4.1, 2.2) -- +(0.9 * 0.9, 0.9 * 0.8);
   
  \foreach \i in {0, 0.02,..., 1} {
    \fill[opacity={exp(-8 * \i)}, green] (5, 3) circle ({3 * \i});
  }
  \mcpoint{5}{3}  
   
  \mcpoint{4.1}{2.2}
  \draw[color=dark, ->, >=stealth, line width=1.25] (2.8, 1.1) -- +(0.9 * 1.3, 0.9 * 1.1);
  \mcpoint{2.8}{1.1}
  \draw[color=dark, ->, >=stealth, line width=1.25] (2.4, 0.2) -- +(0.9 * 0.4, 0.9 * 0.9);
  \mcpoint{2.4}{0.2}
  \draw[color=dark, ->, >=stealth, line width=1.25] (1, 1.2) -- +(0.9 * 1.4, -0.9 * 1.0);
  \mcpoint{1}{1.2}
  \draw[color=dark, ->, >=stealth, line width=1.25] (-0.75, 0) -- +(0.9 * 1.75, 0.9 * 1.2);
  \mcpoint{-0.75}{0}
  \draw[color=dark, ->, >=stealth, line width=1.25] (-1, -1) -- +(0.9 * 0.25, 0.9 * 1);
  \mcpoint{-1}{-1}
  
\end{tikzpicture}
}
\caption{(a) A Markov chain is a sequence of points in parameter
space generated by a Markov transition density (green) that defines 
the probability of a new point given the current point.  (b) Sampling 
from that distribution yields a new state in the Markov chain and a 
new distribution from which to sample.  (c) Repeating this process 
generates a Markov chain that meanders through parameter space.
}
\label{fig:markov_chain_cartoon}
\end{figure*}

A \emph{Markov chain} is a progression of points in parameter space 
generated by sequentially applying a random map known as a 
\emph{Markov transition}.  Alternatively, we can think of a Markov
transition as a conditional probability density, $\mathbb{T} ( q' \mid q )$,
defining to which point, $q'$, we are most likely to jump from the initial
point, $q$ (Figure \ref{fig:markov_chain_cartoon}).

An arbitrary Markov chain will simply wander through parameter space and 
will not be of any particular use in computing expectations.  Something 
very special happens, however, if the Markov transition \emph{preserves} 
the target distribution,
\begin{equation*}
\pi \! \left( q \right) 
= 
\int_{\mathcal{Q}} \mathrm{d} q' \, \pi ( q' ) \, \mathbb{T} ( q \mid q' ).
\end{equation*}
More intuitively, this condition implies that if we generated a
ensemble of samples from the target distribution and applied the
transition then we would get a new ensemble that was still distributed
according to the target distribution.  

\begin{figure*}
\centering
\begin{tikzpicture}[scale=0.35, thick]
  \draw[white] (-6, -3) rectangle (12, 6);
  
  \draw[color=dark, ->, >=stealth, line width=1.25] (-4, 1) -- +(0.9 * 2, 0.9 * 0.75);

  \pgfmathsetmacro{\x}{-2}
  \pgfmathsetmacro{\y}{1.75}
  \pgfmathsetmacro{\delta}{3}
  \foreach \i in {0, 0.02,..., 1} {
    \fill[color=green, opacity={exp(-5 * \i)}]
      ({\x - \delta * \i}, \y) 
      .. controls ({\x - 0.5 * \delta * \i}, \y) 
        and ({\x - 0.25 * \delta * \i}, {\y - 0.25 * \delta * \i}) 
      .. ({\x + 0.25 * \delta * \i}, {\y - 0.75 * \delta * \i})
      .. controls ({\x + \delta * \i}, {\y - 0.5 * \delta * \i}) 
         and ({\x + \delta * \i}, {\y + 0.5 * \delta * \i}) 
       .. ({\x + 0.25 * \delta * \i}, {\y + 0.75 * \delta * \i})
       .. controls ({\x - 0.25 * \delta * \i}, {\y + 0.25 * \delta * \i}) 
         and ({\x - 0.5 * \delta * \i}, \y) .. ({\x - \delta * \i}, \y);
  }
  \mcpoint{-2}{1.75}
  \mcpoint{-4}{1}
  
  \draw[color=dark, ->, >=stealth, line width=1.25] (-5.8, 2) -- +(0.9 * 1.8, -0.9 * 1);
  \mcpoint{-5.8}{2}
  
  \draw[color=dark, ->, >=stealth, line width=1.25] (-7, 1.5) -- +(0.9 * 1.2, 0.9 * 0.5);
  \mcpoint{-7}{1.5}
  
  \draw[color=dark, ->, >=stealth, line width=1.25] (-9.6, 3) -- +(0.95 * 2.6, -0.95 * 1.5);
  \mcpoint{-9.6}{3}
  
  \begin{scope}
    \clip (-6, -3) rectangle (12, 6);
    \foreach \i in {0, 0.05,..., 1} {
      \draw[line width={50 * \i}, opacity={exp(-5 * \i)}, dark] 
        (8, 0) .. controls (8, 1.5 * 15) and (18,  1.5 * 5) .. (18,  1.5 * 5)
        .. controls (23,  1.5 * 5) and (28,  1.5 * 8) .. (28, 0)
        .. controls (28, - 1.5 * 8) and (23, - 1.5 * 3) .. (18, - 1.5 * 3)
        .. controls (13, - 1.5 * 3) and (8, - 1.5 * 5) .. (8, 0);
    }  
  \end{scope}
  
\end{tikzpicture}
\caption{When a Markov transition (green) preserves the target distribution, 
it concentrates towards the typical set (red), no matter where it is applied.  
Consequently, the resulting Markov chain will drift into and then across the 
typical set regardless of its initial state, providing a powerful quantification 
of the typical set from which we can derive accurate expectation estimators.
}
\label{fig:probability_attraction}
\end{figure*}
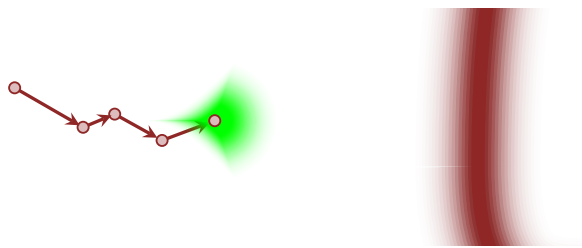

So long as this condition holds, at every initial point the Markov 
transition will concentrate \emph{towards} the typical set.  Consequently, 
no matter where we begin in parameter space the corresponding Markov chain 
will eventually drift into, and then across, the typical set (Figure 
\ref{fig:probability_attraction}).  

Given sufficient time, the history of the Markov chain, 
$\left\{ q_{0}, \ldots, q_{N} \right\}$, denoted \emph{samples} generated by 
the Markov chain, becomes a convenient quantification of the typical set.  In
particular, we can estimate expectations across the typical set, and hence 
expectations across the entire parameter space, by averaging the target function 
over this history,
\begin{equation*}
\hat{f}_{N} = \frac{1}{N} \sum_{n = 0}^{N} f \! \left( q_{n} \right).
\end{equation*}
As we run the Markov chain for longer and longer, it will better explore 
the typical set and, up to some technicalities, these \emph{Markov chain 
Monte Carlo estimators} will converge to the true expectations,
\begin{equation*}
\lim_{N \rightarrow \infty} \hat{f}_{N} = \mathbb{E}_{\pi} \! \left[ f \right].
\end{equation*}

Unfortunately, this asymptotic behavior is of limited use in practice 
because we do not have the infinite computational resources to ensure 
that we can always run a Markov chain long enough to achieve sufficient 
exploration.  In order to develop a robust tool we need to understand 
how Markov chains behave after only a finite number of transitions.

\subsection{Ideal Behavior}

\begin{figure*}
\subfigure[]{
\begin{tikzpicture}[scale=0.25, thick]
  \draw[-,color=white] (-12, 0) to (12, 0);
  
  \begin{scope}
  \clip (-12, -10) rectangle (12, 10);
  
  \foreach \i in {0, 0.05,..., 1} {
    \draw[line width={30 * \i}, opacity={exp(-8 * \i)}, dark] 
      (-10, 0) .. controls (-10, 15) and (-5, 5) .. (0, 5)
      .. controls (5, 5) and (10, 8) .. (10, 0)
      .. controls (10, -8) and (5, -3) .. (0, -3)
      .. controls (-5, -3) and (-10, -5) .. (-10, 0);
  }

  \mcpoint{-10}{-6.5}
  \mcpoint{-9}{-6.75}
  \mcpoint{-8.25}{-5}
  \mcpoint{-7.5}{-4}
  \mcpoint{-7}{-3.5}
  \mcpoint{-6.5}{-3.6}
  
  \end{scope}
  
  \draw[-,color=white] (14, 0) to (38, 0);  
  \node[] at (28,3) {\includegraphics[width=5cm]{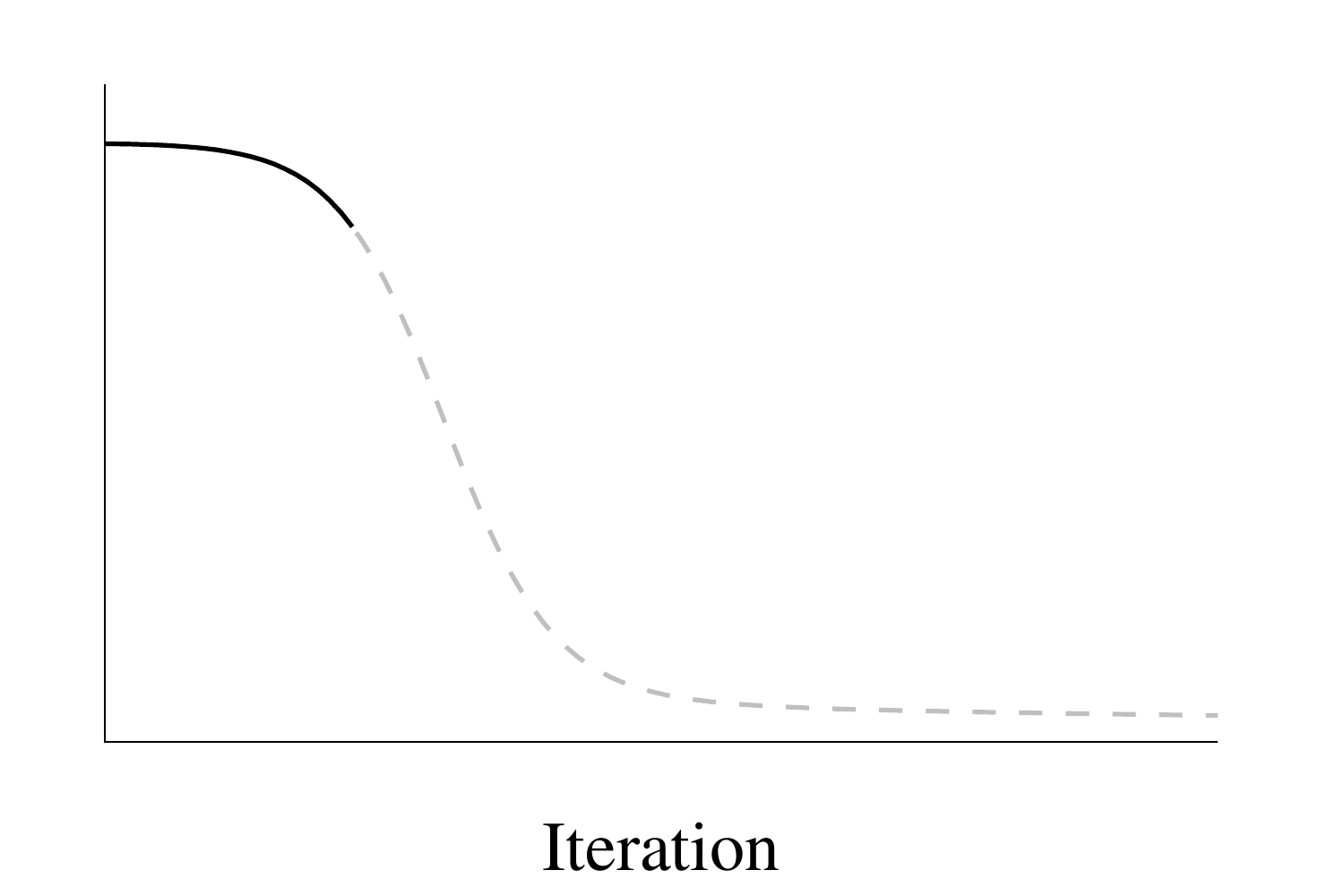}};
  \node[rotate=90] at (15.5,3) { $\left| \mathbb{E} \! \left[ f \right] - \hat{f} \right|$ };
\end{tikzpicture}
}
\subfigure[]{
\begin{tikzpicture}[scale=0.25, thick]
  \draw[-,color=white] (-12, 0) to (12, 0);
  
  \begin{scope}
  \clip (-12, -10) rectangle (12, 10);
  \foreach \i in {0, 0.05,..., 1} {
    \draw[line width={30 * \i}, opacity={exp(-8 * \i)}, dark] 
      (-10, 0) .. controls (-10, 15) and (-5, 5) .. (0, 5)
      .. controls (5, 5) and (10, 8) .. (10, 0)
      .. controls (10, -8) and (5, -3) .. (0, -3)
      .. controls (-5, -3) and (-10, -5) .. (-10, 0);
  }
  
  % Convergence
  \mcpoint{-10}{-6.5}
  \mcpoint{-9}{-6.75}
  \mcpoint{-8.25}{-5}
  \mcpoint{-7.5}{-4}
  \mcpoint{-7}{-3.5}
  \mcpoint{-6.5}{-3.6}
  
  % Mixing
  \mcpoint{-6}{-3}
  %\mcpoint{-5.5}{-3.5}
  %\mcpoint{-5}{-3.75}
  \mcpoint{-4.75}{-3.25}
  %\mcpoint{-5}{-3.5}
  %\mcpoint{-4}{-3.8}
  \mcpoint{-3}{-3.5}
  %\mcpoint{-2.8}{-3}
  %\mcpoint{-2.9}{-3.3}
  \mcpoint{-2}{-3.5}
  %\mcpoint{-1.25}{-3}
  %\mcpoint{-0.5}{-2.75}
  \mcpoint{-0.3}{-3}
  %\mcpoint{0.1}{-2.75}
  %\mcpoint{0.15}{-3.1}
  \mcpoint{0.4}{-3.4}
  %\mcpoint{1.1}{-3.25}
  %\mcpoint{2.7}{-3.45}
  \mcpoint{2.9}{-3.2}
  %\mcpoint{3.1}{-3.7}
  %\mcpoint{4.2}{-3.9}
  \mcpoint{4}{-4.1}
  %\mcpoint{5}{-4}
  %\mcpoint{5.8}{-4.3}
  \mcpoint{6.5}{-4.8}
  %\mcpoint{6.8}{-4.5}
  %\mcpoint{7}{-4.6}
  \mcpoint{6.8}{-4.7}
  %\mcpoint{7.75}{-4.2}
  %\mcpoint{8.5}{-4.25}
  \mcpoint{8.6}{-4.75}
  %\mcpoint{8.8}{-4.5}
  %\mcpoint{9.25}{-3.5}
  \mcpoint{9.65}{-2.75}
  %\mcpoint{9.5}{-2}
  %\mcpoint{9.8}{-3}
  \mcpoint{9.85}{-0.8}
  %\mcpoint{10.1}{-0.4}
  %\mcpoint{10.25}{0.3}
  \mcpoint{10}{0.7}
  %\mcpoint{9.7}{0.9}
  %\mcpoint{10.3}{1.4}
  \mcpoint{9.8}{2.25}
  %\mcpoint{10}{2.5}
  %\mcpoint{10}{3}
  \mcpoint{9.6}{3.75}
  %\mcpoint{9.1}{4.4}
  %\mcpoint{9.1}{4.1}
  \mcpoint{8.5}{4.75}
  %\mcpoint{8.1}{5.75}
  %\mcpoint{7.7}{5.2}
  \mcpoint{7.3}{5.3}
  %\mcpoint{6.6}{5.8}
  %\mcpoint{6}{6}
  \mcpoint{5.25}{5.45}
  %\mcpoint{4.75}{5.75}
  %\mcpoint{3.5}{5.28}
  \mcpoint{5}{5.65}
  %\mcpoint{3}{5.5}
  %\mcpoint{2.75}{5.3}
  \mcpoint{2.25}{4.8}
  %\mcpoint{2}{5.2}
  %\mcpoint{1.75}{4.9}
  \mcpoint{1.5}{5.1}
  %\mcpoint{0.8}{5.2}
  %\mcpoint{0}{4.7}
  \mcpoint{-0.6}{5}
  %\mcpoint{-1.25}{5.2}
  %\mcpoint{-1.55}{5.4}
  \mcpoint{-1.85}{5}
  %\mcpoint{-2.75}{5.5}
  %\mcpoint{-3.5}{6.6}
  \mcpoint{-4.25}{6.5} 
  %\mcpoint{-4.9}{7.2}
  %\mcpoint{-5.2}{7}
  \mcpoint{-5.75}{7.4}
  %\mcpoint{-6}{7.9}
  %\mcpoint{-6.25}{7.8}
  \mcpoint{-6.5}{7.5}  
  %\mcpoint{-6.75}{8} 
  %\mcpoint{-7.5}{8.5}  
  \mcpoint{-8.25}{8}  
  %\mcpoint{-8.65}{7.8}  
  %\mcpoint{-8.4}{7.7}
  \mcpoint{-9.2}{7}  
  %\mcpoint{-9.5}{6.75}  
  %\mcpoint{-10}{6}
  \mcpoint{-9.7}{5.3}
  %\mcpoint{-9.6}{5}  
  %\mcpoint{-9.9}{4.75}
  \mcpoint{-10.1}{4}
  %\mcpoint{-9.7}{3.5}  
  %\mcpoint{-9.8}{3}  
  \mcpoint{-10}{2.2}  
  %\mcpoint{-10.25}{1.9}  
  %\mcpoint{-9.7}{1.8} 
  \mcpoint{-10.2}{1.4}  
  %\mcpoint{-9.8}{0.7}  
  %\mcpoint{-10.2}{0.1}  
  \mcpoint{-9.9}{-0.1} 
  %\mcpoint{-10}{-0.9}
  %\mcpoint{-9.6}{-1.5} 
  \mcpoint{-9.5}{-1.8}  
  %\mcpoint{-9.7}{-2.1}
  %\mcpoint{-9.3}{-2.8}
  \mcpoint{-8.3}{-3}
  %\mcpoint{-7.8}{-3.4}  
  %\mcpoint{-7.6}{-3}
   
  \end{scope}
  
  \draw[-,color=white] (14, 0) to (38, 0);  
  \node[] at (28,3) {\includegraphics[width=5cm]{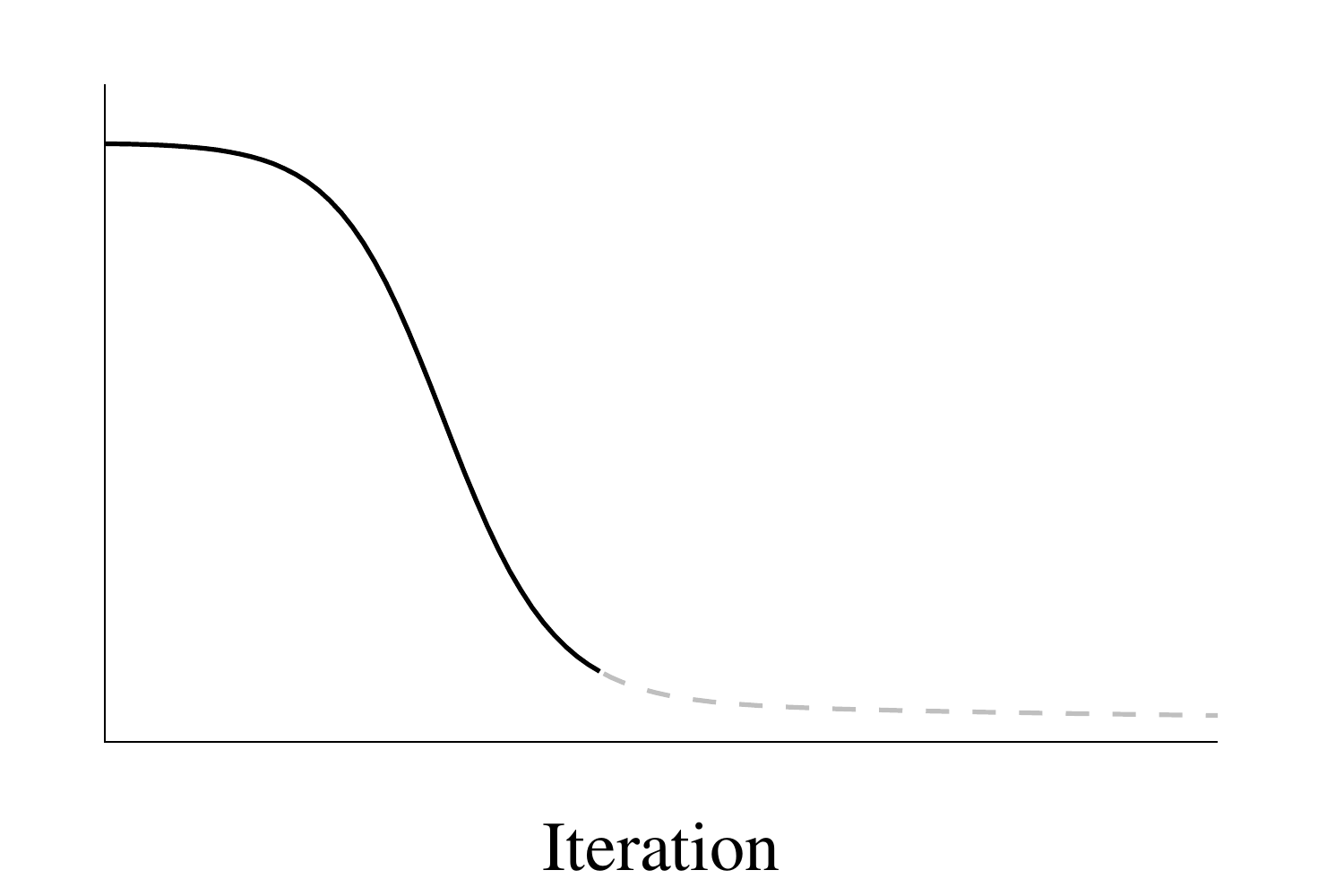}};
  \node[rotate=90] at (15.5,3) { $\left| \mathbb{E} \! \left[ f \right] - \hat{f} \right|$ };
\end{tikzpicture}
}
\subfigure[]{
\begin{tikzpicture}[scale=0.25, thick]
  \draw[-,color=white] (-12, 0) to (12, 0);
  
  \begin{scope}
  \clip (-12, -10) rectangle (12, 10);
  \foreach \i in {0, 0.05,..., 1} {
    \draw[line width={30 * \i}, opacity={exp(-8 * \i)}, dark] 
      (-10, 0) .. controls (-10, 15) and (-5, 5) .. (0, 5)
      .. controls (5, 5) and (10, 8) .. (10, 0)
      .. controls (10, -8) and (5, -3) .. (0, -3)
      .. controls (-5, -3) and (-10, -5) .. (-10, 0);
  }
  
  % Convergence
  \mcpoint{-10}{-6.5}
  \mcpoint{-9}{-6.75}
  \mcpoint{-8.25}{-5}
  \mcpoint{-7.5}{-4}
  \mcpoint{-7}{-3.5}
  \mcpoint{-6.5}{-3.6}
   
  % Mixing
  \mcpoint{-6}{-3}
  \mcpoint{-5.5}{-3.5}
  \mcpoint{-5}{-3.75}
  \mcpoint{-4.75}{-3.25}
  \mcpoint{-5}{-3.5}
  \mcpoint{-4}{-3.8}
  \mcpoint{-3}{-3.5}
  \mcpoint{-2.8}{-3}
  \mcpoint{-2.9}{-3.3}
  \mcpoint{-2}{-3.5}
  \mcpoint{-1.25}{-3}
  \mcpoint{-0.5}{-2.75}
  \mcpoint{-0.3}{-3}
  \mcpoint{0.1}{-2.75}
  \mcpoint{0.15}{-3.1}
  \mcpoint{0.4}{-3.4}
  \mcpoint{1.1}{-3.25}
  \mcpoint{2.7}{-3.45}
  \mcpoint{2.9}{-3.2}
  \mcpoint{3.1}{-3.7}
  \mcpoint{4.2}{-3.9}
  \mcpoint{4}{-4.1}
  \mcpoint{5}{-4}
  \mcpoint{5.8}{-4.3}
  \mcpoint{6.5}{-4.8}
  \mcpoint{6.8}{-4.5}
  \mcpoint{7}{-4.6}
  \mcpoint{6.8}{-4.7}
  \mcpoint{7.75}{-4.2}
  \mcpoint{8.5}{-4.25}
  \mcpoint{8.6}{-4.75}
  \mcpoint{8.8}{-4.5}
  \mcpoint{9.25}{-3.5}
  \mcpoint{9.65}{-2.75}
  \mcpoint{9.5}{-2}
  \mcpoint{9.8}{-3}
  \mcpoint{9.85}{-0.8}
  \mcpoint{10.1}{-0.4}
  \mcpoint{10.25}{0.3}
  \mcpoint{10}{0.7}
  \mcpoint{9.7}{0.9}
  \mcpoint{10.3}{1.4}
  \mcpoint{9.8}{2.25}
  \mcpoint{10}{2.5}
  \mcpoint{10}{3}
  \mcpoint{9.6}{3.75}
  \mcpoint{9.1}{4.4}
  \mcpoint{9.1}{4.1}
  \mcpoint{8.5}{4.75}
  \mcpoint{8.1}{5.75}
  \mcpoint{7.7}{5.2}
  \mcpoint{7.3}{5.3}
  \mcpoint{6.6}{5.8}
  \mcpoint{6}{6}
  \mcpoint{5.25}{5.45}
  \mcpoint{4.75}{5.75}
  \mcpoint{3.5}{5.28}
  \mcpoint{5}{5.65}
  \mcpoint{3}{5.5}
  \mcpoint{2.75}{5.3}
  \mcpoint{2.25}{4.8}
  \mcpoint{2}{5.2}
  \mcpoint{1.75}{4.9}
  \mcpoint{1.5}{5.1}
  \mcpoint{0.8}{5.2}
  \mcpoint{0}{4.7}
  \mcpoint{-0.6}{5}
  \mcpoint{-1.25}{5.2}
  \mcpoint{-1.55}{5.4}
  \mcpoint{-1.85}{5}
  \mcpoint{-2.75}{5.5}
  \mcpoint{-3.5}{6.6}
  \mcpoint{-4.25}{6.5} 
  \mcpoint{-4.9}{7.2}
  \mcpoint{-5.2}{7}
  \mcpoint{-5.75}{7.4}
  \mcpoint{-6}{7.9}
  \mcpoint{-6.25}{7.8}
  \mcpoint{-6.5}{7.5}  
  \mcpoint{-6.75}{8} 
  \mcpoint{-7.5}{8.5}  
  \mcpoint{-8.25}{8}  
  \mcpoint{-8.65}{7.8}  
  \mcpoint{-8.4}{7.7}
  \mcpoint{-9.2}{7}  
  \mcpoint{-9.5}{6.75}  
  \mcpoint{-10}{6}
  \mcpoint{-9.7}{5.3}
  \mcpoint{-9.6}{5}  
  \mcpoint{-9.9}{4.75}
  \mcpoint{-10.1}{4}
  \mcpoint{-9.7}{3.5}  
  \mcpoint{-9.8}{3}  
  \mcpoint{-10}{2.2}  
  \mcpoint{-10.25}{1.9}  
  \mcpoint{-9.7}{1.8} 
  \mcpoint{-10.2}{1.4}  
  \mcpoint{-9.8}{0.7}  
  \mcpoint{-10.2}{0.1}  
  \mcpoint{-9.9}{-0.1} 
  \mcpoint{-10}{-0.9}
  \mcpoint{-9.6}{-1.5} 
  \mcpoint{-9.5}{-1.8}  
  \mcpoint{-9.7}{-2.1}
  \mcpoint{-9.3}{-2.8}
  \mcpoint{-8.3}{-3}
  \mcpoint{-7.8}{-3.4}  
  \mcpoint{-7.6}{-3}
  
  \end{scope}
  
  \draw[-,color=white] (14, 0) to (38, 0);  
  \node[] at (28,3) {\includegraphics[width=5cm]{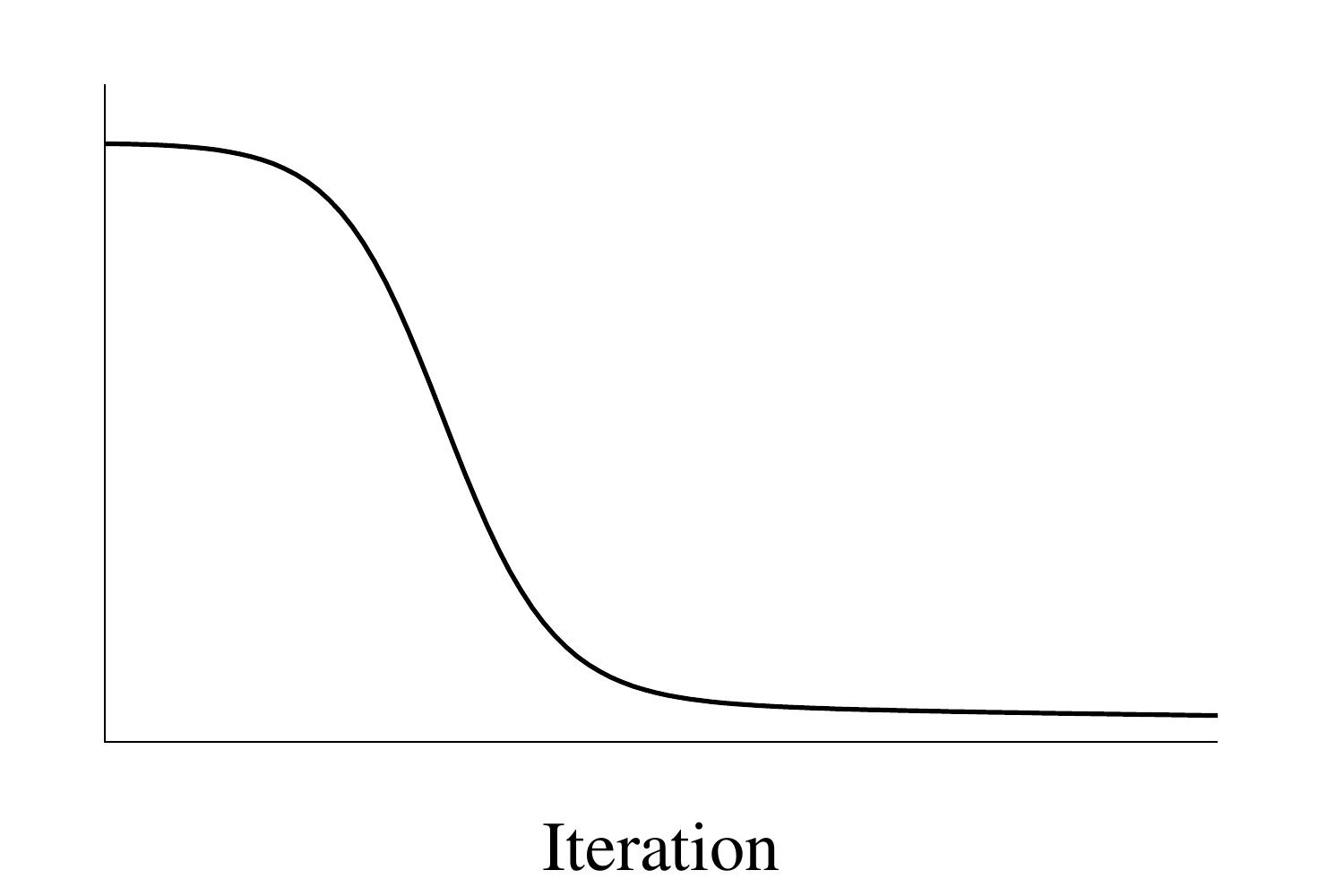}};
  \node[rotate=90] at (15.5,3) { $\left| \mathbb{E} \! \left[ f \right] - \hat{f} \right|$ };
\end{tikzpicture}
}
\caption{Under ideal circumstances, a Markov chain explores the target 
distribution in three phases.  (a) First the Markov chain converges to the 
typical set and estimators suffer from initial but ultimately transient 
biases. (b) Once the Markov chain finds the typical set and makes the first 
sojourn through it, this initial bias rapidly vanishes and the estimators 
become much more accurate.  (c) As the Markov chain continues it explores 
more details of the typical set, gradually reducing the precision of the 
Markov chain Monte Carlo estimators towards zero.}
\label{fig:ideal_mcmc}
\end{figure*}
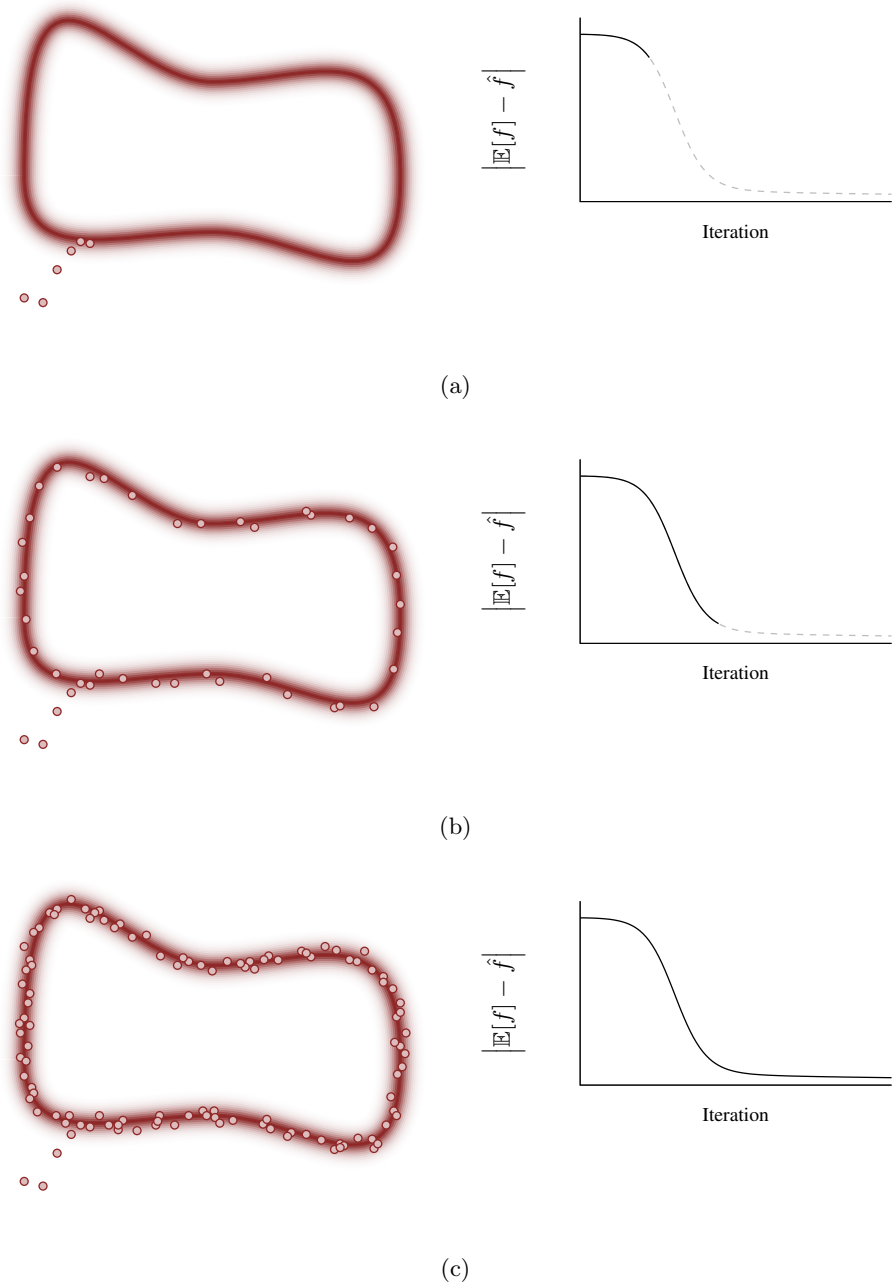

Under ideal conditions, Markov chains explore the target distribution 
in three distinct phases.  In the first phase the Markov chain converges 
towards the typical set from its initial position in parameter space while 
the Markov chain Monte Carlo estimators suffer from strong biases (Figure 
\ref{fig:ideal_mcmc}a).  The second phase begins once the Markov chain 
finds the typical set and persists through the first sojourn across the 
typical set.  This initial exploration is extremely effective and the 
accuracy of Markov chain Monte Carlo estimators rapidly improves as the 
bias from the initial samples is eliminated (Figure \ref{fig:ideal_mcmc}b).  
The third phase consists of all subsequent exploration where the Markov 
chain refines its exploration of the typical set and the precision of the 
Markov chain Monte Carlo estimators improves, albeit at a slower rate 
(Figure \ref{fig:ideal_mcmc}c).

Once the Markov chain has entered into this third phase the Markov chain 
Monte Carlo estimators satisfy a Central Limit Theorem
\begin{equation*}
\hat{f}_{N}^{\mathrm{MCMC}} \sim 
\mathcal{N} \! \left( \mathbb{E}_{\pi} [ f ],
\text{MCMC-SE} \right),
\end{equation*}
where the \emph{Markov Chain Monte Carlo Standard Error} is 
given by
\begin{equation*}
\text{MCMC-SE} \equiv \sqrt{ \frac{ \mathrm{Var}_{\pi} [ f ] }{\mathrm{ESS} } }.
\end{equation*}

The \emph{effective sample size} is defined as
\begin{equation*}
\mathrm{ESS} = 
\frac{N}
{ 1 + 2 \sum_{l = 1}^{\infty} \rho_{l} },
\end{equation*}
where $\rho_{l}$ is the lag-$l$ autocorrelation of $f$ over the history
of the Markov chain.  The effective sample size quantifies the number of 
exact samples from the target distribution necessary to give an equivalent 
estimator precision and hence the effective number of exact samples 
``contained'' in the Markov chain; we can also interpret the effective 
sample size as the total number of sojourns the Markov chain has made 
across the typical set.  In practice the effective sample size can be 
estimated from the Markov chain itself, although care must be taken to 
avoid biases~\citep{Geyer:1992, GelmanEtAl:2014a}.

Because the states of the Markov chain generated during the initial 
convergence phase mostly bias Markov chain Monte Carlo estimators, we 
can drastically improve the precision of these estimators by using only
those samples generated once the Markov chain has begun to explore the
typical set.  Consequently, it is common practice to \emph{warm up} the
Markov chain by throwing away those initial converging samples before 
computing Markov chain Monte Carlo estimators.  Warm-up can also be
extended to allow for any degrees of freedom in the Markov transition 
to be empirically optimized without biasing the subsequent estimators.

\subsection{Pathological Behavior}
\label{sec:pathological_behavior}

Unfortunately, this idealized behavior requires that the Markov 
transition is compatible with the structure of the target distribution.  
When the target distribution exhibits pathological behavior, however,
Markov transitions will have trouble exploring and Markov chain Monte 
Carlo will fail.

\begin{figure*}
\centering
\begin{tikzpicture}[scale=0.3, thick]

\foreach \i in {0, 0.05,..., 1} {
  \begin{scope}
    \clip (-0.005, -9.5) rectangle (11, 6.5);
    \draw[line width={30 * \i}, opacity={exp(-8 * \i)}, dark] 
    (0, 5) .. controls (5, 5) and (10, 8) .. (10, 0)
             .. controls (10, -8) and (5, -3) .. (-2, -10);
  \end{scope}
  
  \begin{scope}
    \clip (0.005, -9.5) rectangle (-11, 6.5);
    \draw[line width={30 * \i}, opacity={exp(-8 * \i)}, dark] 
    (0, 5) .. controls (-5, 5) and (-10, 8) .. (-10, 0)
             .. controls (-10, -8) and (-5, -3) .. (2, -10);
  \end{scope}
}  
\fill[opacity=0.3, green] (0, -8.5) circle (1);
\end{tikzpicture}
\caption{Markov chains typically have trouble exploring regions of the 
typical set with large curvature (green).  Incomplete exploration biases
Markov chain Monte Carlo estimators and spoils critical results such as 
Central Limit Theorems.}
\label{fig:pathological_typical_set}
\end{figure*}
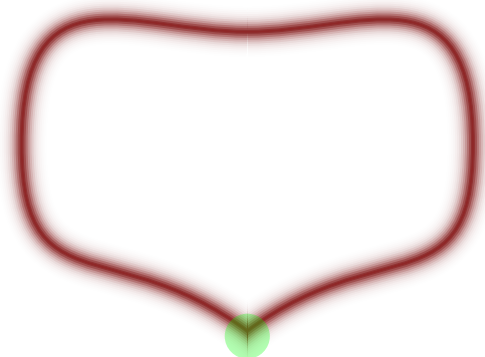

Consider, for example, a target probability distribution where the 
typical set pinches into a region of high curvature (Figure 
\ref{fig:pathological_typical_set}).  Most Markov transitions are not 
able to resolve these details and hence they cannot maneuver into these 
tight regions.  The resulting Markov chains simply ignore them, biasing 
subsequent Markov chain Monte Carlo estimators due to the incomplete 
exploration.  It is as if there are thin but deep cracks across a surface 
in parameter space hiding a significant amount of probability that the 
Markov chains pass right over and miss entirely.

\begin{figure*}
\subfigure[]{
\begin{tikzpicture}[scale=0.20, thick]
  \draw[-,color=white] (-11, 0) to (12, 0);
  
  \foreach \i in {0, 0.05,..., 1} {
    \begin{scope}
      \clip (-0.005, -9.5) rectangle (11, 6.5);
      \draw[line width={30 * \i}, opacity={exp(-8 * \i)}, dark] 
      (0, 5) .. controls (5, 5) and (10, 8) .. (10, 0)
               .. controls (10, -8) and (5, -3) .. (-2, -10);
    \end{scope}
  
    \begin{scope}
      \clip (0.005, -9.5) rectangle (-11, 6.5);
      \draw[line width={30 * \i}, opacity={exp(-8 * \i)}, dark] 
      (0, 5) .. controls (-5, 5) and (-10, 8) .. (-10, 0)
               .. controls (-10, -8) and (-5, -3) .. (2, -10);
    \end{scope}
  }  
  \fill[opacity=0.3, green] (0, -8.5) circle (1);
  
  \draw[<->, >=stealth, line width=0.5] (-11, -9) -- (-8, -9)
  node[right] {$q_{1}$};
  
  \draw[<->, >=stealth, line width=0.5] (-10, -10) -- (-10, -7)
  node[above] {$q_{2}$};
  
  \mcpoint{-8}{5.5}
  \mcpoint{-9}{4.75}
  \mcpoint{-8.7}{4.6}
  \mcpoint{-9.1}{4.5}
  \mcpoint{-9.2}{3.5}
  \mcpoint{-9.9}{3}
  \mcpoint{-9.6}{1.5}
  \mcpoint{-9.6}{1.5}
  \mcpoint{-9.75}{1}
  \mcpoint{-10.2}{0.25}
  \mcpoint{-9.8}{0}
  \mcpoint{-10}{-0.5}
  \mcpoint{-9.75}{-2}
  \mcpoint{-9.9}{-3}
  \mcpoint{-9.25}{-3.5}
  \mcpoint{-9}{-4.25}  
  \mcpoint{-8}{-4.5}
  \mcpoint{-8.25}{-4.75}
  \mcpoint{-7.75}{-5}
  \mcpoint{-6.5}{-5.5}
  \mcpoint{-5.75}{-5.25}
  \mcpoint{-5}{-6}
  \mcpoint{-4.75}{-6.25}
  \mcpoint{-3.8}{-6}
  \mcpoint{-3.25}{-6.5}
  \mcpoint{-2.75}{-6.75}
  \mcpoint{-2}{-7.5}
  \mcpoint{-1.75}{-6.8}
  \mcpoint{-1.5}{-7.2}

  \draw[-,color=white] (12, 0) to (38, 0);  
  \node[] at (26,-2) {\includegraphics[width=4.8cm]{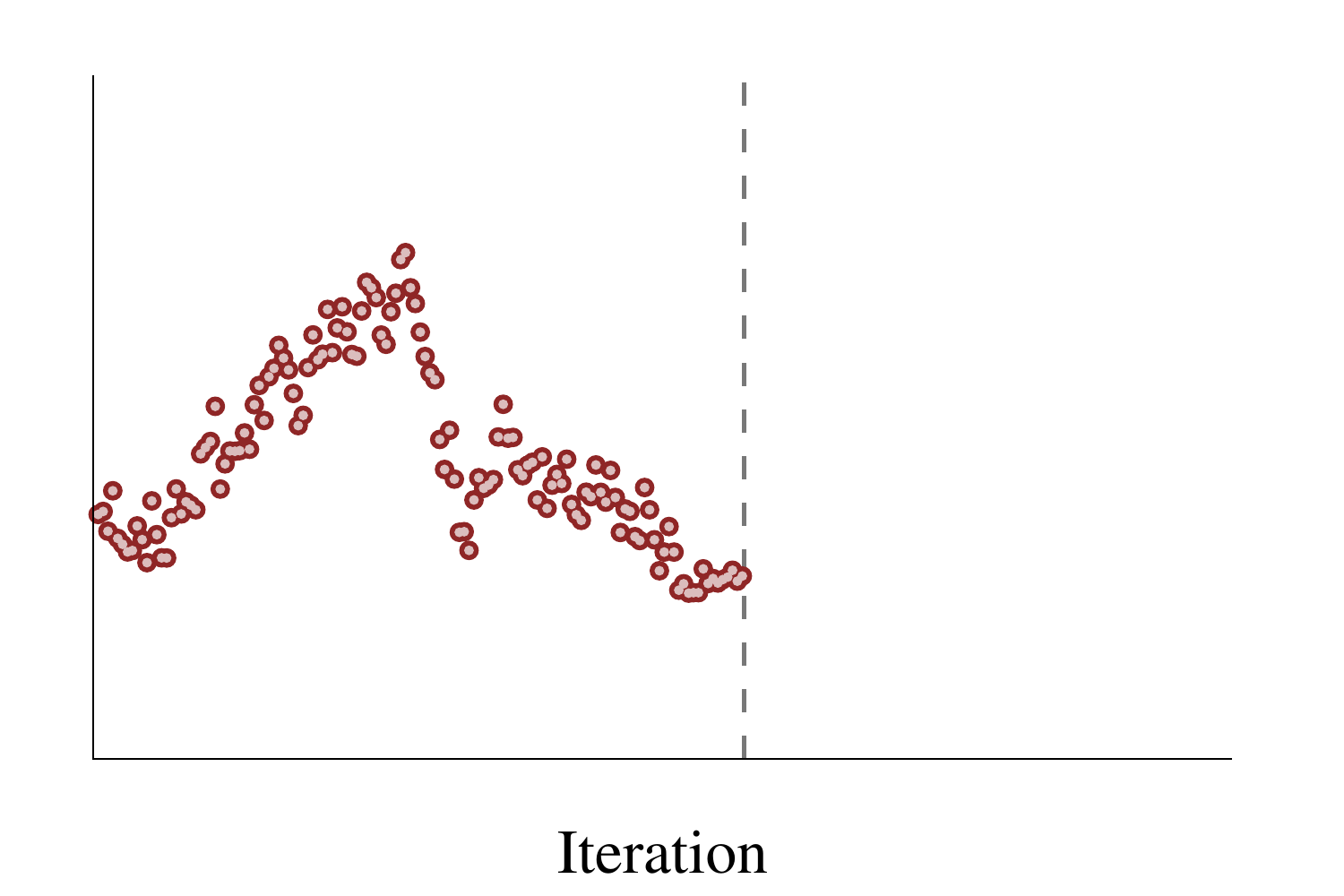}};
  \node[rotate=90] at (14, -1) { $q_{2}$ };
  
  \draw[-,color=white] (39, 0) to (64, 0);  
  \node[] at (53,-2) {\includegraphics[width=4.8cm]{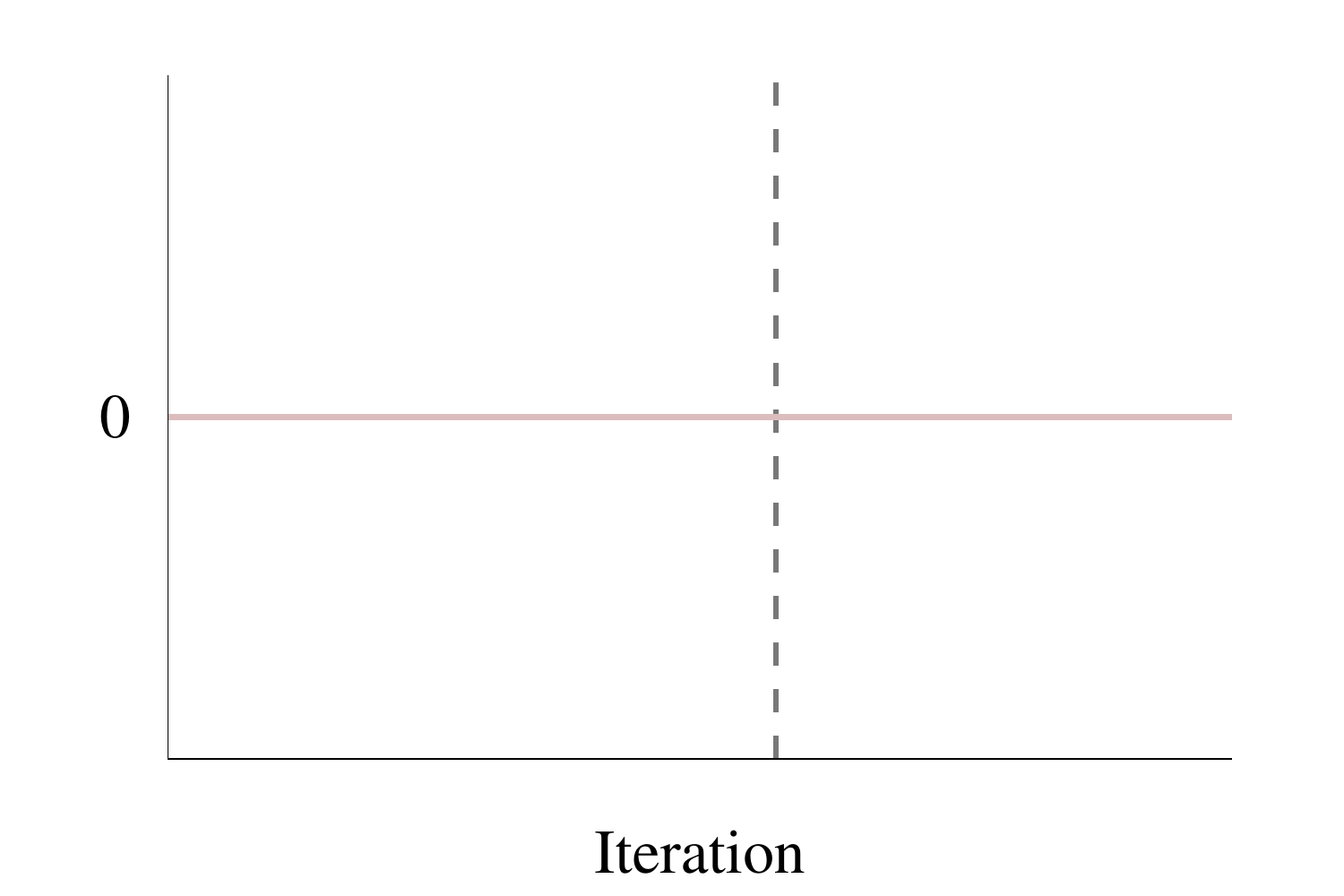}};
  \node[rotate=90] at (41, -1) { $\left| \mathbb{E} \! \left[ q_{2} \right] - \hat{q_{2}} \right|$ };
  
  \draw[-,color=black, line width=0.5] (44, 4)
                                       .. controls (46, 1.5) and (52, 1.5) 
                                       .. (54.9, 1.5);
\end{tikzpicture}
}
\subfigure[]{
\begin{tikzpicture}[scale=0.20, thick]
  \draw[-,color=white] (-11, 0) to (12, 0);
  
  \foreach \i in {0, 0.05,..., 1} {
    \begin{scope}
      \clip (-0.005, -9.5) rectangle (11, 6.5);
      \draw[line width={30 * \i}, opacity={exp(-8 * \i)}, dark] 
      (0, 5) .. controls (5, 5) and (10, 8) .. (10, 0)
               .. controls (10, -8) and (5, -3) .. (-2, -10);
    \end{scope}
  
    \begin{scope}
      \clip (0.005, -9.5) rectangle (-11, 6.5);
      \draw[line width={30 * \i}, opacity={exp(-8 * \i)}, dark] 
      (0, 5) .. controls (-5, 5) and (-10, 8) .. (-10, 0)
               .. controls (-10, -8) and (-5, -3) .. (2, -10);
    \end{scope}
  }  
  \fill[opacity=0.3, green] (0, -8.5) circle (1);
  
  \draw[<->, >=stealth, line width=0.5] (-11, -9) -- (-8, -9)
  node[right] {$q_{1}$};
  
  \draw[<->, >=stealth, line width=0.5] (-10, -10) -- (-10, -7)
  node[above] {$q_{2}$};
  
  \mcpoint{-8}{5.5}
  \mcpoint{-9}{4.75}
  \mcpoint{-8.7}{4.6}
  \mcpoint{-9.1}{4.5}
  \mcpoint{-9.2}{3.5}
  \mcpoint{-9.9}{3}
  \mcpoint{-9.6}{1.5}
  \mcpoint{-9.6}{1.5}
  \mcpoint{-9.75}{1}
  \mcpoint{-10.2}{0.25}
  \mcpoint{-9.8}{0}
  \mcpoint{-10}{-0.5}
  \mcpoint{-9.75}{-2}
  \mcpoint{-9.9}{-3}
  \mcpoint{-9.25}{-3.5}
  \mcpoint{-9}{-4.25}  
  \mcpoint{-8}{-4.5}
  \mcpoint{-8.25}{-4.75}
  \mcpoint{-7.75}{-5}
  \mcpoint{-6.5}{-5.5}
  \mcpoint{-5.75}{-5.25}
  \mcpoint{-5}{-6}
  \mcpoint{-4.75}{-6.25}
  \mcpoint{-3.8}{-6}
  \mcpoint{-3.25}{-6.5}
  \mcpoint{-2.75}{-6.75}
  \mcpoint{-2}{-7.5}
  \mcpoint{-1.75}{-6.8}
  \mcpoint{-1.5}{-7.2}
  
  \mcpoint{-0.5}{-7.5}
  \mcpoint{0}{-7.6}
  \mcpoint{1}{-7.7}

  \draw[-,color=white] (12, 0) to (38, 0);  
  \node[] at (26,-2) {\includegraphics[width=4.8cm]{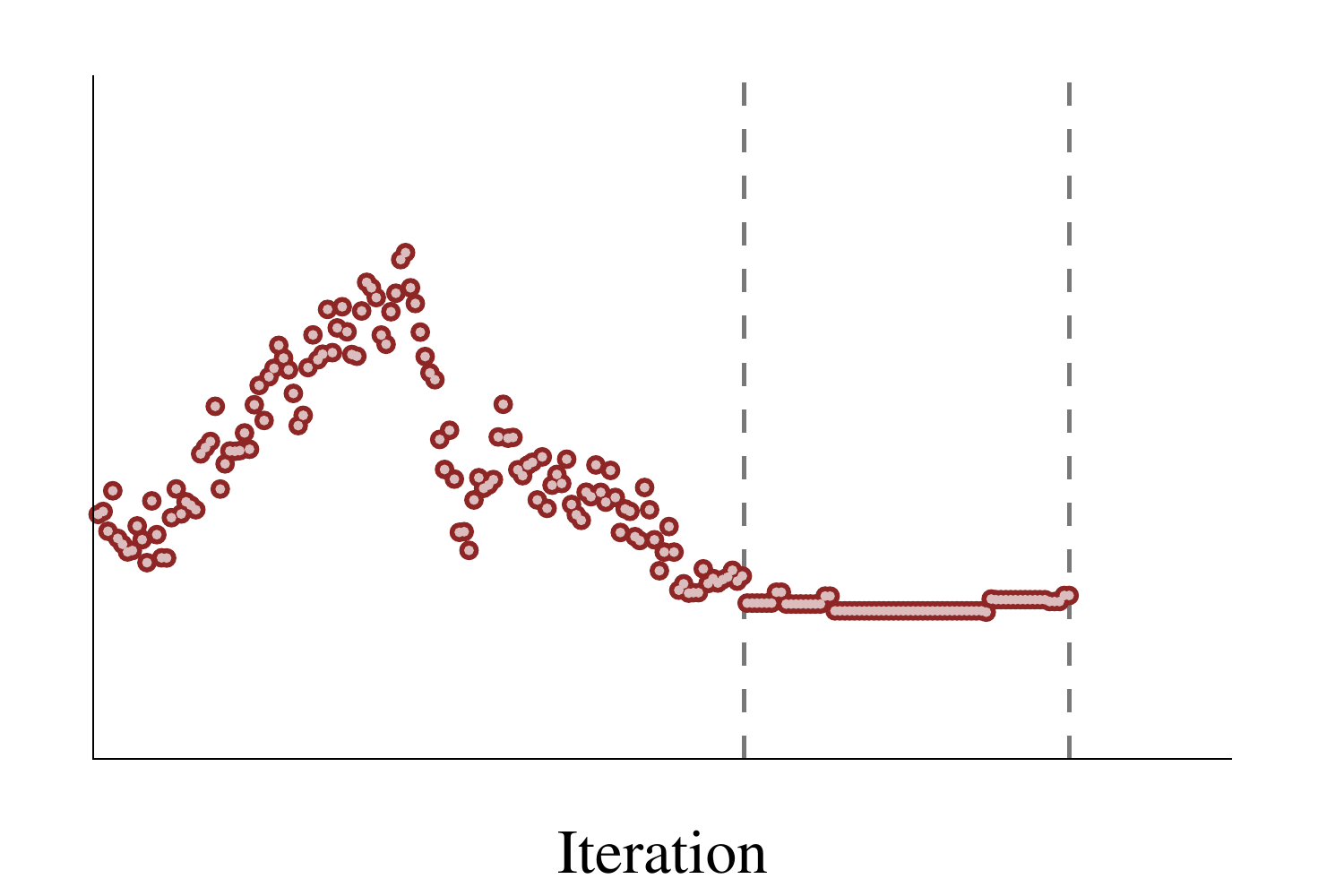}};
  \node[rotate=90] at (14, -1) { $q_{2}$ };
  
  \draw[-,color=white] (39, 0) to (64, 0);  
  \node[] at (53,-2) {\includegraphics[width=4.8cm]{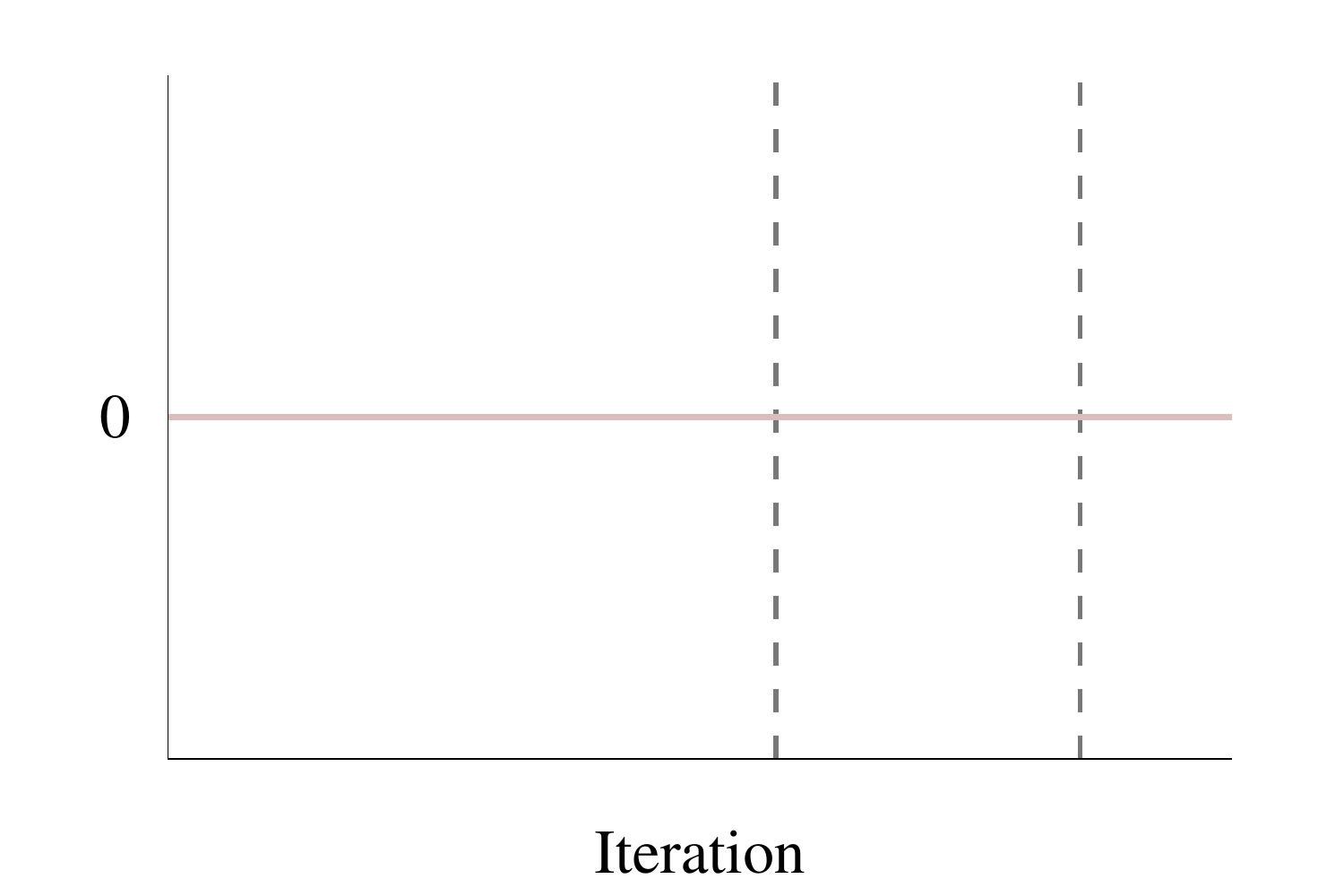}};
  \node[rotate=90] at (41, -1) { $\left| \mathbb{E} \! \left[ q_{2} \right] - \hat{q_{2}} \right|$ };
  
  \draw[-,color=black, line width=0.5] (44, 4)
                                       .. controls (46, 1.5) and (52, 1.5) 
                                       .. (54.9, 1.5);
  \draw[-,color=black, line width=0.5] (54.9, 1.5) 
                                       .. controls (57.8, 1.5) and (58.25, -1.5) 
                                       .. (60.25, -3.5);
\end{tikzpicture}
}
\subfigure[]{
\begin{tikzpicture}[scale=0.20, thick]
  \draw[-,color=white] (-11, 0) to (12, 0);
  
  \foreach \i in {0, 0.05,..., 1} {
    \begin{scope}
      \clip (-0.005, -9.5) rectangle (11, 6.5);
      \draw[line width={30 * \i}, opacity={exp(-8 * \i)}, dark] 
      (0, 5) .. controls (5, 5) and (10, 8) .. (10, 0)
               .. controls (10, -8) and (5, -3) .. (-2, -10);
    \end{scope}
  
    \begin{scope}
      \clip (0.005, -9.5) rectangle (-11, 6.5);
      \draw[line width={30 * \i}, opacity={exp(-8 * \i)}, dark] 
      (0, 5) .. controls (-5, 5) and (-10, 8) .. (-10, 0)
               .. controls (-10, -8) and (-5, -3) .. (2, -10);
    \end{scope}
  }  
  \fill[opacity=0.3, green] (0, -8.5) circle (1);
  
  \draw[<->, >=stealth, line width=0.5] (-11, -9) -- (-8, -9)
  node[right] {$q_{1}$};
  
  \draw[<->, >=stealth, line width=0.5] (-10, -10) -- (-10, -7)
  node[above] {$q_{2}$};
  
  \mcpoint{-8}{5.5}
  \mcpoint{-9}{4.75}
  \mcpoint{-8.7}{4.6}
  \mcpoint{-9.1}{4.5}
  \mcpoint{-9.2}{3.5}
  \mcpoint{-9.9}{3}
  \mcpoint{-9.6}{1.5}
  \mcpoint{-9.6}{1.5}
  \mcpoint{-9.75}{1}
  \mcpoint{-10.2}{0.25}
  \mcpoint{-9.8}{0}
  \mcpoint{-10}{-0.5}
  \mcpoint{-9.75}{-2}
  \mcpoint{-9.9}{-3}
  \mcpoint{-9.25}{-3.5}
  \mcpoint{-9}{-4.25}  
  \mcpoint{-8}{-4.5}
  \mcpoint{-8.25}{-4.75}
  \mcpoint{-7.75}{-5}
  \mcpoint{-6.5}{-5.5}
  \mcpoint{-5.75}{-5.25}
  \mcpoint{-5}{-6}
  \mcpoint{-4.75}{-6.25}
  \mcpoint{-3.8}{-6}
  \mcpoint{-3.25}{-6.5}
  \mcpoint{-2.75}{-6.75}
  \mcpoint{-2}{-7.5}
  \mcpoint{-1.75}{-6.8}
  \mcpoint{-1.5}{-7.2}
  
  \mcpoint{-0.5}{-7.5}
  \mcpoint{0}{-7.6}
  \mcpoint{1}{-7.7}
  
  \mcpoint{9}{4.7}
  \mcpoint{9.1}{4.25}
  \mcpoint{9.8}{3}
  \mcpoint{9.9}{1.5}
  \mcpoint{9.75}{-0.5}
  \mcpoint{9.9}{-1}
  \mcpoint{9.6}{-3.5}
  \mcpoint{7.8}{-4.5}
  \mcpoint{7}{-5}
  \mcpoint{6.5}{-5.5}
  \mcpoint{4.25}{-6.5}
  \mcpoint{2.1}{-6.5}
  
  \draw[-,color=white] (12, 0) to (38, 0);  
  \node[] at (26,-2) {\includegraphics[width=4.8cm]{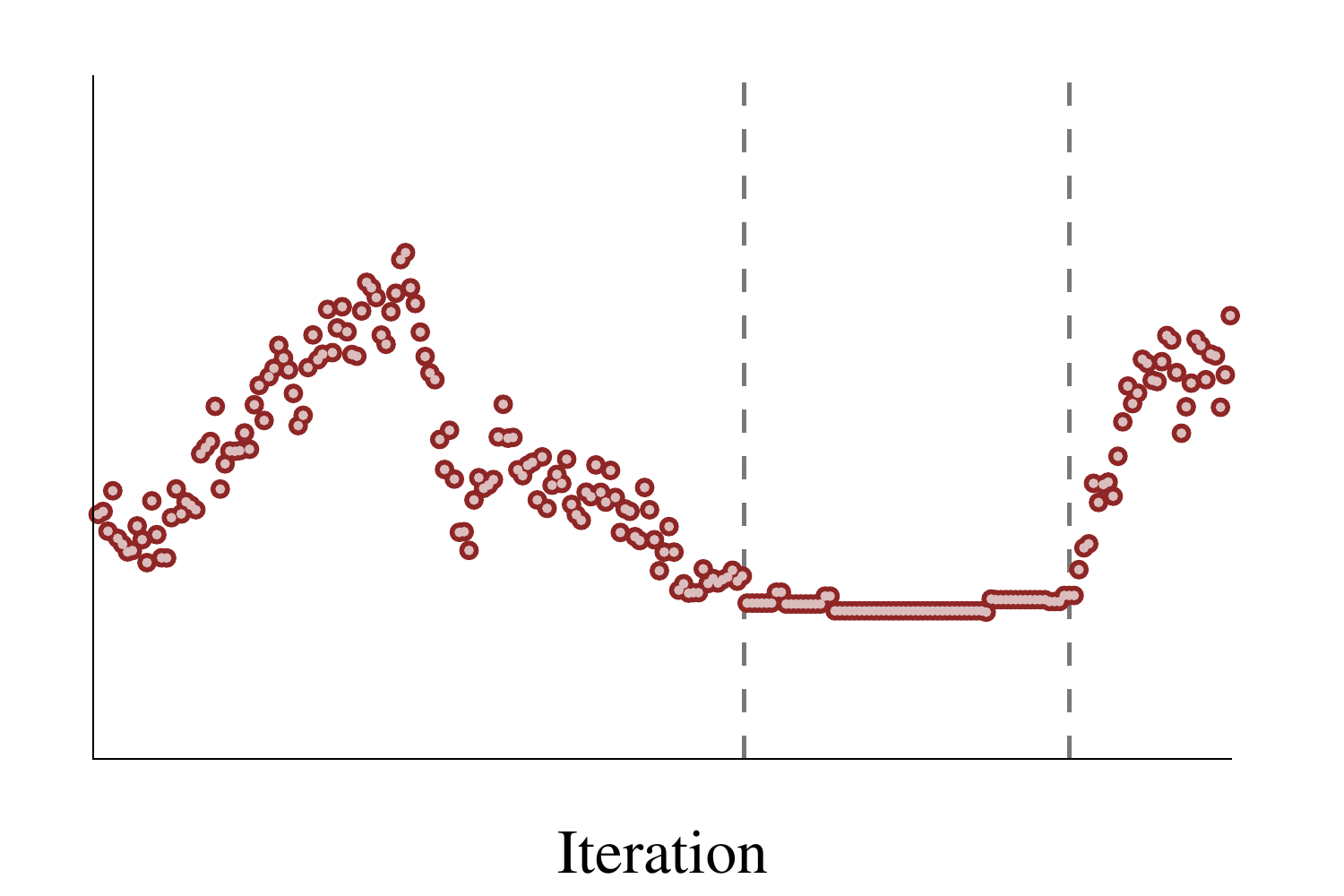}};
  \node[rotate=90] at (14, -1) { $q_{2}$ };
  
  \draw[-,color=white] (39, 0) to (64, 0);  
  \node[] at (53,-2) {\includegraphics[width=4.8cm]{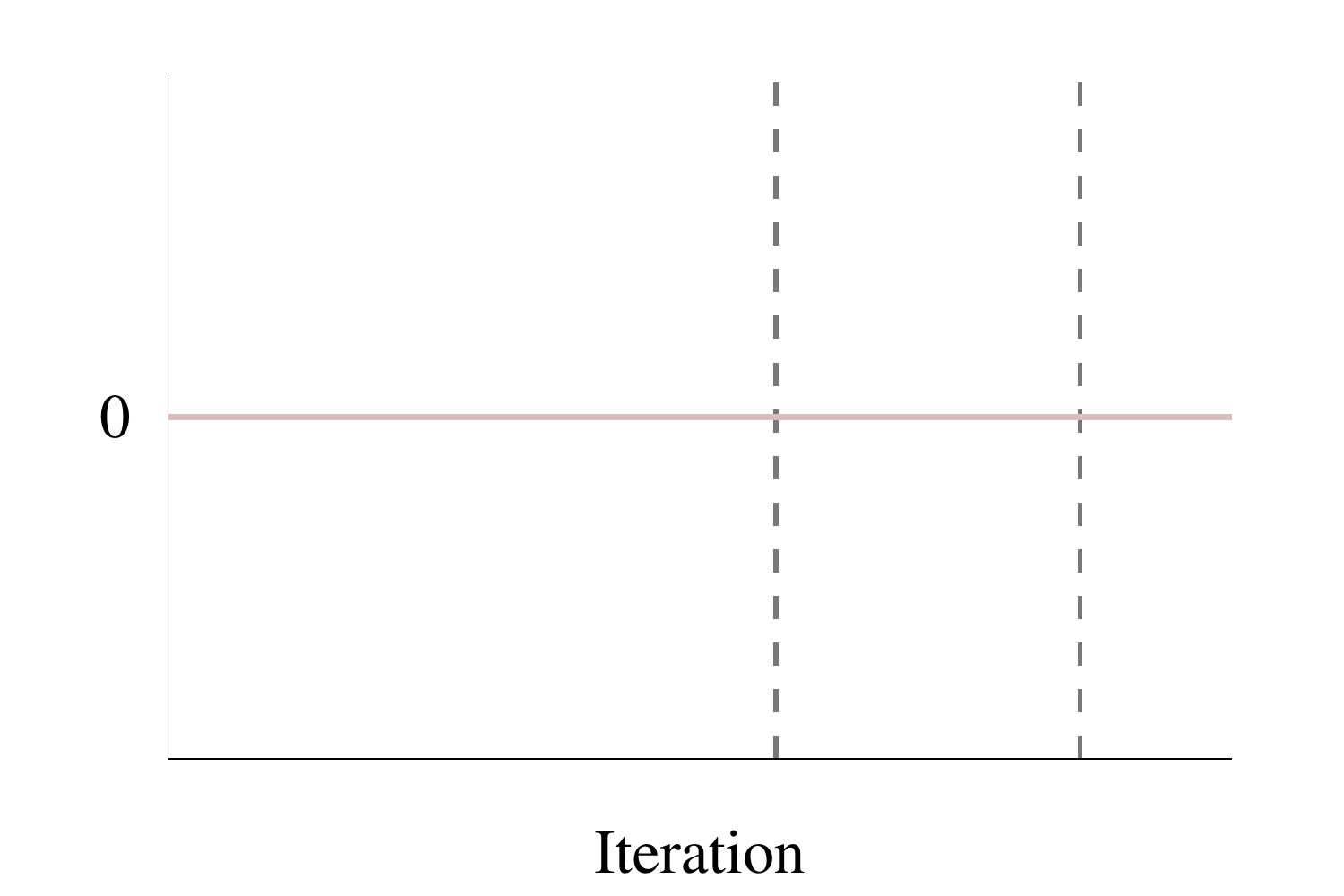}};
  \node[rotate=90] at (41, -1) { $\left| \mathbb{E} \! \left[ q_{2} \right] - \hat{q_{2}} \right|$ };
  
  \draw[-,color=black, line width=0.5] (44, 4)
                                       .. controls (46, 1.5) and (52, 1.5) 
                                       .. (54.9, 1.5);
  \draw[-,color=black, line width=0.5] (54.9, 1.5) 
                                       .. controls (57.8, 1.5) and (58.25, -1.5) 
                                       .. (60.25, -3.5);
  \draw[-,color=black, line width=0.5] (60.25, -3.5)
                                       .. controls (61, -4.25) and (62, -1.25)
                                       .. (62.975, -1);
\end{tikzpicture}
}
\caption{In practice, pathological regions of the typical set usually
cause Markov chains to get ``stuck''.  (a) Initially the Markov chain 
explores well-behaved regions of the typical set, avoiding the 
pathological neighborhood entirely and biasing Markov chain Monte Carlo 
estimators.  (b) If the Markov chain is run long enough then it will 
get stuck near the boundary of the pathological region, slowly correcting 
the Markov chain Monte Carlo estimators.  (c) Eventually the Markov 
chain escapes to explore the rest of the typical set.  This process 
repeats, causing the resulting estimators to oscillate around the true 
expectations and inducing strong biases unless the chain is improbably 
stopped at exactly the right time.}
\label{fig:pathological exploration}
\end{figure*}
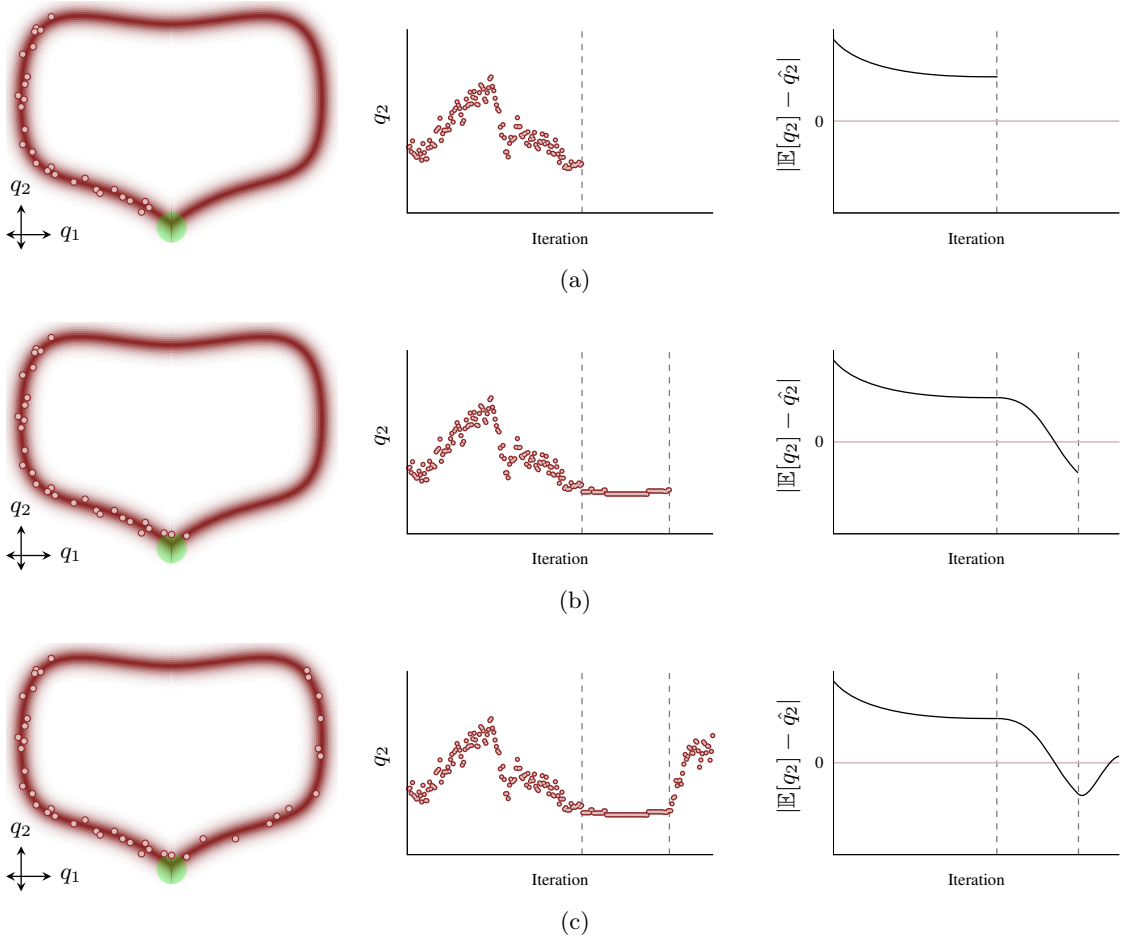

Because Markov chains have to recover the exact expectations asymptotically, 
they have to somehow compensate for not being able to explore these regions.  
Typically the Markov chain accomplishes this by getting stuck near the 
boundary of the pathological region: as the chain hovers around the 
pathological region the estimators are drawn down almost as if the Markov 
chain were exploring the pathological region.  Eventually this causes the 
estimators to overcorrect, inducing a bias in the opposite direction, at 
which point the Markov chain escapes to explore the rest of the typical set 
(Figure \ref{fig:pathological exploration}).  

Ultimately this behavior results in estimators that strongly oscillate 
around the true expectations.  Asymptotically the oscillations average 
out to give the true values, but that balance is fragile. Terminating 
the Markov chain after only a finite time almost always destroys this 
balance, and resulting estimators will suffer from substantial biases.

Whether or not features of the target distribution become pathological 
in a given application of Markov chain Monte Carlo depends on how exactly 
the given Markov transition interacts with those features.  Some 
transitions are generically more robust to certain features than others, 
and some can achieve robust performance by carefully tuning degrees of
freedom in the transition.  Regardless, great care has to be taken to 
ensure that a Markov transition is sufficiently robust to be used in a 
given application.

Formally, the sufficient condition that guarantees the idealized 
behavior, most importantly a Central Limit Theorem for the Markov chain 
Monte Carlo estimators, is known as 
\emph{geometric ergodicity}~\citep{RobertsEtAl:2004}.  Unfortunately, 
geometric ergodicity is extremely difficult to verify theoretically for 
all but the simplest problems and we must instead resort to empirical 
diagnostics.  The most powerful of these is the split $\hat{R}$ 
statistic~\citep{GelmanEtAl:2014a}, which quantifies the variation 
of an ensemble of Markov chains initialized from different points in 
parameter space.  The pathologies that frustrate geometric ergodicity induce 
inconsistencies amongst the individual chains in the ensemble, and hence 
large values of split $\hat{R}$.  Consequently, when split $\hat{R}$ is not 
near the nominal value of $1$, we should be suspicious of geometric ergodicity 
being satisfied and hence the practical utility of any resulting estimators.

\subsection{The Metropolis-Hastings Algorithm}

Given a Markov transition that targets the desired distribution, Markov 
chain Monte Carlo defines a generic strategy for quantifying the typical 
set.  Constructing such a transition, however, is itself a nontrivial problem.  
Fortunately there are various procedures for automatically constructing 
appropriate transitions for any given target distribution, with the foremost
amongst these the \emph{Metropolis-Hastings} algorithm
\citep{MetropolisEtAl:1953, Hastings:1970}.

The Metropolis-Hastings algorithm is comprised of two steps: a proposal 
and a correction.  The proposal is any stochastic perturbation of the initial 
state while the correction rejects any proposals that stray too far away from 
the typical set of the target distribution.  More formally, let 
$\mathbb{Q} ( q' \mid q )$ be the probability density defining each proposal.  
The probability of accepting a given proposal is then given by
\begin{equation*}
a \! \left( q' \mid q \right)
=
\min \! \left(1, 
\frac{ \mathbb{Q} ( q \mid q' ) \, \pi ( q' ) }
{ \mathbb{Q} ( q' \mid q ) \, \pi ( q )  } \right).
\end{equation*}

The original Markov chain Monte Carlo algorithm, and one still commonly
in use today, utilizes a Gaussian distribution as its proposal mechanism,
\begin{equation*}
\mathbb{Q} ( q' \mid q ) = \mathcal{N} ( q' \mid q, \Sigma ),
\end{equation*}
an algorithm to which I will refer to as \emph{Random Walk Metropolis}.
Because the proposal mechanism is symmetric under the exchange
of the initial and proposed points, the proposal density cancels in the
acceptance probability, leaving the simple form
\begin{equation*}
a ( q' \mid q )
=
\min \! \left(1, 
\frac{ \pi ( q' ) }
{ \pi ( q ) } \right).
\end{equation*}

Random Walk Metropolis is not only simple to implement, it also has a 
particularly nice intuition.  The proposal distribution is biased towards
large volumes, and hence the tails of the target distribution, while the 
Metropolis correction rejects those proposals that jump into neighborhoods 
where the density is too small.  The combined procedure then 
preferentially selects out those proposals that fall into neighborhoods 
of high probability mass, concentrating towards the typical set as 
desired.

Because of its conceptual simplicity and the ease in which it can be 
implemented by practitioners, Random Walk Metropolis is still popular in 
many applications.  Unfortunately, that seductive simplicity hides a 
performance that scales poorly with increasing dimension and complexity 
of the target distribution.

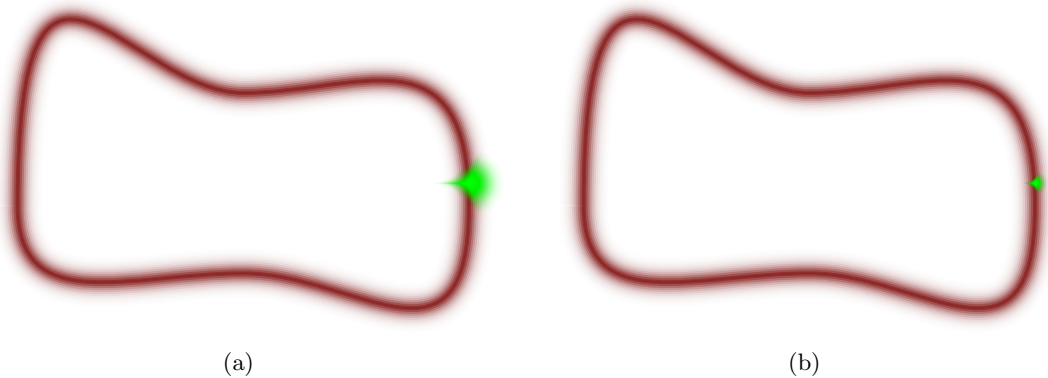
\begin{figure*}
\centering
\subfigure[]{
\begin{tikzpicture}[scale=0.3, thick]
  \begin{scope}
    \clip (-12, -6) rectangle (12, 10);
    \foreach \i in {0, 0.05,..., 1} {
      \draw[line width={30 * \i}, opacity={exp(-8 * \i)}, dark] 
        (-10, 0) .. controls (-10, 15) and (-5, 5) .. (0, 5)
        .. controls (5, 5) and (10, 8) .. (10, 0)
        .. controls (10, -8) and (5, -3) .. (0, -3)
        .. controls (-5, -3) and (-10, -5) .. (-10, 0);
    }  
  \end{scope}
  
  \pgfmathsetmacro{\x}{10}
  \pgfmathsetmacro{\y}{1}
  \pgfmathsetmacro{\delta}{2}
  \foreach \i in {0, 0.02,..., 1} {
    \fill[color=green, opacity={exp(-5 * \i)}]
      ({\x - \delta * \i}, \y) 
      .. controls ({\x - 0.5 * \delta * \i}, \y) 
        and ({\x - 0.25 * \delta * \i}, {\y - 0.25 * \delta * \i}) 
      .. ({\x + 0.25 * \delta * \i}, {\y - 0.75 * \delta * \i})
      .. controls ({\x + \delta * \i}, {\y - 0.5 * \delta * \i}) 
         and ({\x + \delta * \i}, {\y + 0.5 * \delta * \i}) 
       .. ({\x + 0.25 * \delta * \i}, {\y + 0.75 * \delta * \i})
       .. controls ({\x - 0.25 * \delta * \i}, {\y + 0.25 * \delta * \i}) 
         and ({\x - 0.5 * \delta * \i}, \y) .. ({\x - \delta * \i}, \y);
  }
\end{tikzpicture}
}
\subfigure[]{
\begin{tikzpicture}[scale=0.3, thick]
  \begin{scope}
    \clip (-12, -6) rectangle (12, 10);
    \foreach \i in {0, 0.05,..., 1} {
      \draw[line width={30 * \i}, opacity={exp(-8 * \i)}, dark] 
        (-10, 0) .. controls (-10, 15) and (-5, 5) .. (0, 5)
        .. controls (5, 5) and (10, 8) .. (10, 0)
        .. controls (10, -8) and (5, -3) .. (0, -3)
        .. controls (-5, -3) and (-10, -5) .. (-10, 0);
    }  
  \end{scope}
  
  \pgfmathsetmacro{\x}{10}
  \pgfmathsetmacro{\y}{1}
  \pgfmathsetmacro{\delta}{0.75}
  \foreach \i in {0, 0.02,..., 1} {
    \fill[color=green, opacity={exp(-5 * \i)}]
      ({\x - \delta * \i}, \y) 
      .. controls ({\x - 0.5 * \delta * \i}, \y) 
        and ({\x - 0.25 * \delta * \i}, {\y - 0.25 * \delta * \i}) 
      .. ({\x + 0.25 * \delta * \i}, {\y - 0.75 * \delta * \i})
      .. controls ({\x + \delta * \i}, {\y - 0.5 * \delta * \i}) 
         and ({\x + \delta * \i}, {\y + 0.5 * \delta * \i}) 
       .. ({\x + 0.25 * \delta * \i}, {\y + 0.75 * \delta * \i})
       .. controls ({\x - 0.25 * \delta * \i}, {\y + 0.25 * \delta * \i}) 
         and ({\x - 0.5 * \delta * \i}, \y) .. ({\x - \delta * \i}, \y);
  }
\end{tikzpicture}
}
\caption{In high dimensions, the Random Walk Metropolis proposal density
(green) is strongly biased towards the outside of the typical set where 
the target density, and hence the Metropolis acceptance probability 
vanishes.  (a) If the proposal variances are large then the proposals 
will stray too far away from the typical set and are rejected.  (b) 
Smaller proposal variances stay within the typical set and hence are 
accepted, but the resulting transition density concentrates tightly 
around the initial point.  Either way we end up with a Markov chain 
that explores the typical set very, very slowly.
}
\label{fig:random_walk_proposals}
\end{figure*}

The geometrical perspective introduced in Section \ref{sec:exploring} 
proves particularly powerful in illuminating these issues.  As the 
dimension of the target distribution increases, the volume exterior to 
the typical set overwhelms the volume interior to the typical set, and 
almost every Random Walk Metropolis proposal will produce a point on the 
outside of the typical set, towards the tails 
(Figure \ref{fig:random_walk_proposals}a).  The density of these points, 
however, is so small, that the acceptance probability becomes negligible.
In this case almost all of the proposals will be rejected and the resulting 
Markov chain will only rarely move.  We can induce a larger acceptance 
probability by shrinking the size of the proposal to stay within the 
typical set (Figure \ref{fig:random_walk_proposals}b), but those small 
jumps will move the Markov chain extremely slowly.

Regardless of how we tune the covariance of the Random Walk Metropolis 
proposal or the particular details of the target distribution, the 
resulting Markov chain will explore the typical set extremely slowly in 
all but the lowest dimensional spaces.  In the worst case this exploration 
will be so slow that we can't even complete a single sojourn across the 
typical set using our finite computational resources, and the resulting 
Markov chain Monte Carlo estimators will be highly biased.  Even if we 
can safely reach the mixing regime, however, the slow exploration will 
yield large autocorrelations and extremely imprecise estimators.

Consequently, if want to scale Markov chain Monte Carlo to the 
high-dimensional probability distributions of practical interest then 
we need a better way of exploring the typical set.  In particular, we 
need to better exploit the geometry of the typical set itself.

\section{The Foundations of Hamiltonian Monte Carlo}

\begin{figure*}
\centering
\begin{tikzpicture}[scale=0.35, thick]

  \foreach \i in {0, 0.05,..., 1} {
    \draw[line width={30 * \i}, opacity={exp(-8 * \i)}, dark] 
      (-10, 0) .. controls (-10, 15) and (-5, 5) .. (0, 5)
      .. controls (5, 5) and (10, 8) .. (10, 0)
      .. controls (10, -8) and (5, -3) .. (0, -3)
      .. controls (-5, -3) and (-10, -5) .. (-10, 0);
  }  

  \fill[color=green] (-5, 7.175) circle (5pt);
  
  \begin{scope}
    \clip (-11, 0) rectangle (-5, 10);
    \draw[->, >=stealth, >=stealth, color=green, line width=2] 
    (0, 5) .. controls (-5, 5) and (-10, 15) .. (-10, 0);
  \end{scope}
\end{tikzpicture}
\caption{Most Markov transitions are diffusive, concentrating around the
initial point such that the corresponding Markov chains linger in small
neighborhoods of the typical set for long periods of time.  In order to
maximize the utility of our computational resources we need \emph{coherent}
Markov transitions that are able to glide across the typical set towards
new, unexplored neighborhoods.
}
\label{fig:coherent_exploration}
\end{figure*}
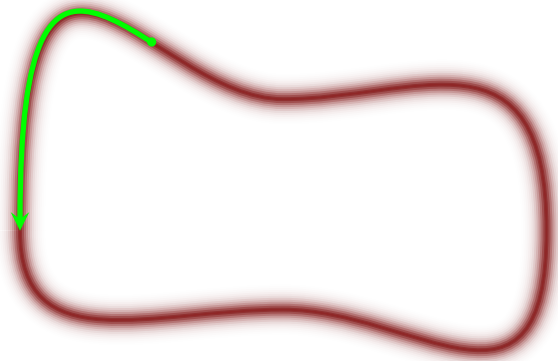

The guess-and-check strategy of Random Walk Metropolis is doomed to fail 
in high-dimensional spaces where there are an exponential number of 
directions in which to guess but only a singular number of directions 
that stay within the typical set and pass the check.  In order to make 
large jumps away from the initial point, and into new, unexplored regions 
of the typical set, we need to exploit information about the geometry of 
the typical set itself.  Specifically, we need transitions that can follow 
those contours of high probability mass, coherently gliding through the 
typical set (Figure \ref{fig:coherent_exploration}).

Hamiltonian Monte Carlo is the unique procedure for automatically 
generating this coherent exploration for sufficiently well-behaved target 
distributions.  In this section I will first introduce some intuition to 
motivate how we can generate the desired exploration by carefully exploiting 
the differential structure of the target probability density.  I will then 
discuss the procedure more formally, ending with the complete construction 
of the Hamiltonian Markov transition.

\subsection{Informing Effective Markov Transitions}
\label{sec:effective_transitions}

\begin{figure*}
\centering
\begin{tikzpicture}[scale=0.3, thick]
  \foreach \i in {0, 0.05,..., 1} {
    \draw[line width={30 * \i}, opacity={exp(-6 * \i)}, dark] 
      (0, 0) circle (8.25);
  }  
  \node[] at (0, 0) {\includegraphics[width=6.857cm]{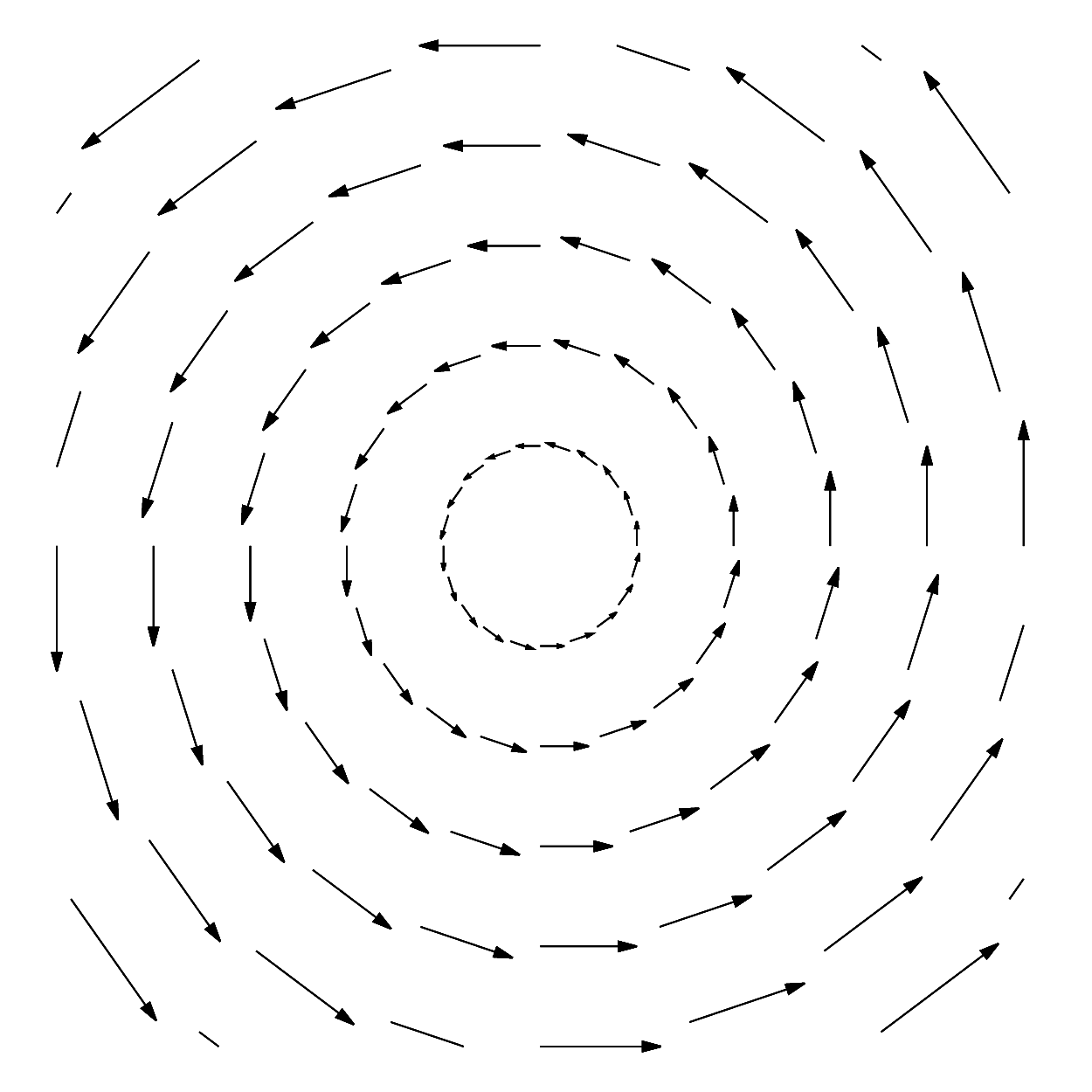}};
  
  \fill[color=green] (0, -8.25) circle (6pt);
  \draw[line width=2.5, green, ->, >=stealth] (0, -8.25) arc(-90:30:8.25);
\end{tikzpicture}
\caption{A vector field is the assignment of a direction at every point in 
parameter space.  When those directions are aligned with the typical set we 
can follow them like guide posts, generating coherent exploration of the 
target distribution.
}
\label{fig:vector_fields}
\end{figure*}

How can we distill the geometry of the typical set into information about 
how to move through it?  When the sample space is continuous, a natural way 
of encoding this direction information is with a \emph{vector field} aligned 
with the typical set (Figure \ref{fig:vector_fields}).  A vector field is 
the assignment of a direction at every point in parameter space, and if those 
directions are aligned with the typical set then they act as a guide through 
this neighborhood of largest target probability.

In other words, instead of fumbling around parameter space with random, 
uninformed jumps, we can follow the direction assigned to each at point for 
a small distance.  By construction this will move us to a new point in the 
typical set, where we will find a new direction to follow.  Continuing this 
process traces out a coherent trajectory through the typical set that 
efficiently moves us far away from the initial point to new, unexplored 
regions of the typical set as quickly as possible.

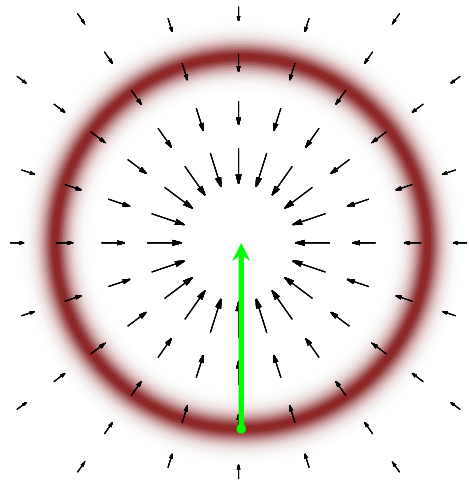
\begin{figure*}
\centering
\begin{tikzpicture}[scale=0.3, thick]
  \foreach \i in {0, 0.05,..., 1} {
    \draw[line width={30 * \i}, opacity={exp(-6 * \i)}, dark] 
      (0, 0) circle (8.25);
  }  
  \node[] at (0, 0) {\includegraphics[width=6.857cm]{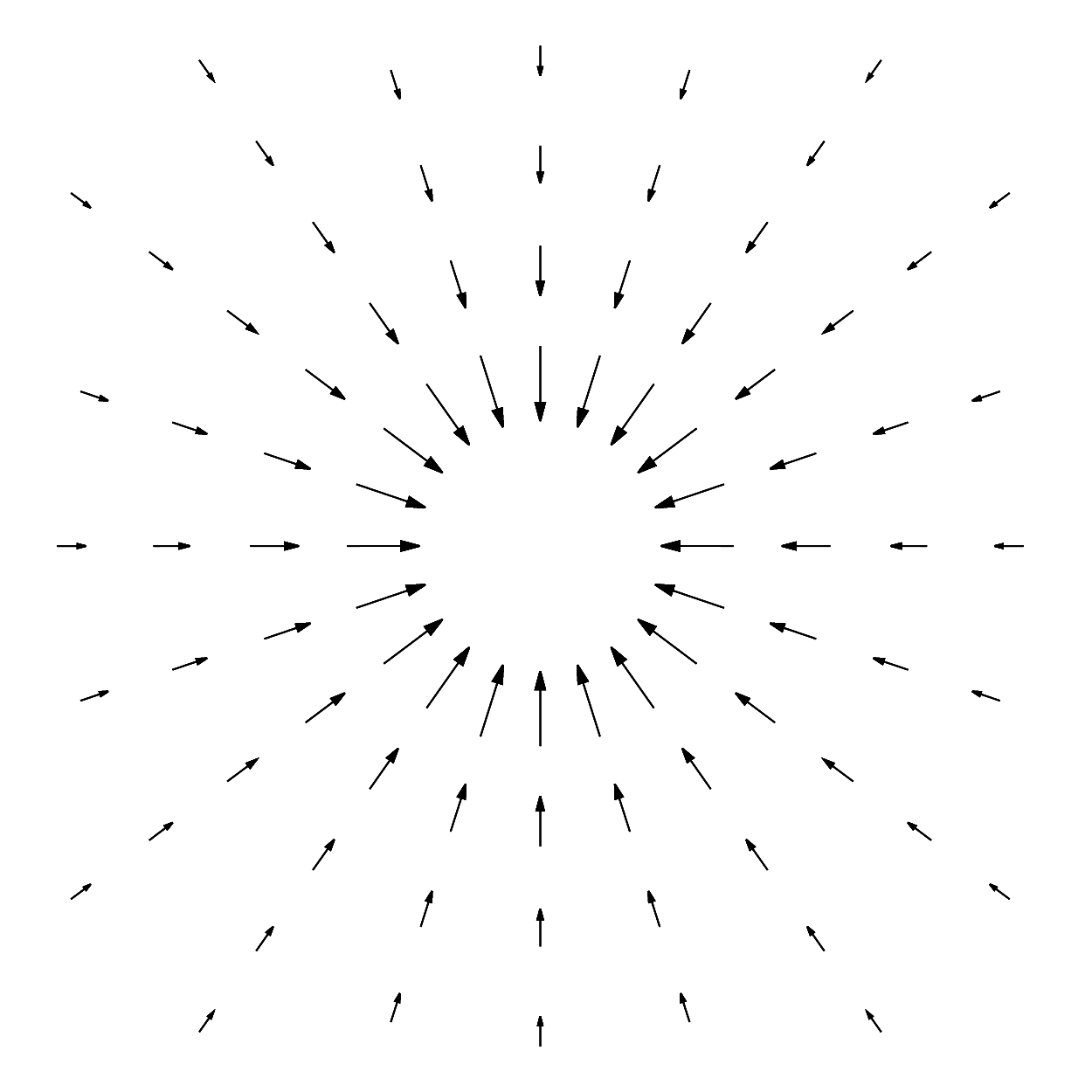}};
  
  \fill[color=green] (0, -8.25) circle (6pt);
  \draw[->, >=stealth, >=stealth, line width=2, green] (0, -8.25) -- (0, 0);
  
\end{tikzpicture}
\caption{The gradient of the target probability density function encodes 
information about the geometry of the typical set, but not enough to guide us 
through the typical set by itself.  Following along the gradient instead pulls
us away from the typical set and towards the mode of the target density.  In 
order to generate motion through the typical set we need to introduce additional 
structure that carefully twists the gradient into alignment with the typical set.
}
\label{fig:gradient}
\end{figure*}

We are still left, however, with the problem of constructing a vector field 
aligned with the typical set using only information that we can extract from 
the target distribution.  The natural information that we have yet to exploit 
is the differential structure of the target distribution which we can query 
through the \emph{gradient} of the target probability density function.  In 
particular, the gradient defines a vector field in parameter space sensitive 
to the structure of the target distribution (Figure \ref{fig:gradient}).

Unfortunately, that sensitivity is not sufficient as the gradient will never 
be aligned with the typical set.  Following the guidance of the gradient pulls 
us away from the typical set and towards the mode of the target density.  This 
behavior, however, isn't necessarily surprising.  Because the target density 
depends on the choice of parameterization, so too will its gradient.  
Consequently the gradient can direct us towards only parameterization-sensitive 
neighborhoods like that around the mode, and not the parameterization-invariant 
neighborhoods within the typical set.

To utilize the information in the gradient we need to complement it with 
additional geometric constraints, carefully removing the dependence on any 
particular parameterization while twisting the directions to align with the 
typical set.  Auspiciously, there is an elegant procedure for doing exactly 
this in a field of mathematics known as differential geometry.  Because 
differential geometry is a challenging subject that is rarely taught in 
applied statistics curricula, however, building an understanding of the 
details and subtleties of that procedure is no easy task.

Fortunately, there is a convenient equivalence that we can employ to build 
an intuition for this procedure without delving into the technical details.  
The same differential geometry that we need to use to correct the density 
gradients also happens to be the mathematics that describes classical physics.  
In other words, for every probabilistic system there is a mathematically 
equivalent \emph{physical} system about which we can more easily reason.

\begin{figure*}
\centering
\begin{tikzpicture}[scale=0.3, thick]
  \foreach \i in {0, 0.05,..., 1} {
    \draw[line width={30 * \i}, opacity={exp(-6 * \i)}, dark] 
      (0, 0) circle (8.25);
  }  
  \node[] at (0, 0) {\includegraphics[width=6.857cm]{gradient_field.eps}};
  \node[] at (0, 0) {\includegraphics[width=1.371cm]{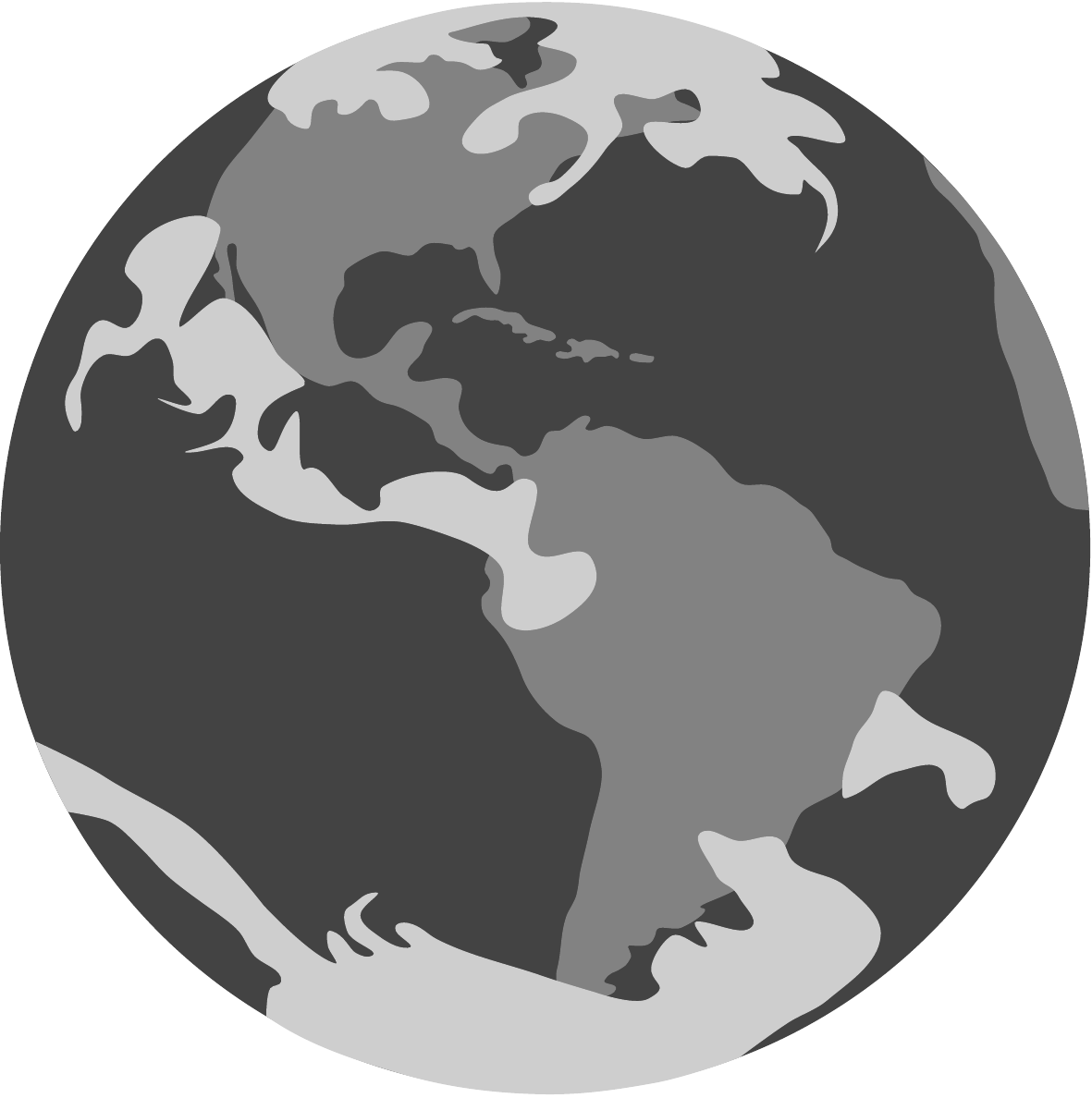}};
  
  \node[] at (0, -8.25) {\includegraphics[width=2.142cm]{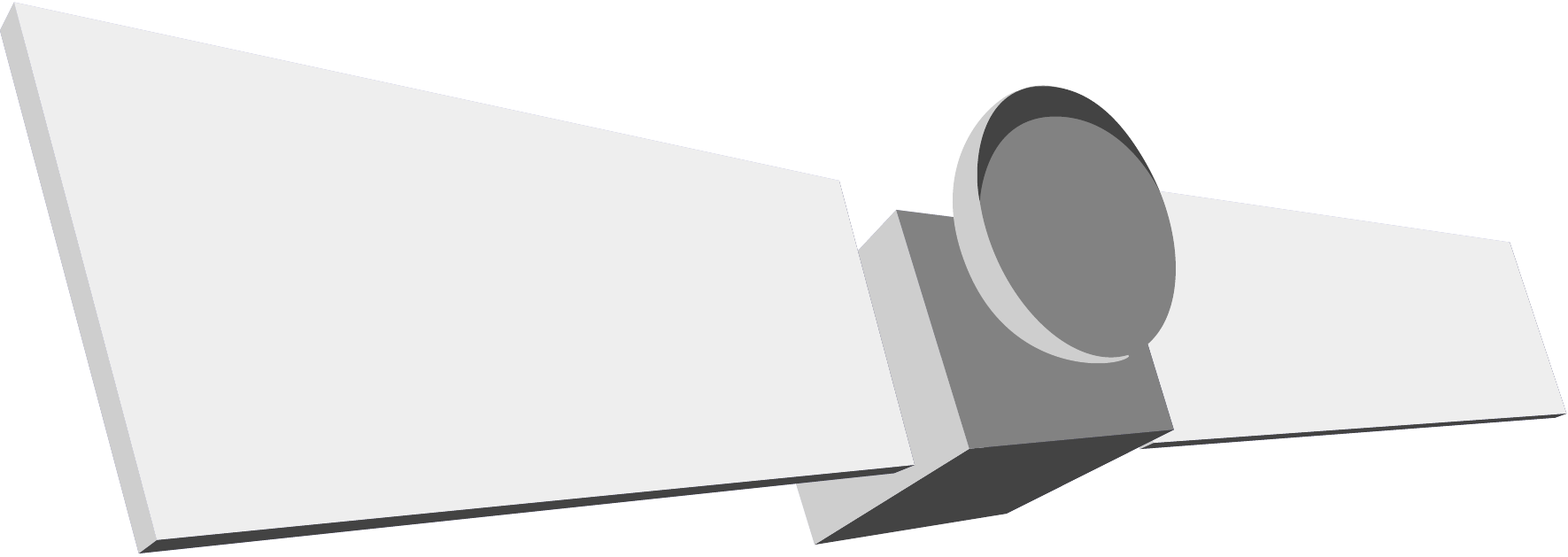}};
  
\end{tikzpicture}
\caption{The exploration of a probabilistic system is mathematically equivalent 
to the exploration of a physical system.  For example, we can interpret the 
mode of the target density as a massive planet and the gradient of the target 
density as that planet's gravitational field.  The typical set becomes the 
space around the planet through which we want a test object, such as a satellite, 
to orbit.
}
\label{fig:analogy}
\end{figure*}

For example, instead of trying to reason about a mode, a gradient, and a 
typical set, we can equivalently reason about a planet, a gravitational 
field, and an orbit (Figure \ref{fig:analogy}).  The probabilistic endeavor 
of exploring the typical set then becomes a physical endeavor of placing a 
satellite in a stable orbit around the hypothetical planet.

\begin{figure*}
\centering
\begin{tikzpicture}[scale=0.3, thick]
  \foreach \i in {0, 0.05,..., 1} {
    \draw[line width={30 * \i}, opacity={exp(-6 * \i)}, dark] 
      (0, 0) circle (8.25);
  }  
  \node[] at (0, 0) {\includegraphics[width=6.857cm]{gradient_field.eps}};
  \node[] at (0, -3.65) {\includegraphics[width=0.857cm,angle=180]{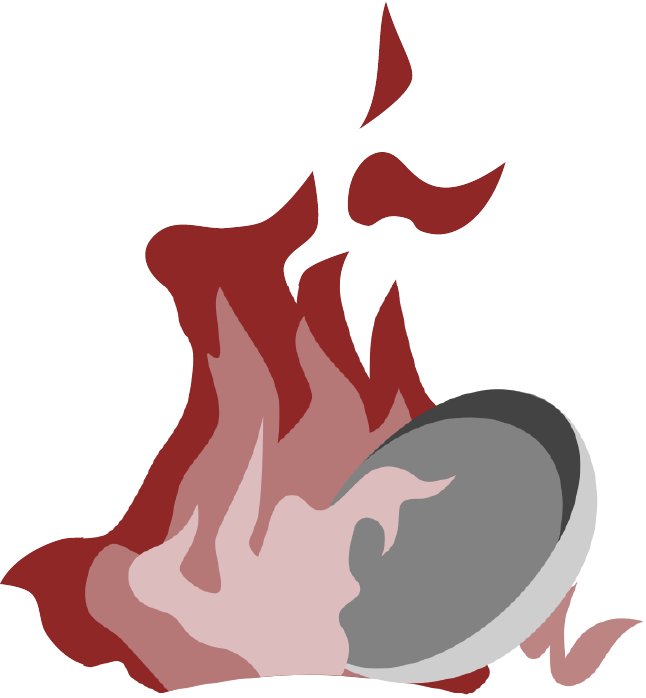}};
  \node[] at (0, 0) {\includegraphics[width=1.371cm]{earth.eps}};
  
  \fill[color=green] (0, -8.25) circle (6pt);
  \draw[->, >=stealth, >=stealth, line width=2, green] (0, -8.25) -- (0, -5);
  
\end{tikzpicture}
\caption{The analogous physical system suffers from the same pathologies as 
the motivating probabilistic system.  In particular, a satellite at rest will 
fall under the planet's gravity and crash into the surface of the planet, just 
as any gradient-driven trajectory will crash into the mode.
}
\label{fig:analogy_no_momentum}
\end{figure*}

Because these are just two different perspectives of the same mathematical 
system, they will suffer from the same pathologies.  Indeed, if we place a 
satellite at rest out in space it will fall in the gravitational field and 
crash into the surface of the planet, just as  naive gradient-driven 
trajectories crash into the mode (Figure \ref{fig:analogy_no_momentum}).  
From either the probabilistic or physical perspective we are left with a
catastrophic outcome.

\begin{figure*}
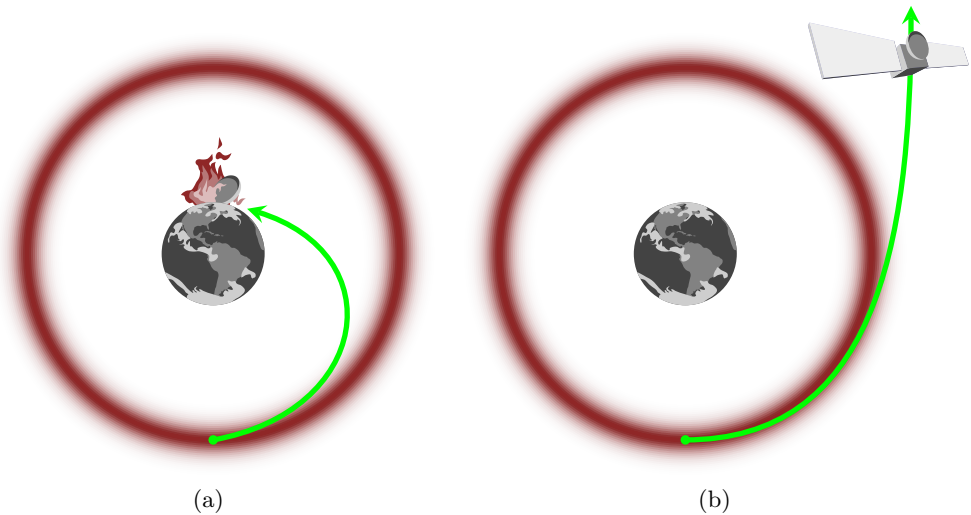

\centering
\subfigure[] {
\begin{tikzpicture}[scale=0.3, thick]
  \foreach \i in {0, 0.05,..., 1} {
    \draw[line width={30 * \i}, opacity={exp(-6 * \i)}, dark] 
      (0, 0) circle (8.25);
  }  
  %\node[] at (0, 0) {\includegraphics[width=8cm]{gradient_field.eps}};
  \node[] at (0, 3.65) {\includegraphics[width=0.857cm, angle=0]{satellite_crashed.eps}};
  \node[] at (0, 0) {\includegraphics[width=1.371cm]{earth.eps}};
  
  \fill[color=green] (0, -8.25) circle (6pt);
  \draw[->, >=stealth, >=stealth, line width=2, green] 
  (0, -8.25) .. controls (7, -7) and (8, 0) .. (1.5, 2);
  
\end{tikzpicture}
}
\subfigure[] {
\begin{tikzpicture}[scale=0.3, thick]
  \foreach \i in {0, 0.05,..., 1} {
    \draw[line width={30 * \i}, opacity={exp(-6 * \i)}, dark] 
      (0, 0) circle (8.25);
  }  
 % \node[] at (0, 0) {\includegraphics[width=8cm]{gradient_field.eps}};
  \node[] at (0, 0) {\includegraphics[width=1.371cm]{earth.eps}};
  
  \fill[color=green] (0, -8.25) circle (6pt);
  \draw[->, >=stealth, >=stealth, line width=2, green] 
  (0, -8.25) .. controls (8.25, -8.25) and (10, 0) .. (10, 11);
  
  \node[] at (9, 9) {\includegraphics[width=2.142cm]{satellite.eps}};
  
\end{tikzpicture}
}
\caption{(a) Without enough transverse momentum to balance against the gravitational 
attraction of the planet, a satellite will still crash into the planet. (b) On other 
other hand, if the satellite is given too much momentum then the gravitational 
attraction will be too weak to capture the satellite in a stable orbit, which will 
instead abandon the planet for the depths of space.
}
\label{fig:analogy_bad_momentum}
\end{figure*}

The physical picture, however, provides an immediate solution: although objects 
at rest will crash into the planet, we can maintain a stable orbit by endowing 
our satellite with enough \emph{momentum} to counteract the gravitational 
attraction.  We have to be careful, however, in how exactly we add momentum to 
our satellite.  If we add too little momentum transverse to the gravitational 
field, for example, then the gravitational attraction will be too strong and 
the satellite will still crash into the planet 
(Figure \ref{fig:analogy_bad_momentum}a).  On the other hand, if we add too much 
momentum then the gravitational attraction will be too weak to capture the 
satellite at all and it will instead fly out into the depths of space 
(Figure \ref{fig:analogy_bad_momentum}b).

\begin{figure*}
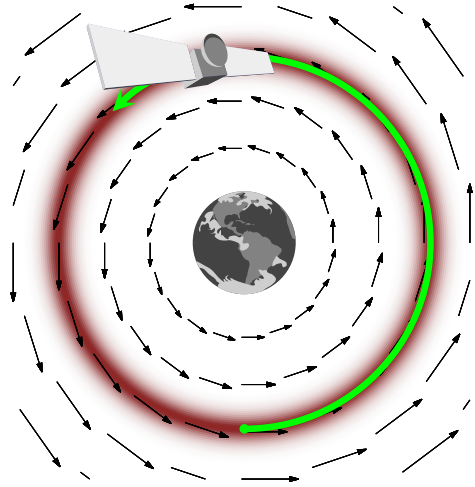

\centering
\begin{tikzpicture}[scale=0.3, thick]
  \foreach \i in {0, 0.05,..., 1} {
    \draw[line width={30 * \i}, opacity={exp(-6 * \i)}, dark] 
      (0, 0) circle (8.25);
  }  
  \node[] at (0, 0) {\includegraphics[width=6.857cm]{vector_field.eps}};
  \node[] at (0, 0) {\includegraphics[width=1.371cm]{earth.eps}};
  
  \fill[color=green] (0, -8.25) circle (6pt);
  \draw[line width=2.5, green, ->, >=stealth] (0, -8.25) arc(-90:135:8.25);
  
  \node[] at ({8.25 * cos(110)}, {8.25 * sin(110) + 0.5}) 
  {\includegraphics[width=2.5cm]{satellite.eps}};
\end{tikzpicture}
\caption{When we introduce exactly the right amount of momentum to the physical 
system, the equations describing the evolution of the satellite define a vector 
field aligned with the orbit.  The subsequent evolution of the system will then
trace out orbital trajectories.
}
\label{fig:analogy_perfect_momentum}
\end{figure*}

If we add just the right amount of momentum, however, then the momentum will 
exactly balance against the gravitational force, and the corresponding dynamics 
of the system will be \emph{conservative}.  As the satellite falls towards the 
planet the momentum grows until it is large enough to propel the satellite away 
from the planet.  Similarly, if the satellite drifts away from the planet then 
the momentum shrinks and the satellite slows, allowing gravity more time to pull 
it back towards the planet.  This careful exchange balances along the desired 
orbit, ensuring that the subsequent evolution of the satellite will generate 
exactly the trajectories that we need (Figure \ref{fig:analogy_perfect_momentum}).  

Pulling this physical intuition back into the probabilistic perspective, the 
key to twisting the gradient vector field into a vector field aligned with 
the typical set, and hence one capable of generating efficient exploration, is 
to expand our original probabilistic system with the introduction of auxiliary 
momentum parameters.  As in the physical system, however, we can't just add those 
momentum parameters arbitrarily.  They need to be endowed with a probabilistic 
structure that ensures conservative dynamics.  

Remarkably, there is only one procedure for introducing auxiliary momentum with 
such a probabilistic structure: Hamiltonian Monte Carlo.

\subsection{Phase Space and Hamilton's Equations}

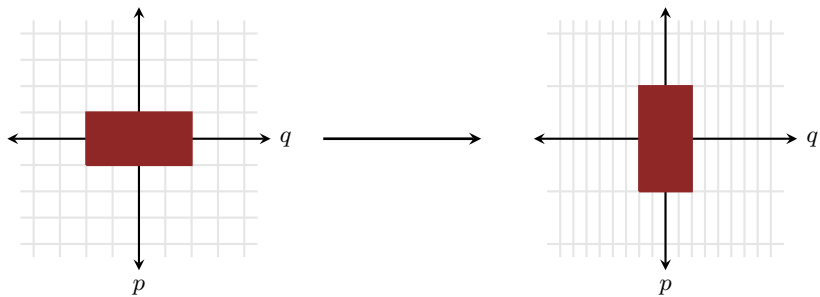
\begin{figure*}
\centering
\begin{tikzpicture}[scale=0.35, thick]
  % Left
  \foreach \i in {-19, -18, ..., -11} {
    \draw[gray90]  (\i, -4.5) -- (\i, 4.5);
  }  

  \foreach \i in {-4, -3, ..., 4} {
    \draw[gray90]  (-19.5, \i) -- (-10.5, \i);
  }  

  \draw[<->, >=stealth] (-15, 5) -- (-15, -5)
  node[below] { $p$ }; 
  
  \draw[<->, >=stealth] (-20, 0) -- (-10, 0)
  node[right] { $q$ };  
  
  \draw[fill, dark] (-17, -1) -- (-17, 1) -- (-13, 1) -- (-13, -1) -- (-17, -1);
  
  % Center
  \draw[->, >=stealth, >=stealth, line width=1] (-8, 0) -- (-2, 0);
  
  % Right
  \foreach \i in {1, 1.5, ..., 9} {
    \draw[gray90]  (\i, -4.5) -- (\i, 4.5);
  }  

  \foreach \i in {-4, -2, ..., 4} {
    \draw[gray90]  (0.5, \i) -- (9.5, \i);
  }  

  \draw[<->, >=stealth] (5, 5) -- (5, -5)
  node[below] { $p$ }; 
  
  \draw[<->, >=stealth] (0, 0) -- (10, 0)
  node[right] { $q$ };  
  
  \draw[fill, dark] (4, -2) -- (4, 2) -- (6, 2) -- (6, -2) -- (4, -2);
\end{tikzpicture}
\caption{A defining feature of conservative dynamics is the preservation of 
volume in position-momentum phase space.  For example, although dynamics might 
compress volumes in position space, the corresponding volume in momentum space 
expands to compensate and ensure that the total volume is invariant.  Such 
volume-preserving mappings are also known as \emph{shear} transformations.
}
\label{fig:liouville}
\end{figure*}

Conservative dynamics in physical systems requires that volumes are exactly 
preserved.  As the system evolves, any compression or expansion in position 
space must be compensated with a respective expansion or compression in 
momentum space to ensure that the volume of any neighborhood in 
position-momentum \emph{phase space} is unchanged (Figure \ref{fig:liouville}).

In order to mimic this behavior in our probabilistic system we need to 
introduce auxiliary momentum parameters, $p_{n}$, to complement each dimension 
of our target parameter space,
\begin{equation*}
q_{n} \rightarrow \left( q_{n}, p_{n} \right),
\end{equation*}
expanding the $D$-dimensional  parameter space into a $2D$-dimensional phase 
space.  Moreover, these auxiliary momentum have to be dual to the target 
parameters, transforming in the opposite way under any reparameterization so 
that phase space volumes are invariant.  

\begin{figure}
\centering
\begin{tikzpicture}[scale=0.25, thick]
  \node[] at (0, 0) {\includegraphics[height=4.166cm, width=8.333cm]
  {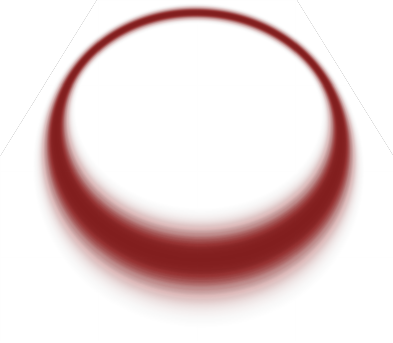}};

  \draw[line width=2, light, dashed, ->, >=stealth] 
  (0, 0) -- (0, -3.9);
  \draw[line width=2, light, dashed, ->, >=stealth] 
  (6.7, 11.1) -- (6.7, 7.1);

  \fill[color=light] (0, -4.4) circle (8pt);
  \draw[line width=2, green, ->, >=stealth] 
  (0, -4.4) .. controls (10, -4.4) and (18, 2.5) .. (6.7, 6.7);
  
  \fill[color=dark] (0, 0) circle (8pt);
  \draw[line width=2, dark, ->, >=stealth] 
  (0, 0) .. controls (10, 2) and (18, 15) .. (6.7, 11.1);
\end{tikzpicture}
\caption{By constructing a probability distribution on phase space that 
marginalizes to the target distribution, we ensure that the typical set 
on phase space projects to the typical set of the target distribution.
In particular, if we can construct trajectories that efficiently explore 
the joint distribution (black) they will project to trajectories that 
efficiently explore the target distribution (green).}
\label{fig:projected_trajectory}
\end{figure}
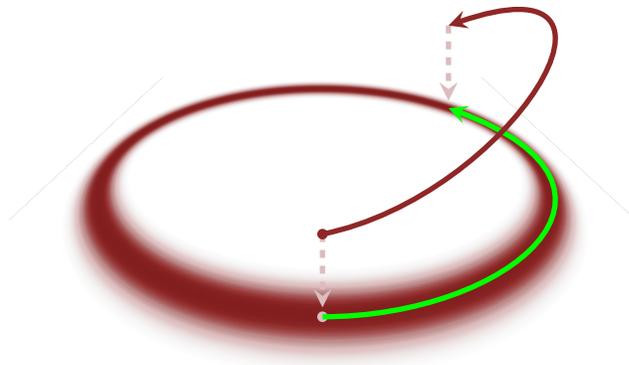

Having expanded the target parameter space to phase space, we can now lift 
the target distribution onto a joint probability distribution on phase space 
called the \emph{canonical distribution}.  We do this with the choice of a 
conditional probability distribution over the auxiliary momentum,
\begin{equation*}
\pi \! \left( q, p \right) = \pi \! \left( p \mid q \right) \pi \! \left( q \right),
\end{equation*}
which ensures that if we marginalize out the momentum we immediately recover 
our target distribution.  More importantly, it guarantees that any trajectories 
exploring the typical set of the phase space distribution will project to 
trajectories exploring the typical set of the target distribution 
(Figure \ref{fig:projected_trajectory}).

Because of the duality of the target parameters and the auxiliary momentum, 
the corresponding probability densities also transform oppositely to each other.  
In particular, the canonical density $\pi \! \left( q, p \right)$ does not 
depend on a particular choice of parameterization, and we can write it in
terms of an invariant \emph{Hamiltonian} function, $H \! \left(q, p \right)$,
\begin{equation*}
\pi \! \left( q, p \right) = e^{- H (q, p) }.
\end{equation*}
Because $H \! \left(q, p \right)$ is independent of the details of any 
parameterization, it captures the invariant probabilistic structure of the phase 
space distribution and, most importantly, the geometry of its typical set.  
Appealing to the physical analogy, the value of the Hamiltonian at any point in 
phase space is called the \emph{energy} at that point.

Because of the decomposition of the joint density, the Hamiltonian,
\begin{equation*}
H \! \left( q, p \right) \equiv - \log \pi \! \left(q, p \right),
\end{equation*}
itself decomposes into two terms,
\begin{align*}
H \! \left( q, p \right) 
&= 
- \log \pi \! \left(p \mid q \right)
- \log \pi \! \left( q \right)
\\
&\equiv
\quad\;\; K \! \left( p, q \right) \quad
+ 
\; V \! \left( q \right).
\end{align*}
Once again leveraging the physical analogy, the term corresponding to the 
density over the auxiliary momentum, $K \! \left( p, q \right)$ is called 
the \emph{kinetic energy}, while the term corresponding to the density of 
the target distribution, $V \! \left( q \right)$ is known as the 
\emph{potential energy}.  The potential energy is completely determined by 
the target distribution while the kinetic energy is unconstrained and must 
be specified by the implementation.

Because the Hamiltonian captures the geometry of the typical set, it should 
we should be able to use it to generate a vector field oriented with the
typical set of the canonical distribution and hence the trajectories that 
we are after.  Indeed, the desired vector field can be generated from a
given Hamiltonian with \emph{Hamilton's equations},
\begin{align*}
\frac{ \mathrm{d} q }{ \mathrm{d} t }
&=
+\frac{ \partial H}{ \partial p}
=
\frac{ \partial K}{ \partial p}
\\
\frac{ \mathrm{d} p }{ \mathrm{d} t }
&=
-\frac{ \partial H}{ \partial q}
=
- \frac{ \partial K}{ \partial q} - \frac{ \partial V}{ \partial q}.
\end{align*}

Recognizing $\partial V / \partial q$ as the gradient of the logarithm
of the target density, we see that Hamilton's equations fulfill exactly 
the intuition introduced in Section \ref{sec:effective_transitions}.  By 
channeling the gradient through the momentum instead of the target parameters 
directly, Hamilton's equations twist differential information to align with 
the typical set of canonical distribution.  Following the Hamiltonian vector 
field for some time, $t$, generates trajectories, $\phi_{t} (q, p)$, that 
rapidly move through phase space while being constrained to the typical set.  
Projecting these trajectories back down onto the target parameter space finally 
yields the efficient exploration of the target typical set for which we are 
searching.

\subsection{The Idealized Hamiltonian Markov Transition}

In order to utilize these Hamiltonian trajectories to construct an efficient 
Markov transition, we need a mechanism for introducing momentum to a given 
point in the target parameter space.  Fortunately, this is straightforward 
if we exploit the probabilistic structure that we have already endowed to the 
system.

To lift an initial point in parameter space into one on phase space we simply 
sample from the conditional distribution over the momentum,
\begin{equation*}
p \sim \pi \! \left( p \mid q \right).
\end{equation*}
Assuming that the initial point was in the typical set of the target 
distribution, sampling the momentum directly from the conditional distribution 
guarantees that the lift will fall into the typical set on phase space.

Once on phase space we can explore the joint typical set by integrating Hamilton's 
equations for some time,
\begin{equation*}
\left( q, p \right) \rightarrow \phi_{t} \! \left( q, p \right).
\end{equation*}
By construction these trajectories coherently push the Markov transition away 
from the initial point, and neighborhoods that we have already explored, while 
staying confined to the joint typical set.

Because of the carefully chosen structure of the joint distribution, these 
trajectories project down to trajectories that explore the target distribution.  
Consequently, after integrating Hamilton's equations we can return to the target 
parameter space by simply projecting away the momentum,
\begin{equation*}
\left(q, p \right) \rightarrow q.
\end{equation*}

\begin{figure}
\centering
\includegraphics[width=3in]{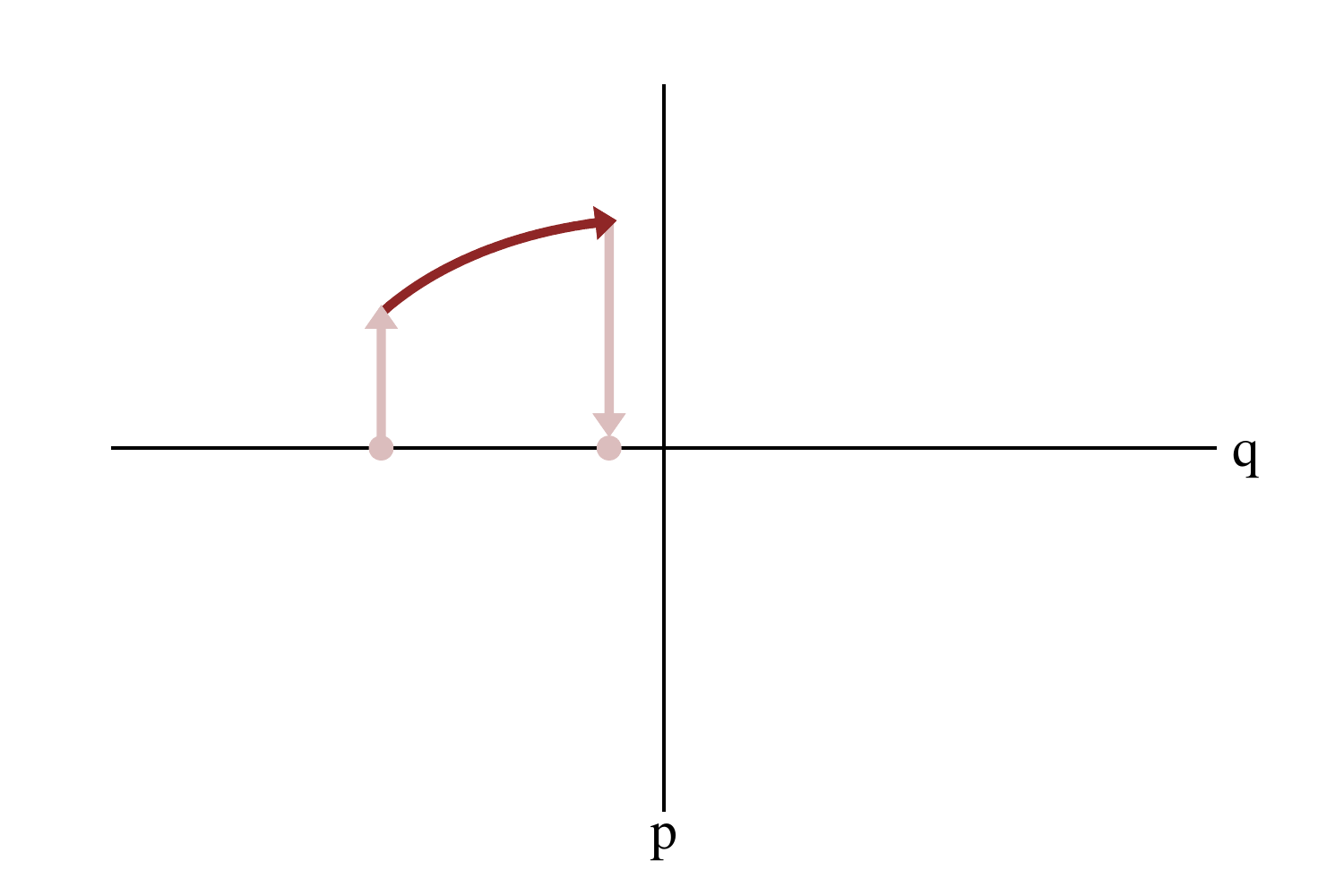}
\caption{Every Hamiltonian Markov transition is comprised of a random lift from 
the target parameter space onto phase space (light red), a deterministic 
Hamiltonian trajectory through phase space (dark red), and a projection back down 
to the target parameter space (light red).}
\label{fig:hmc_transition_cartoon}
\end{figure}

Composing these three steps together yields a Hamiltonian Markov transition
composed of random trajectories that rapidly explore the target distribution,
exactly as desired (Figure \ref{fig:hmc_transition_cartoon}).  Out in the tails 
of the target distribution, the momentum sampling and projection steps allow the 
resulting Markov chain to rapidly fall in toward the typical set.  Once the 
chain has reached the typical set, the Hamiltonian trajectories ensure extremely 
efficiently exploration.

\section{Efficient Hamiltonian Monte Carlo}

An immediate complication with this foundational construction is that it does 
not define a unique Markov transition but rather an infinity of them.  Every 
choice of kinetic energy and integration time yields a new Hamiltonian 
transition that will interact differently with a given target distribution.  
Unfortunately, these interactions will usually lead to suboptimal performance 
and we are left with a delicate tuning problem.  When these degrees of freedom 
are well-chosen, the resulting implementation of Hamiltonian Monte Carlo will 
perform well on even the challenging, high-dimensional problems of applied
interest.  When they are poorly-chosen, however, the performance can suffer 
dramatically.

In order to be able to optimize the application of the Hamiltonian Monte Carlo 
method and ensure robust performance, we need to understand exactly how these 
degrees of freedom interact with the target distribution.  Although this seems 
like a daunting task, we can  facilitate it by exploiting the latent geometry 
of Hamiltonian Monte Carlo itself.  In particular, the analysis is make much
easier by considering a different view of phase space.

\subsection{The Natural Geometry of Phase Space}
\label{sec:microcanonical_geometry}

\begin{figure}
\centering
\includegraphics[width=2.9in]{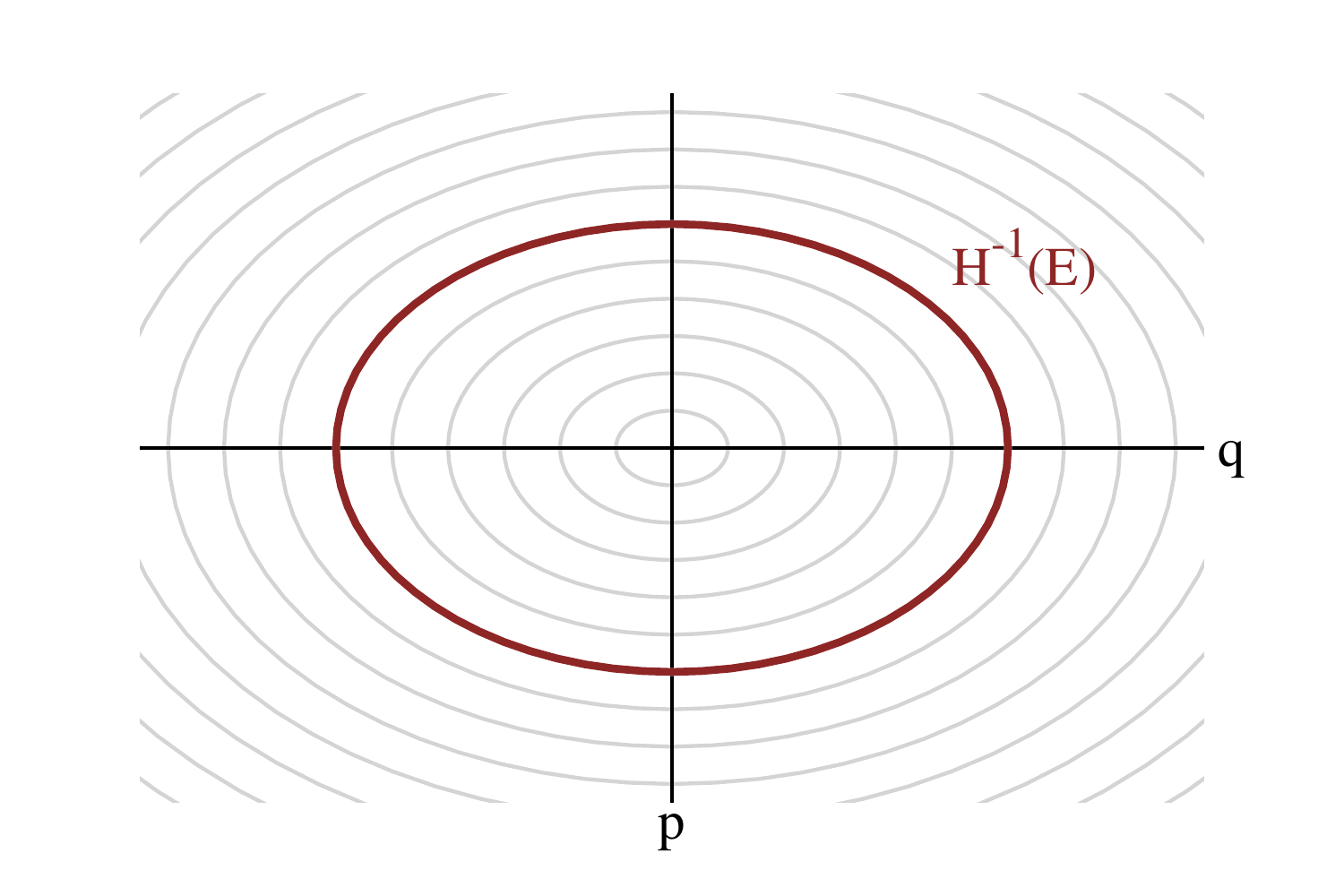}
\caption{Phase space naturally decomposes into level sets of the Hamiltonian, 
$H^{-1} \! \left( E \right)$.  Instead of specifying a point in phase space with 
its position and momentum, we can specify it with an energy, $E$, and its position 
on the corresponding level set, $\theta_{E} \in H^{-1} \! (E)$.}
\label{fig:foliation}
\end{figure}

One of the characteristic properties of Hamilton's equations is that they 
conserve the value of the Hamiltonian.  In other words, every Hamiltonian 
trajectory is confined to an energy \emph{level set},
\begin{equation*}
H^{-1} \! \left( E \right) = \left\{ q, p \mid H \! \left( q, p \right) = E \right\},
\end{equation*}
which, save for some ignorable exceptions, are all $(2D - 1)$-dimensional,
compact surfaces in phase space.  In fact, once we've removed any singular 
level sets, the entirety of phase space neatly decomposes, or \emph{foliates} 
into concentric level sets (Figure \ref{fig:foliation}).  Consequently, we can 
specify any point in phase space by first specifying the energy of the level 
set it falls on, $E$, and the position within that level set, $\theta_{E}$ 
(Figure \ref{fig:foliation}).

Correspondingly the canonical distribution on phase space admits a 
\emph{microcanonical decomposition},
\begin{equation*}
\pi \! \left(q, p \right) 
= 
\pi \left( \theta_{E} \mid E \right) \pi \! \left( E \right),
\end{equation*}
across this foliation.  The conditional distribution over each level set, 
$\pi \left( \theta_{E} \mid E \right)$, is called the \emph{microcanonical 
distribution}, while the distribution across the level sets, 
$\pi \! \left( E \right)$, is called the \emph{marginal energy distribution}.

\begin{figure}
\centering
\subfigure[] {\includegraphics[width=2.9in]{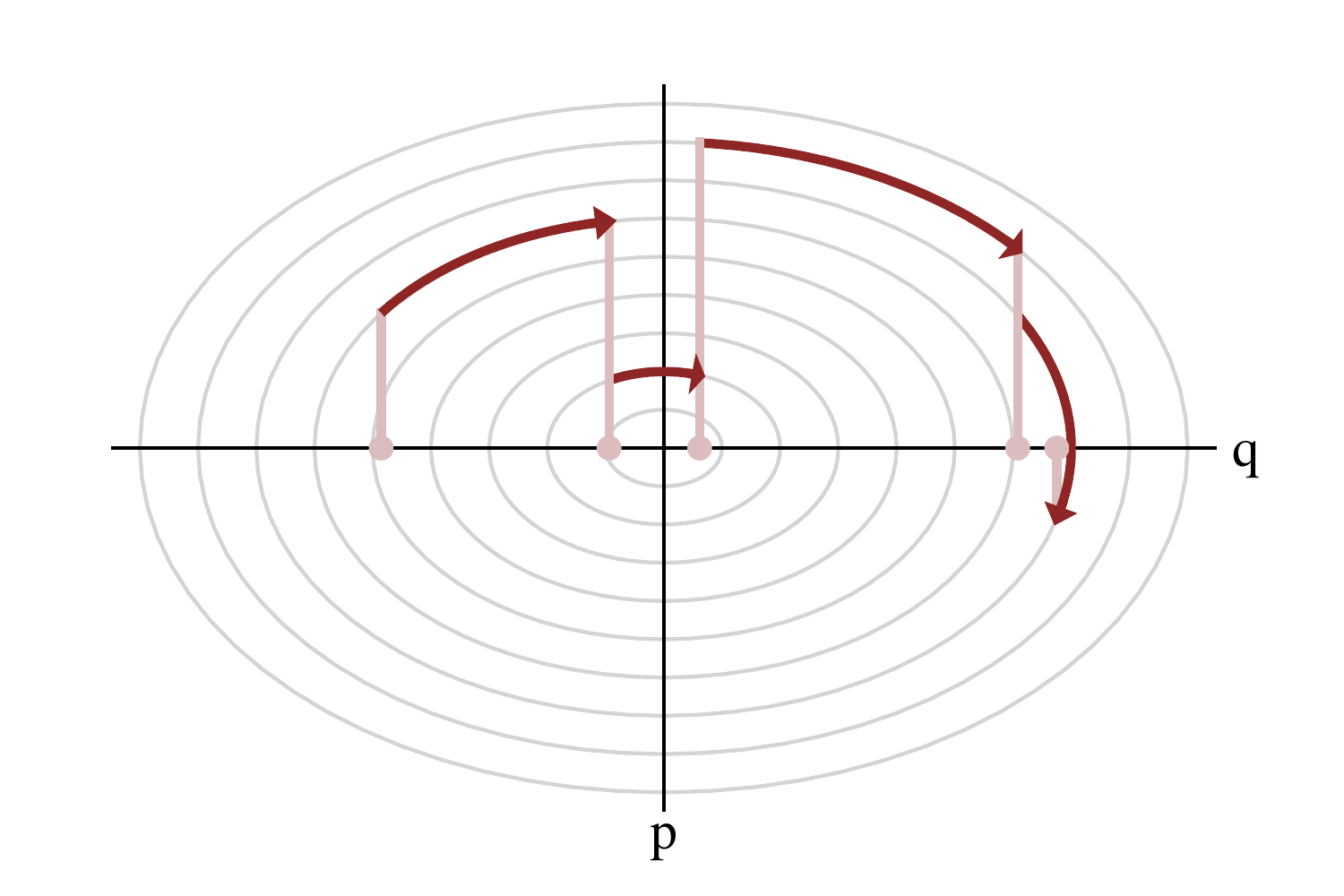} }
\subfigure[] {\includegraphics[width=2.9in]{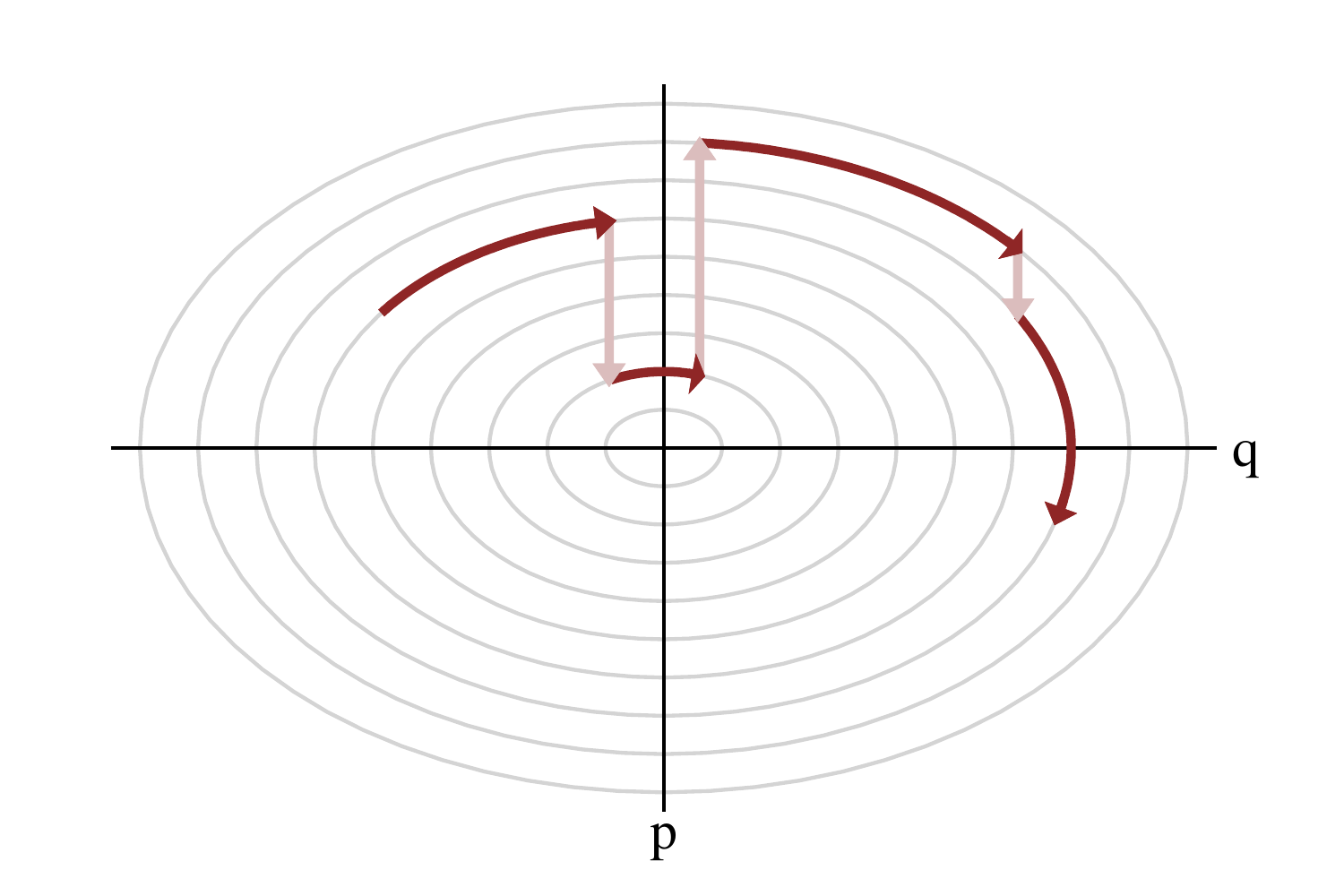} }
\caption{(a) Each Hamiltonian Markov transition lifts the initial state onto a 
random level set of the Hamiltonian, which can then be explored with a Hamiltonian 
trajectory before projecting back down to the target parameter space.  (b) If we 
consider the projection and random lift steps as a single momentum resampling step, 
then the Hamiltonian Markov chain alternates between deterministic trajectories 
along these level sets (dark red) and a random walk across the level sets (light red).}
\label{fig:hmc_chain_cartoon}
\end{figure}

Because they are derived from the same geometry, this microcanonical 
decomposition is particularly well-suited to analyzing the Hamiltonian 
transition.  To see this more clearly, consider a Hamiltonian Markov chain 
consisting of multiple transitions (Figure \ref{fig:hmc_chain_cartoon}a).  
Each Hamiltonian trajectory explores a level set while the intermediate 
projections and lifts define a random jump between the level sets themselves.  
Consequently, the entire Hamiltonian Markov chain decouples into two distinct 
phases: deterministic exploration of individual level sets and a stochastic 
exploration between the level sets themselves (Figure \ref{fig:hmc_chain_cartoon}b).

This decoupling makes it particularly convenient to analyze the efficiency of 
each phase, and hence the efficiency of the overall Hamiltonian Markov 
transition.  For example, the efficacy of the deterministic exploration is 
determined by how long the Hamiltonian trajectories are integrated and,
consequently, how completely they explore the corresponding level sets.  The
cost of this phase, however, is ultimately proportional to the total integration 
time.  The integration time needed to explore just enough of each level set, 
and hence the overall efficiency of the deterministic exploration, depends on 
the geometry of the energy level sets.  The more uniform and regular the level 
sets, the faster the trajectories will explore for a given integration time.

\begin{figure}
\centering
\subfigure[]{ \includegraphics[width=2.5in]{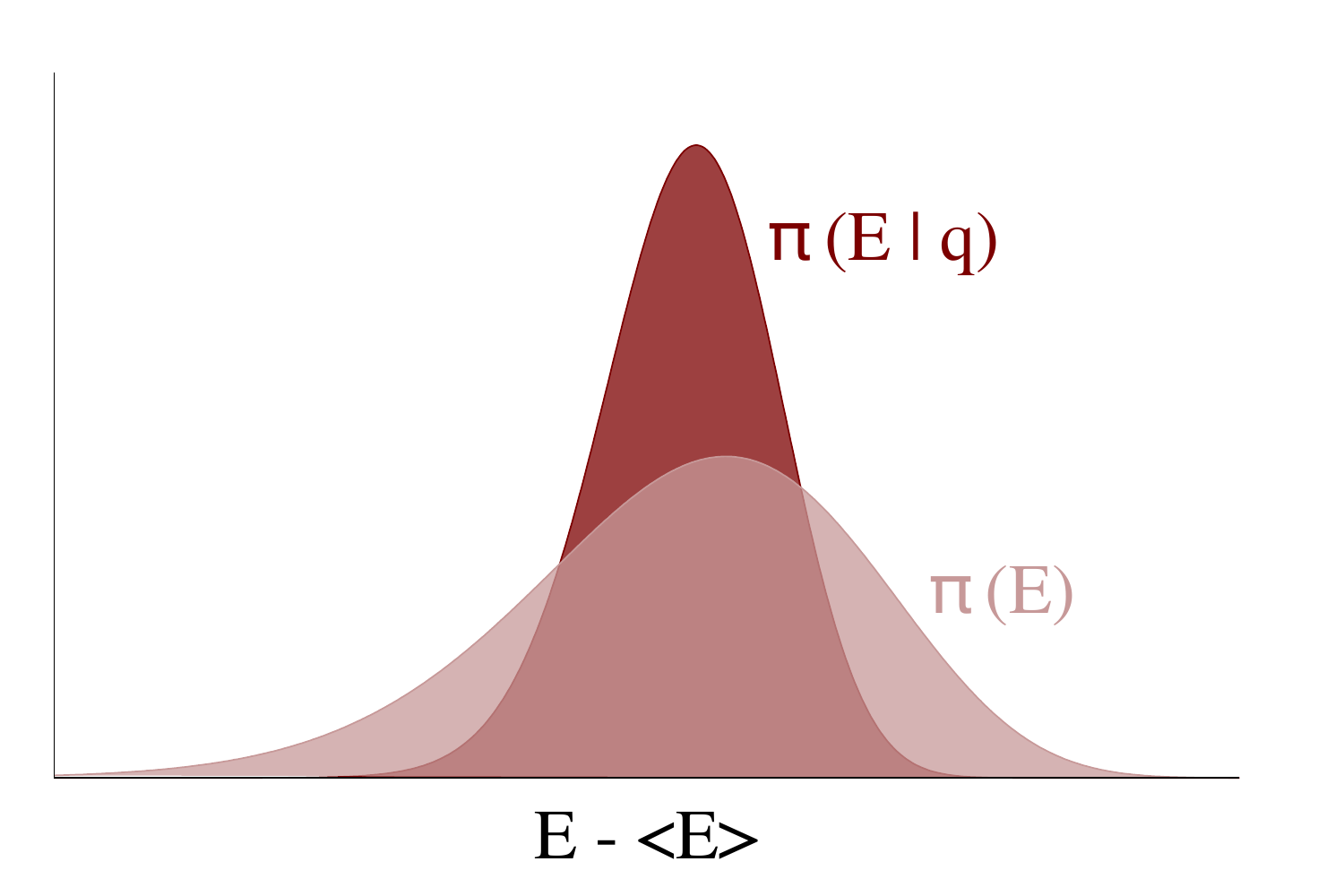} }
\subfigure[]{ \includegraphics[width=2.5in]{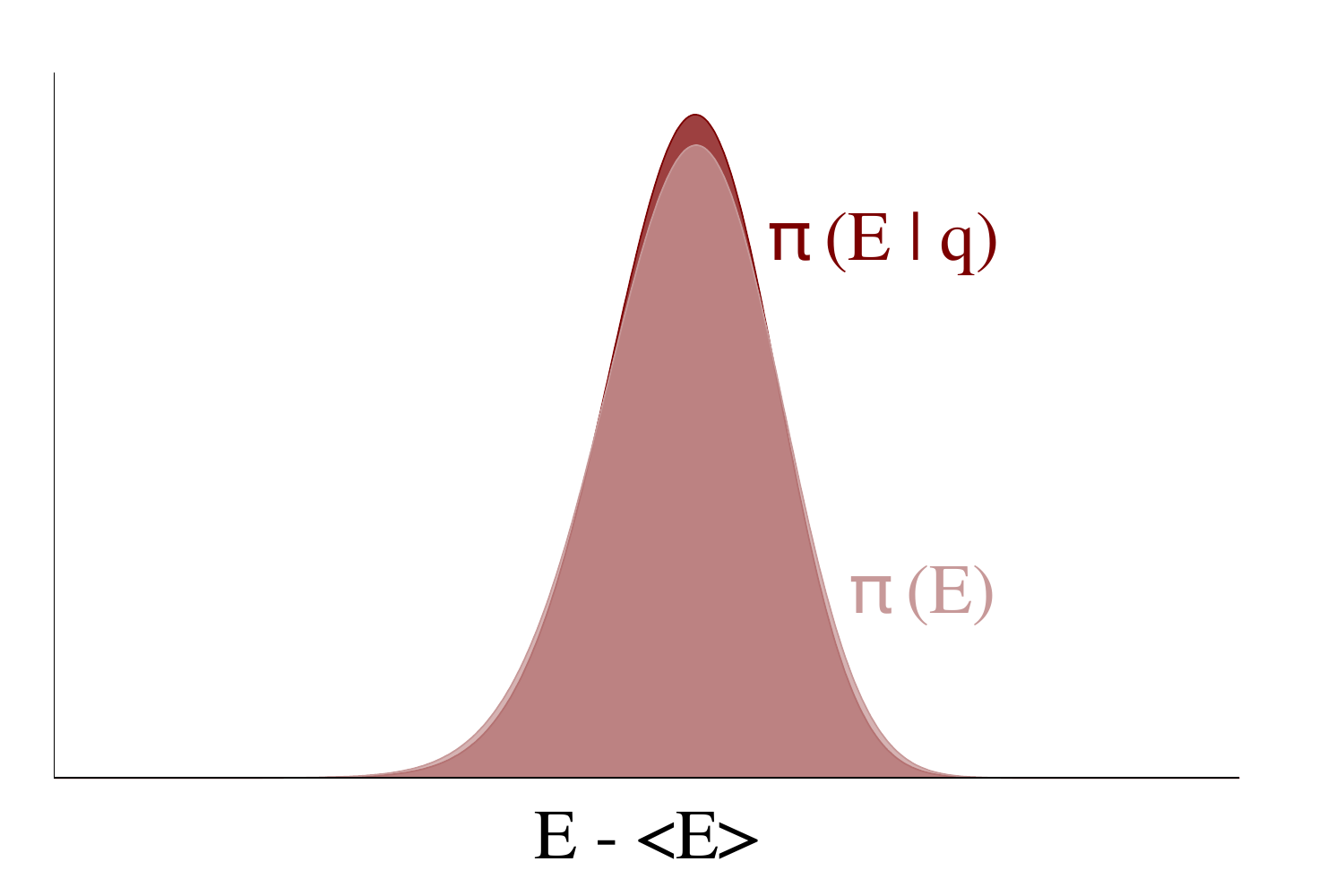} }
\caption{The momentum resampling in a Hamiltonian Markov transition randomly 
changes the energy, inducing a random walk between level sets.  (a) When the 
energy transition distribution, $\pi (E \mid q)$ is narrow relative to the 
marginal energy distribution, $\pi(E)$, this random walk will explore the 
marginal energy distribution only very slowly, requiring many expensive 
transitions to survey all of the relevant energies.  (b) On the other hand, 
when the two distributions are well-matched the random walk will explore the 
marginal energy distribution extremely efficiently.}
\label{fig:energy_marginals}
\end{figure}

Similarly, the performance of the stochastic exploration is determined by how 
quickly the random walk can diffuse across the energies typical to the marginal 
energy distribution.  Writing $\pi (E \mid q)$ as the transition distribution 
of energies induced by a momentum resampling at a given position, $q$, the 
diffusion speed depends on how heavy-tailed the marginal energy distribution 
is relative to $\pi (E \mid q)$.  For example, if this energy transition 
distribution is narrow relative to the marginal energy distribution 
(Figure \ref{fig:energy_marginals}a), then the random walk will proceed very 
slowly, taking many costly transitions to completely explore the target 
distribution.  If the energy transition distribution is similar to the marginal 
energy distribution (Figure \ref{fig:energy_marginals}b), however, then we will
generate nearly-independent samples from the marginal energy distribution at 
every transition, rapidly surveying the relevant energies with maximal efficiency.

By analyzing how these algorithmic degrees of freedom in the Hamiltonian Markov 
transition interact with the target distribution to determine the microcanonical 
geometry, we can determine how they affect the performance of the resulting 
Hamiltonian Markov chain.  In particular, we can construct criteria that identify 
the optimal choices of these degrees of freedom, which then motivate the effective 
tuning of the method, even for the complex target distributions encountered in 
practice.

\subsection{Optimizing the Choice of Kinetic Energy}
\label{sec:optimal_kinetic}

The first substantial degree of freedom in the Hamiltonian Monte Carlo
method that we can tune is the choice of the conditional probability 
distribution over the momentum or, equivalently, the choice of a kinetic 
energy function.  Along with the target distribution, this choice completes 
the probabilistic structure on phase space which then determines the 
geometry of the microcanonical decomposition.  Consequently, the ideal 
kinetic energy will interact with the target distribution to ensure that 
the energy level sets are as uniform as possible while the energy transition 
distribution matches the marginal energy distribution as well as possible.

Unfortunately, there is an infinity of possible kinetic energies and it 
would be impossible to search through them all to find an optimal choice 
for any given target distribution.  It is much more practical to search 
through a restricted family of kinetic energies, especially if that family 
is built from the structure of the problem itself.

\subsubsection{Euclidean-Gaussian Kinetic Energies}
\label{sec:euclidean_gaussian}

For example, in many problems the sample space is endowed with a Euclidean 
metric, $g$, that allows us to measure, amongst other quantities, the 
distance between any two points.  In a given parameterization, $g$ is
represented with a $D \times D$ matrix from which we can compute distances
as
\begin{equation*}
\Delta ( q, q' ) =  ( q - q' )^{T} \cdot g \cdot ( q - q' ),
\end{equation*}
Moreover, we can construct an entire family of modified Euclidean metrics, 
$M$, by scaling and then rotating this natural metric,
\begin{equation*}
M = R \cdot S \cdot g \cdot S^{T} \cdot R^{T},
\end{equation*}
where $S$ is a diagonal scaling matrix and $R$ is an orthogonal rotation
matrix. 

Any such Euclidean structure on the target parameter space immediately 
induces an inverse structure on the momentum space, allowing us to measure
distances between momenta,
\begin{equation*}
\Delta ( p, p' ) =  (p - p')^{T} \cdot M^{-1} \cdot ( p - p' ).
\end{equation*}

Finally, distances in momentum space allow us to construct many common
probability distributions over the momentum, such as a Gaussian 
distribution centered at $0$,
\begin{equation*}
\pi \! \left( p \mid q \right) 
=
\mathcal{N} \! \left( p \mid 0, M \right).
\end{equation*}
This particular choice defines a \emph{Euclidean-Gaussian} kinetic energy,
\begin{equation*}
K \! \left(q, p \right)
=
\frac{1}{2} p^{T} \cdot M^{-1} \cdot p
+ \log \left| M \right| + \mathrm{const}.
\end{equation*}
In the physical perspective the Euclidean metric is known as the \emph{mass 
matrix}, a term that has consequently become common in the Hamiltonian Monte 
Carlo literature.

Because the Euclidean structure over the momentum is dual to the Euclidean 
structure over the parameters, its interactions with the target distribution 
are straightforward to derive.  Applying the transformation
$p' = \sqrt{ M^{-1} } p$ simplifies the kinetic energy, but remember that
we have to apply the \emph{opposite} transformation to the parameters,
$q' = \sqrt{ M } q$, to preserve the Hamiltonian geometry.  Consequently, 
a choice of $M^{-1}$ effectively rotates and then rescales the target 
parameter space, potentially correlating or de-correlating the target 
distribution and correspondingly warping the energy level sets.  

In particular, as the inverse Euclidean metric more closely resembles 
the covariance of the target distribution it de-correlates the target 
distribution, resulting in energy level sets that are more and more 
uniform and hence easier to explore.  We can then readily optimize over 
the family of Euclidean-Gaussian kinetic energies by setting the inverse 
Euclidean metric to the target covariances,
\begin{equation*}
M^{-1} = 
\mathbb{E}_{\pi} [ \left( q - \mu \right) \left( q - \mu \right)^{T} ].
\end{equation*}

In practice we can compute an empirical estimate of the target covariance 
using the Markov chain itself in an extended warm-up phase.  After first 
converging to the typical set we run the Markov chain using a default 
Euclidean metric for a short window to build up an initial estimate of 
the target covariance, then update the metric to this estimate before 
running the now better-optimized chain to build up an improved estimate.  
A few iterations of this adaptation will typical yield an accurate 
estimate of the target covariance and hence a near-optimal metric.

\subsubsection{Riemannian-Gaussian Kinetic Energies}

Unless the target distribution is exactly Gaussian, however, no global
rotation and rescaling will yield completely uniform level sets; locally 
the level sets can still manifest strong curvature that slows the 
exploration of the Hamiltonian trajectories.  To improve further we 
need to introduce a \emph{Riemannian} metric which, unlike the Euclidean
metric, varies as we move through parameter space.  A Riemannian structure 
allows us to construct a Gaussian distribution over the momentum whose 
covariance depends on our current position in parameter space,
\begin{equation*}
\pi \! \left( p \mid q \right) 
=
\mathcal{N} \! \left( p \mid 0, \Sigma \! \left( q \right) \right),
\end{equation*}
which then defines a \emph{Riemannian-Gaussian} kinetic energy,
\begin{equation*}
K \! \left(q, p \right)
=
\frac{1}{2} p^{T} \cdot \Sigma^{-1} \! \left( q \right) \cdot p
+ \frac{1}{2} \log \left| \Sigma \! \left( q \right) \right| 
+ \mathrm{const}.
\end{equation*}
The resulting implementation of Hamiltonian Monte Carlo is known
as \emph{Riemannian Hamiltonian Monte Carlo}~\citep{GirolamiEtAl:2011}.

The variation of the inverse Riemannian metric allows it to make local
corrections to the target distribution that vary depending on where we
are in parameter space.  If the metric resembles the Hessian of the 
target distribution then these local corrections will rectify the 
spatially-varying correlations of the target distribution, ensuring 
extremely uniform level sets and efficient exploration.  Technical care, 
however, must be taken when trying to mimic the Hessian to ensure that 
the inverse Riemannian metric, and hence the entire kinetic energy is 
well-behaved; the SoftAbs metric~\citep{Betancourt:2013b} accomplishes 
this with a heuristic regularization that works well in practice, although 
formalizing this procedure is an active topic of research~\citep{HolmesEtAl:2014}.  

An additional benefit of the Riemannian-Gaussian kinetic energy is that 
the variation of the log determinant, 
$\frac{1}{2} \log \left| \Sigma \! \left( q \right) \right|$, modifies 
the marginal energy distribution.  If the metric is well-chosen then this
modification can bring the marginal energy distribution closer to the energy 
transition distribution, significantly improving the performance of the
Hamiltonian transition when targeting complex 
distributions~\citep{BetancourtEtAl:2015}.

\subsubsection{Non-Gaussian Kinetic Energies}

In theory we are not limited to Gaussian distributions over the momentum --
a Euclidean or Riemannian structure allows us to construct any distribution
with quadratic sufficient statistics.  In particular, why should we not 
consider momentum distributions with particularly heavy or light tails?
Although there is not much supporting theory, empirically non-Gaussian 
kinetic energies tend to perform poorly, especially in high-dimensional 
problems.  Some intuition for the superiority of Gaussian kinetic energies 
may come from considering the asymptotics of the marginal energy distribution.  
As we target higher and higher dimensional models, the marginal energy 
distribution becomes a convolution of more and more parameters and, under 
relatively weak conditions, it tends to follow a central limit theorem.  
When the marginal energy distribution converges towards a Gaussian, only
a Gaussian distribution over the momentum will yield the optimal energy 
transition.

\subsection{Optimizing the Choice of Integration Time}
\label{sec:optimal_integration_time}

The choice of a kinetic energy completely specifies the microcanonical 
geometry and, consequently, the shape of the energy level sets.  How 
effectively each Hamiltonian trajectory explores those level sets is then
determined entirely by the choice of integration times across phase space,
$T \left(q, p \right)$.  Intuitively, if we integrate for only a short time 
then we don't take full advantage of the coherent exploration of the 
Hamiltonian trajectories and we will expect performance to suffer.  On the 
other hand, because the level sets are topologically compact in well-behaved 
problems, trajectories will eventually return to previously explored 
neighborhoods and integrating too long can suffer from diminishing returns.

This intuition is formalized in the notion of 
\emph{dynamic ergodicity}~\citep{Betancourt:2016}.  Here we consider 
the \emph{orbit}, $\phi$, of a trajectory, consisting of all points that 
the trajectory will reach as the integration time is increased to infinity.  
The orbit might encompass the entire level set or it could be limited to a
just subset of the level set, but in either case any trajectory will explore 
the microcanonical distribution restricted to its orbit.  Dynamic ergodicity 
guarantees that a uniform sample from a trajectory will more closely resemble 
a sample from this restricted microcanonical distribution as the integration 
time is increased and the trajectory grows.  In other words, as the 
integration time grows the temporal expectation over the trajectory converges 
to the spatial expectation over its orbit.

\begin{figure}
\centering
\subfigure[]{ \includegraphics[width=2.5in]{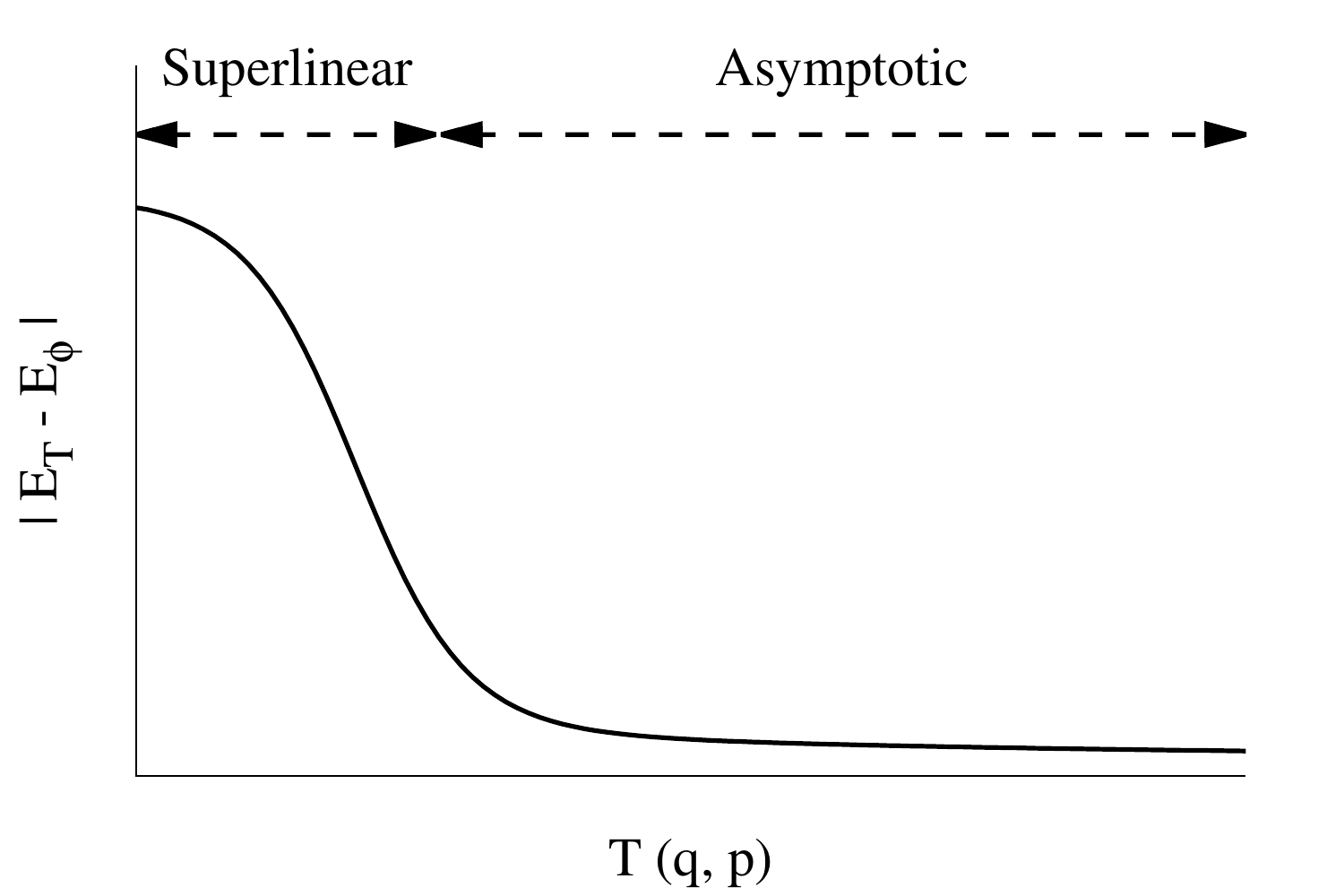} } \\
\subfigure[]{ \includegraphics[width=2.5in]{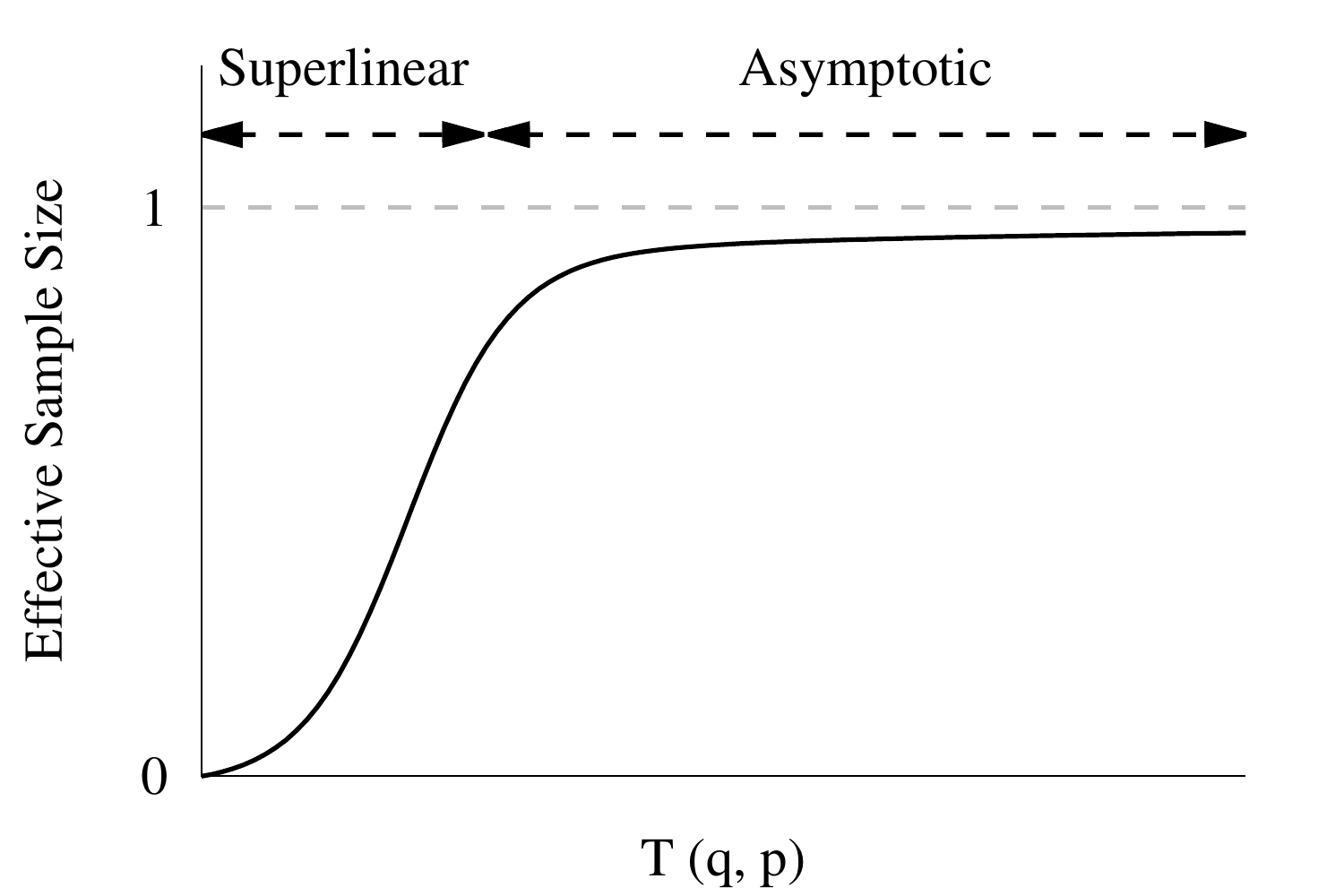} }
\subfigure[]{ \includegraphics[width=2.5in]{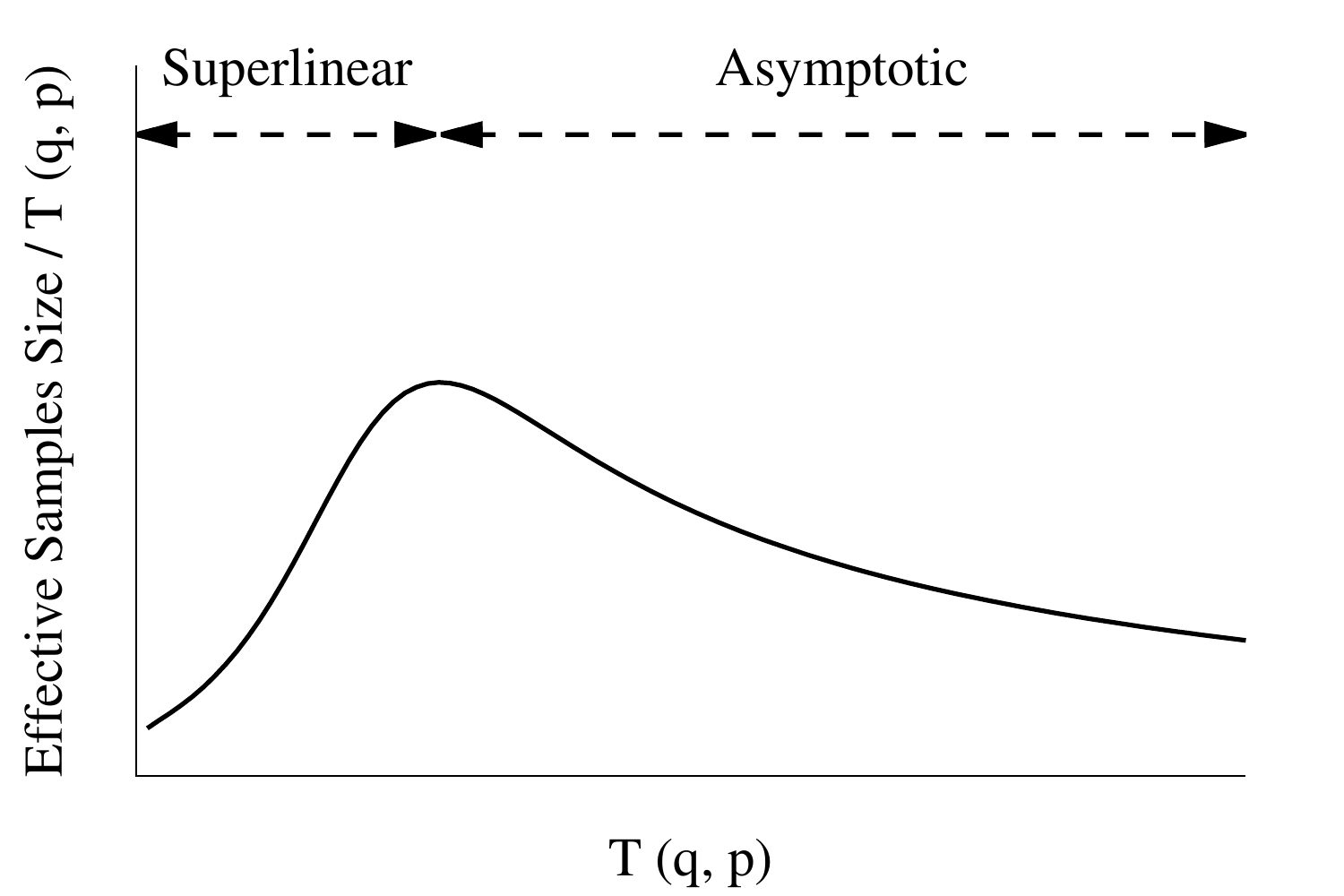} }
\caption{(a) Temporal averages along a Hamiltonian trajectory, $\mathbb{E}_{T}$,
converge to the corresponding spatial expectation over its orbit, 
$\mathbb{E}_{\phi}$, as the integration time, $T$, increases. (b) Correspondingly, 
a uniform sample from the trajectory converges to a sample from the microcanonical 
distribution restricted to the trajectory's orbit, here represented by the 
effective sample size.  Typically this convergence is initially rapid and 
superlinear before settling into an asymptotic regime where the convergence continues 
only with the square of the integration time.  (c) Because the cost of generating each 
trajectory scales with the integration time, those integration times that identify the 
transition between these two regimes will yield optimal performance.}
\label{fig:microcanonical_convergence}
\end{figure}

The performance of the Hamiltonian transition, however, depends on the 
\emph{rate} at which these expectations converge.  Given typical regularity 
conditions, the temporal expectation will initially converges toward the spatial 
expectation quite rapidly (Figure \ref{fig:microcanonical_convergence}a, b),
consistent with our intuition that coherent exploration is extremely effective.  
Eventually, however, that convergence slows, and we enter an asymptotic 
regime where any benefit of exploration comes at an ever increasing cost.
The optimal integration time straddles these two regimes, exploiting the
coherent exploration early on but not wasting computation on the diminishing
returns of long integration times (Figure \ref{fig:microcanonical_convergence}c).

\begin{figure}
\centering
\subfigure[]{ \includegraphics[width=2.5in]{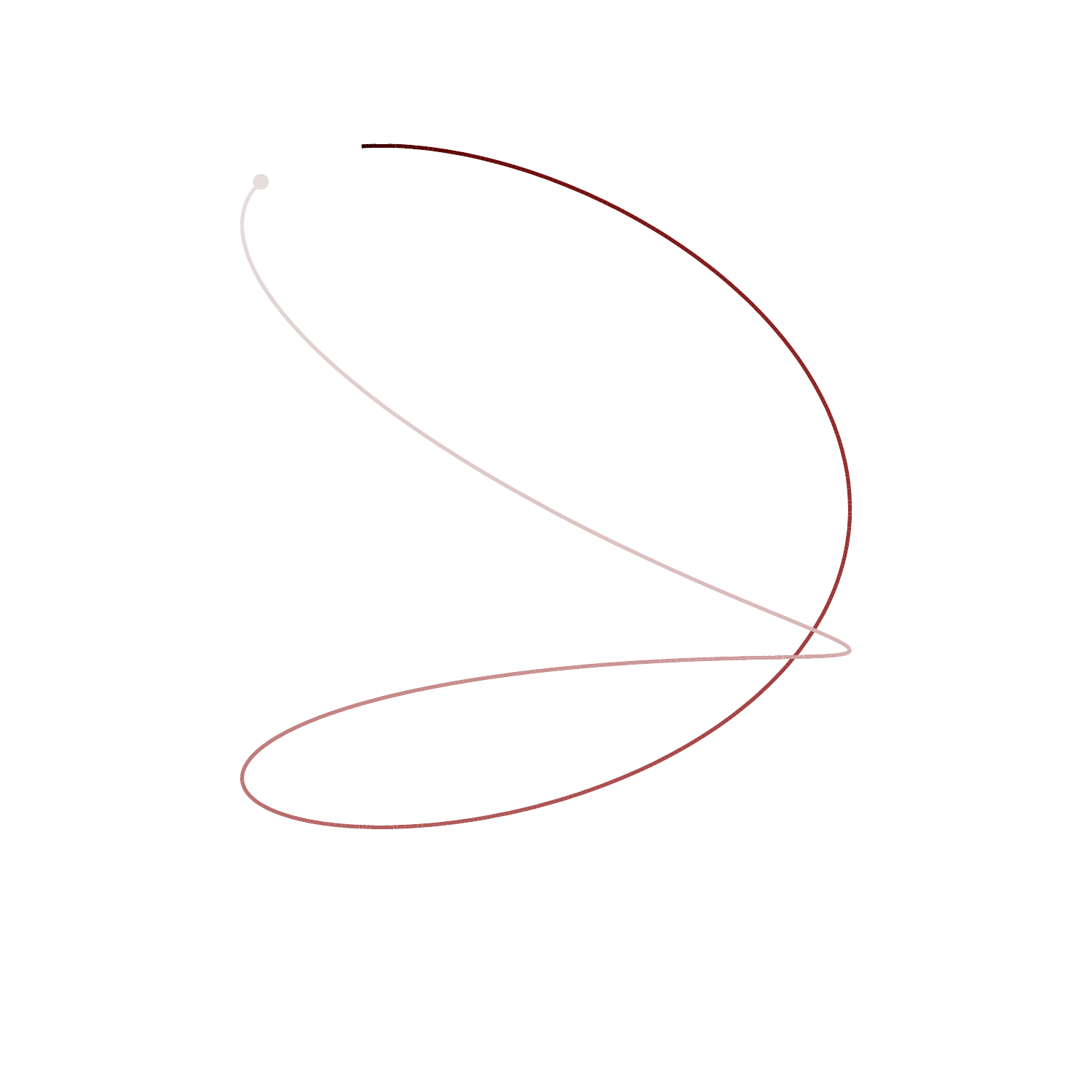} } 
\subfigure[]{ \includegraphics[width=2.5in]{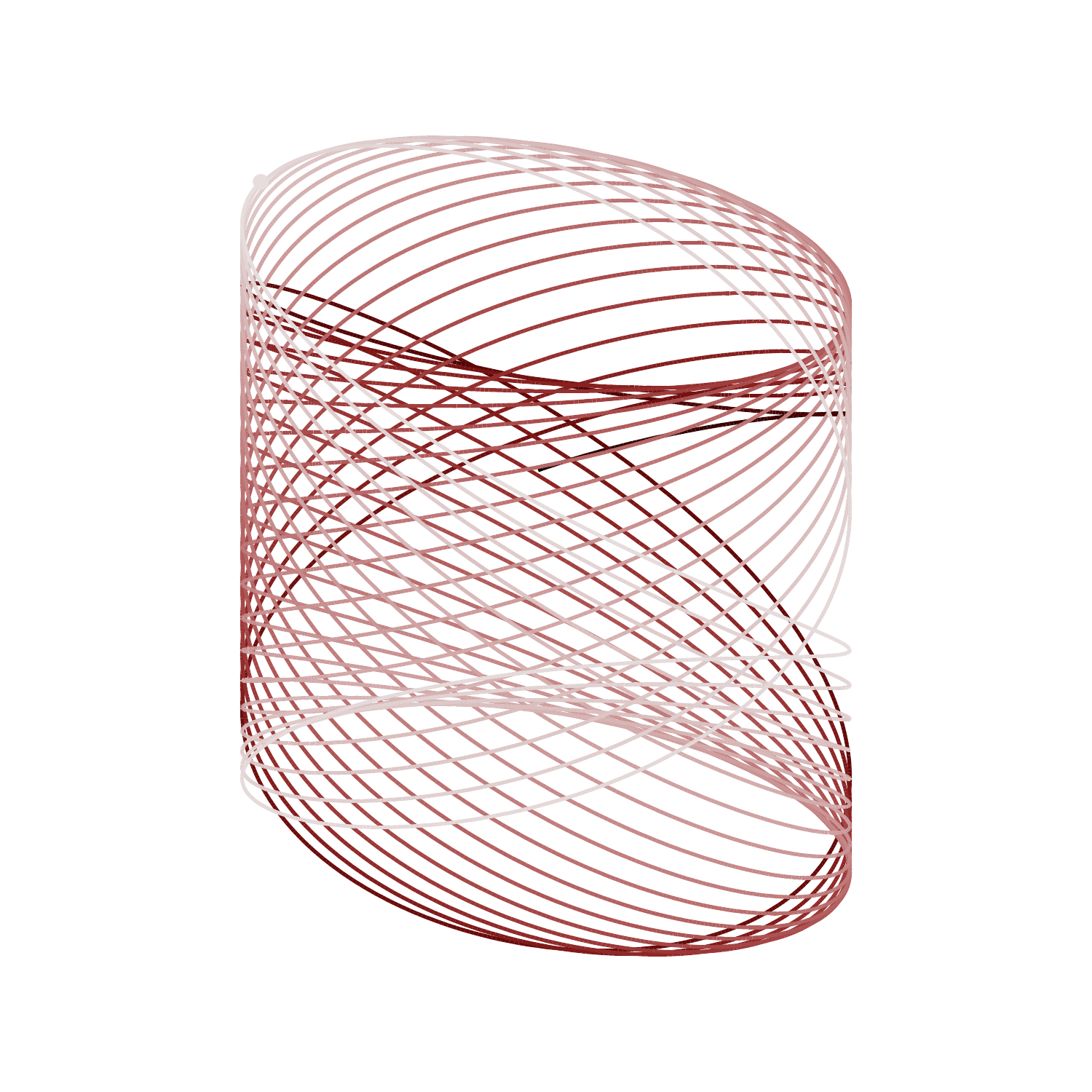} }
\caption{(a) Because a Hamiltonian trajectory's first sojourn through its orbit 
continuously encounters new, un-surveyed neighborhoods, the corresponding
exploration is extremely efficient.  (b) Longer trajectories return to these 
neighborhoods, refining this initial exploration and yielding better, but much
slower, convergence.}
\label{fig:microcanonical_convergence_cartoon}
\end{figure}

This optimization criterion also has a helpful geometric interpretation.  The 
superlinear regime corresponds to the first sojourn around the orbit of the 
trajectory, where every new step forwards is new and informative (Figure 
\ref{fig:microcanonical_convergence_cartoon}a).  Eventually, however, the 
trajectory returns to neighborhoods it has already explored and enters into 
the asymptotic regime (Figure \ref{fig:microcanonical_convergence_cartoon}b).  
This additional exploration refines the exploration of the orbit, improving 
the accuracy of the temporal expectation only very slowly. 

\begin{figure}
\centering
\includegraphics[width=3in]{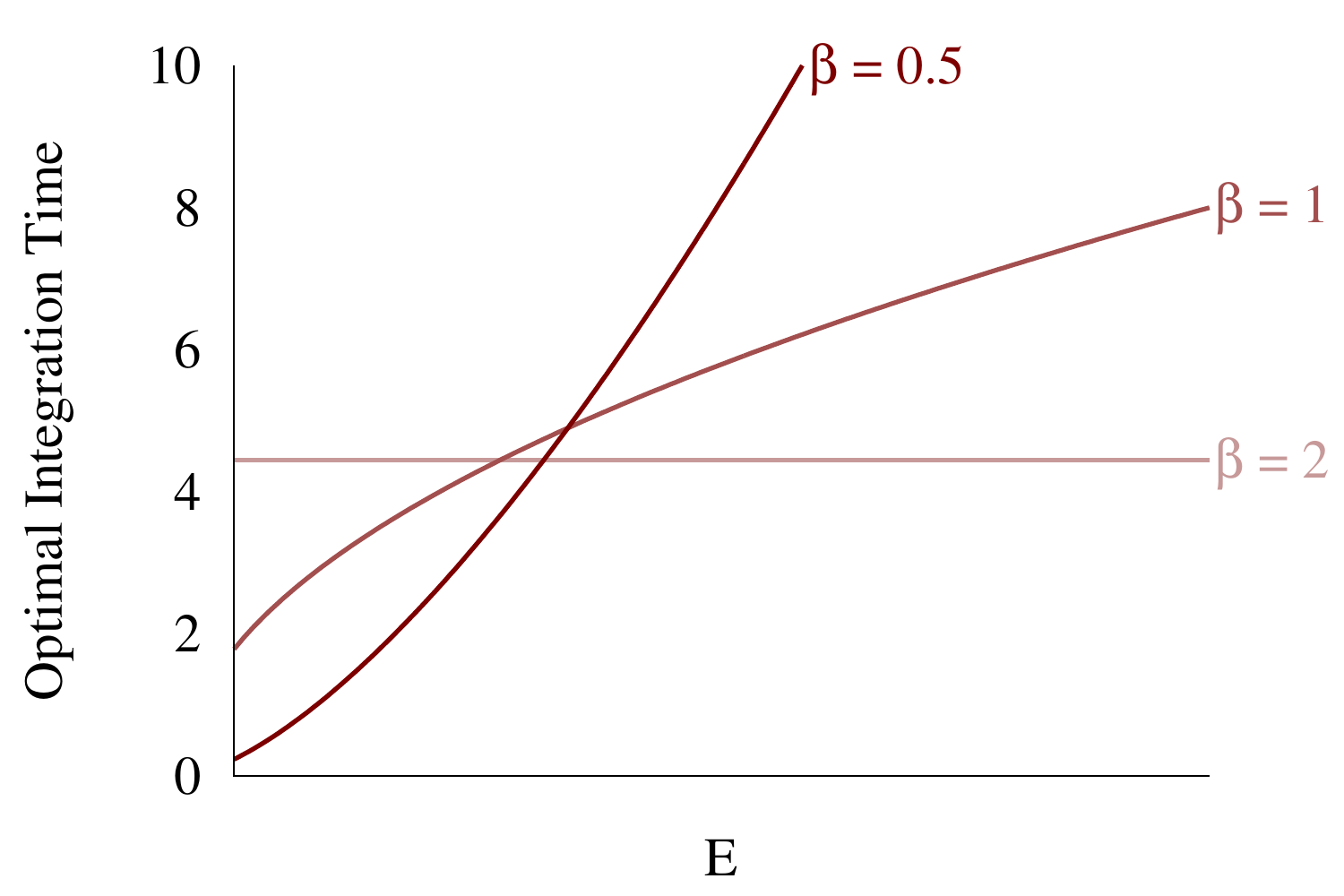}
\caption{The spatial variation of optimal integration times is evident in a 
one-dimensional example where the optimal integration times can be calculated 
analytically and are found to scale with the energy of the initial point in 
phase space.  Only for a Gaussian target distribution, $\beta = 2$, are the 
optimal integration times constant for all energies.  The optimal integration 
time for heavy-tailed target distributions, $\beta < 2$, grows larger as we 
move deeper into the tails with higher energies.  A given static integration 
time might suffice for low energies, but it will yield trajectories that are 
much too short for any effective exploration of the higher energy level sets.}
\label{fig:opt_int_time}
\end{figure}

In general, this optimal integration time will vary strongly depending on which
trajectory we are considering: no single integration time will perform well 
everywhere.  We can make this explicit in one dimension where the optimal
integration times can be identified analytically.  For example, for the family of
target densities
\begin{equation*}
\pi_{\beta} \! \left( q \right) \propto e^{- \left| q \right|^{\beta} },
\end{equation*}
with the Euclidean-Gaussian kinetic energy
\begin{equation*}
\pi (p \mid q) = \mathcal{N} \! \left( 0, 1 \right),
\end{equation*}
the optimal integration time scales with the energy of the level set containing
the trajectory,
\begin{equation*}
T_{\mathrm{optimal}} \! \left( q, p \right) 
\propto 
\left(  H \! \left( q, p \right) \right)^{\frac{2 - \beta}{2 \beta}}.
\end{equation*}
In particular, when the target distribution is heavy-tailed, $\beta < 2$, the 
optimal integration time will quickly grow as we move from trajectories exploring 
the bulk to trajectories exploring the tails (Figure \ref{fig:opt_int_time}).  
Consequently, the exploration generated by trajectories with any \emph{static} 
integration time will decay and the Hamiltonian Markov chain will slow to a crawl.

Hence if we want to fully exploit these Hamiltonian trajectories then we need 
to identify the optimal integration time \emph{dynamically}, as we generate the 
trajectories themselves.  How exactly do we identify when a trajectory has reached 
its optimal length?  One heuristic is the 
No-U-Turn termination criterion~\citep{HoffmanEtAl:2014, Betancourt:2013a} which,
like the kinetic energies discussed in Section \ref{sec:optimal_kinetic}, utilizes 
a Euclidean or Riemannian structure on the target parameter space.  The explicit
form of the No-U-Turn termination criterion is introduced in Appendix
\ref{sec:term_criteria}.

In some simple cases the near-optimality of the No-U-Turn criterion can be shown 
rigorously, but it has proven a empirical success on an incredibly diverse set of 
target distributions encountered in applied problems.  More recently proposed 
possibilities are exhaustive termination criteria~\citep{Betancourt:2016} which 
utilize the microcanonical geometry itself to identify the optimal stopping time.  
Exhaustive termination criteria can be more robust than the No-U-Turn termination 
criterion, but they require careful tuning which is an open topic of research.

\section{Implementing Hamiltonian Monte Carlo in Practice}

With careful tuning, the Hamiltonian Monte Carlo method defines a powerful 
Markov transition capable of performing well over a large class of target 
distributions, at least in theory.  Unfortunately, there are almost no 
Hamiltonian transitions that are immediately applicable \emph{in practice}.

The main obstruction to implementing the Hamiltonian Monte Carlo method is 
generating the Hamiltonian trajectories themselves.  Aside from a few trivial 
examples, we cannot solve Hamilton's equations exactly and any implementation 
must instead solve them numerically.  Numerical inaccuracies, however, can 
quickly compromise the utility of even the most well-tuned Hamiltonian 
transition.  Formally, integrating along the vector field defined by 
Hamilton's equations is equivalent to solving a system of ordinary 
differential equations on phase space.  The more accurately we can numerically 
solve this system, the more  effective our implementation will be.

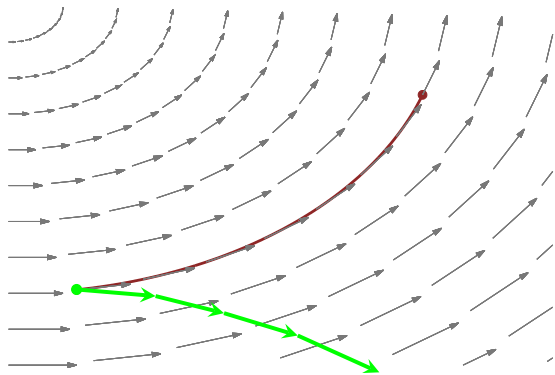
\begin{figure*}
\centering
\begin{tikzpicture}[scale=0.35, thick]
  \node[] at (0, 0) {\includegraphics[width=8cm]{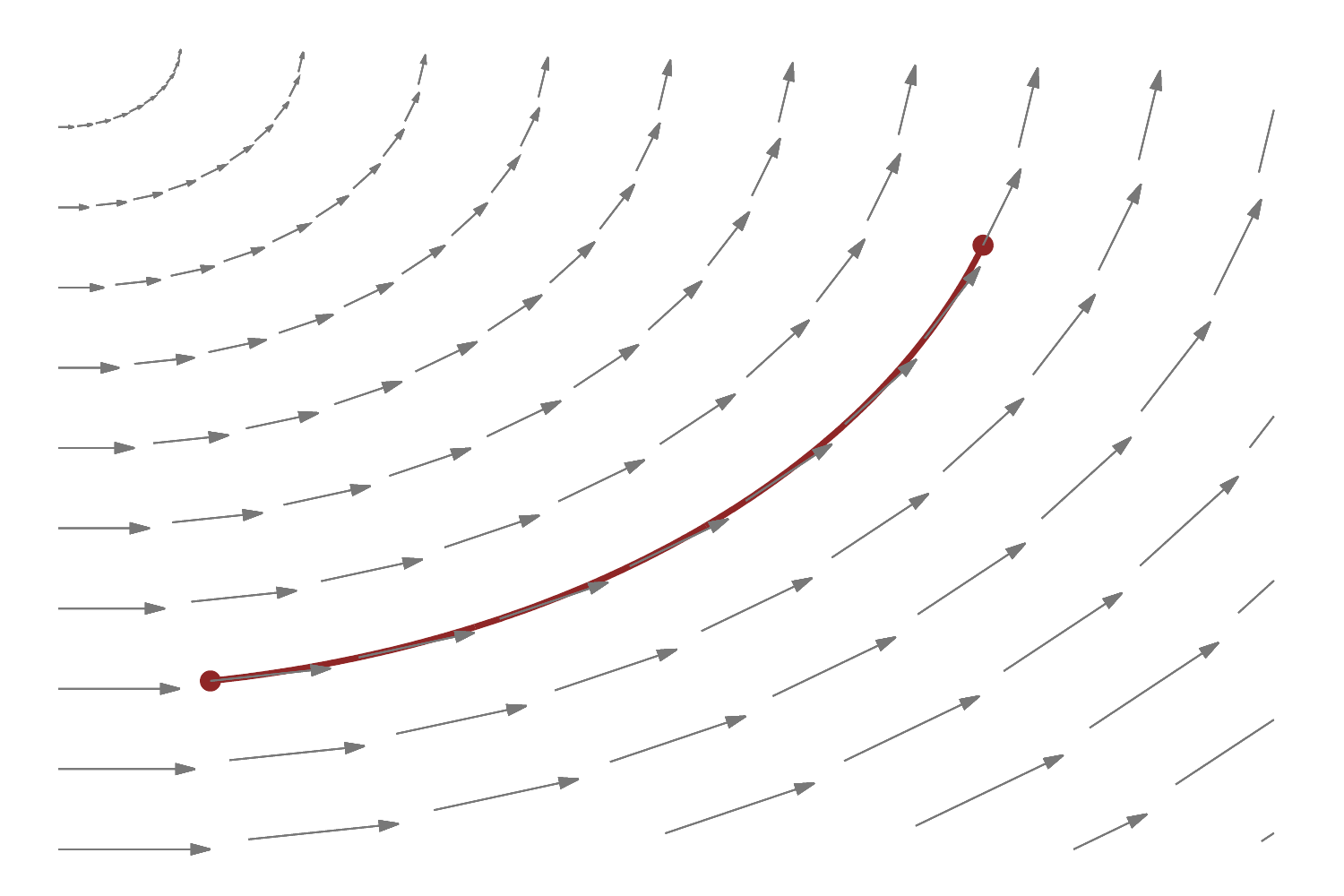}};
  \fill[color=green] (-7.85, -3.95) circle (6pt);
  \draw[->, >=stealth, >=stealth, line width=1.5, green] 
  (-7.85, -3.95) -- +(3, -0.25);
  \draw[->, >=stealth, >=stealth, line width=1.5, green] 
  (-4.85, -4.2) -- +(2.6, -0.65);
  \draw[->, >=stealth, >=stealth, line width=1.5, green] 
  (-2.25, -4.85) -- +(2.8, -0.85);
  \draw[->, >=stealth, >=stealth, line width=1.5, green] 
  (0.55, -5.7) -- +(3.1, -1.4);
\end{tikzpicture}
\caption{The approximate solutions of most numerical integrators tend to 
drift away from the exact solutions.  As the system is integrated longer 
and longer, errors add coherently and push the numerical trajectory away 
from the exact trajectory.
}
\label{fig:naive_integrator}
\end{figure*}

While there is an abundance of ordinary differential equations solvers, or 
numerical integrators, available in popular computational libraries, most of 
those solvers suffer from an unfortunate \emph{drift}.  As we numerically 
solve longer and longer trajectories the error in the solvers adds coherently, 
pushing the approximate trajectory away from the true trajectory and the typical 
set that we want to explore (Figure \ref{fig:naive_integrator}).  Because the 
magnitude of this drift rapidly increases with the dimension of phase space, 
the utility of these generic numerical integrators is limited to approximating 
only short Hamiltonian trajectories that inefficiently explore the energy level 
sets.

Fortunately, we can use the geometry of phase space itself to construct
an extremely powerful family of numerical solvers, known as 
\emph{symplectic integrators}~\citep{LeimkuhlerEtAl:2004, HairerEtAl:2006},
that are robust to phenomena like drift and enable high-performance
implementations of the Hamiltonian Monte Carlo method.  In this section I 
will present the practical properties of symplectic integrators and how we 
can correct for the small errors that they introduce.  We will conclude with 
a discussion of how to choose the best symplectic integrator for a given 
problem.

This section will follow the conceptual presentation of the review, but
given the importance of the material a more thorough discussion of the
technical details is available in Appendix \ref{sec:technical_details}.

\subsection{Symplectic Integrators}

\begin{figure*}
\centering
\begin{tikzpicture}[scale=0.35, thick]
  \node[] at (0, 0) {\includegraphics[width=8cm]{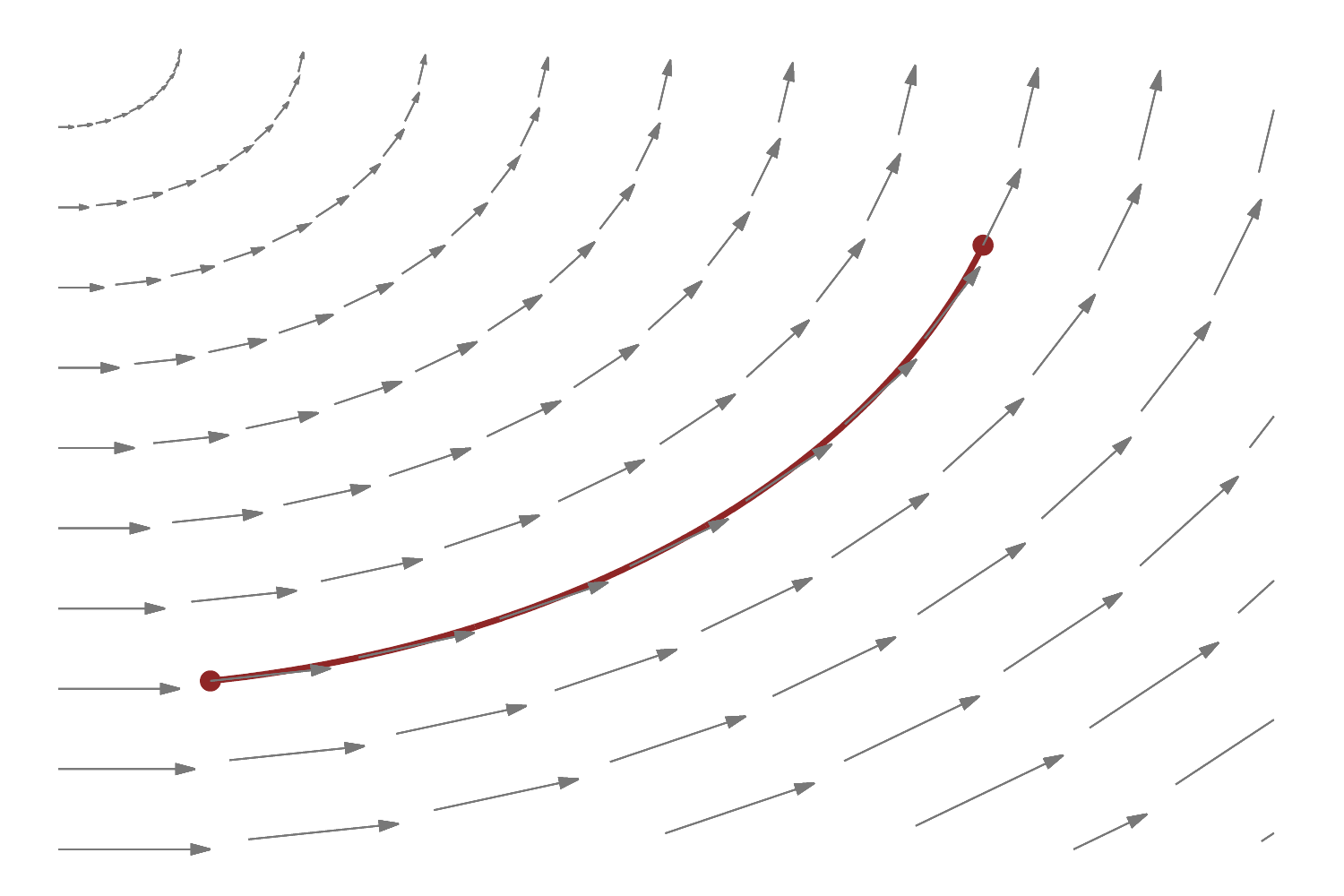}};
  \fill[color=green] (-7.85, -3.95) circle (6pt);
  \draw[->, >=stealth, >=stealth, line width=1.5, green] 
  (-7.85, -3.95) -- +(3, -0.25);
  \draw[->, >=stealth, >=stealth, line width=1.5, green] 
  (-4.85, -4.2) -- +(2.75, 0.5);
  \draw[->, >=stealth, >=stealth, line width=1.5, green] 
  (-2.1, -3.7) -- +(2.25, 1.75);
  \draw[->, >=stealth, >=stealth, line width=1.5, green] 
  (0.15, -1.95) -- +(2.75, 0.75);
  \draw[->, >=stealth, >=stealth, line width=1.5, green] 
  (2.9, -1.2) -- +(2.5, 1.75);
  \draw[->, >=stealth, >=stealth, line width=1.5, green] 
  (5.4, 0.55) -- +(0.25, 2.25);
\end{tikzpicture}
\caption{Symplectic integrators generate numerical trajectories that are
incompressible like the exact Hamiltonian trajectory they approximate.
Consequently their approximation error cannot add up coherently to pull 
the numerical trajectories away from the exact trajectories.  Instead 
the numerical trajectories oscillate around the exact level set, even as 
we integrate for longer and longer times.
}
\label{fig:symplectic_integrator}
\end{figure*}

Symplectic integrators are powerful because the numerical trajectories they 
generate exactly preserve phase space volume, just like the Hamiltonian 
trajectories they are approximating.  This incompressibility limits how much 
the error in the numerical trajectory can deviate from the energy of the exact 
trajectory.  Consequently, the numerical trajectories cannot drift away from
the exact energy level set, instead oscillating near it even for long integration
times (Figure \ref{fig:symplectic_integrator}).

Conveniently for implementations, symplectic integrators are also 
straightforward to implement in practice.  For example, if the 
probabilistic distribution of the momentum is chosen to be independent of 
position, as with the Euclidean-Gaussian kinetic energy, then we can employ 
the deceptively simple \emph{leapfrog integrator}.  Given a time 
discretization, or step size, $\epsilon$, the leapfrog integrator simulates 
the exact trajectory as
\begin{algorithmic}
\STATE $q_{0} \gets q, p_{0} \gets p$
\FOR{ $0 \leq n < \llcorner \, T / \epsilon \, \lrcorner$ }
\vspace{1mm}
\STATE $p_{n + \frac{1}{2}} \gets
               p_{n} - \frac{\epsilon}{2} \, 
               \frac{ \partial V }{ \partial q} \! \left( q_{n} \right)$
\vspace{1mm}
\STATE $q_{n + 1} \gets q_{n} + \epsilon \, p_{n + \frac{1}{2}}$
\vspace{1mm}
\STATE $p_{n + 1} \gets 
               p_{n + \frac{1}{2}} 
               - \frac{\epsilon}{2} \, 
               \frac{ \partial V }{ \partial q} \! \left( q_{n + 1} \right)$
\vspace{1mm}
\ENDFOR.
\end{algorithmic}
This simple but precise interleaving of discrete momentum and position updates 
ensures exact volume preservation on phase space, and hence the accurate 
numerical trajectories we need to realize the potential of a Hamiltonian 
transition.

\begin{figure*}
\centering
\begin{tikzpicture}[scale=0.35, thick]
  \node[] at (0, 0) {\includegraphics[width=8cm]{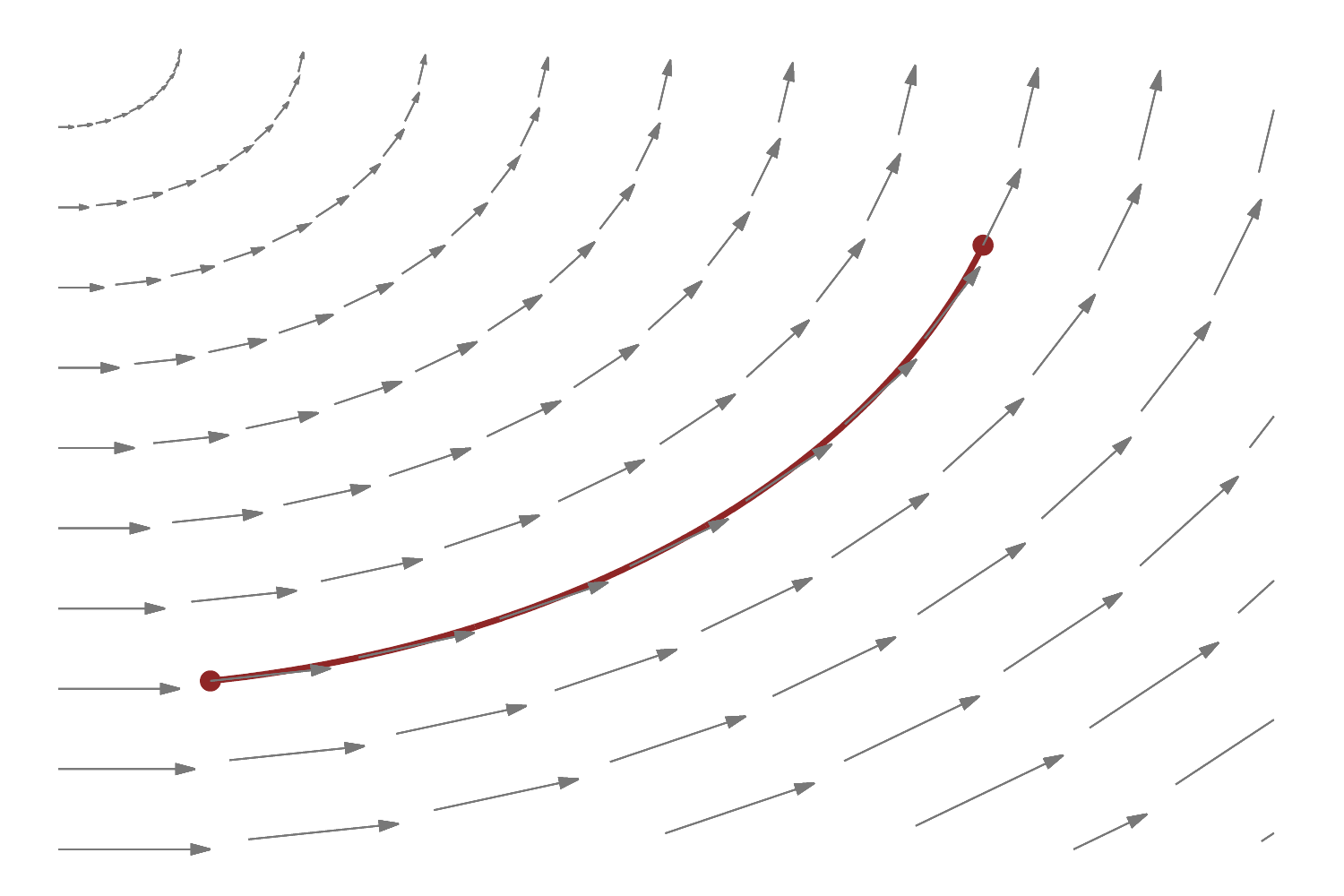}};
  \fill[color=green] (-7.85, -3.95) circle (6pt);
  \draw[->, >=stealth, >=stealth, line width=1.5, green] 
  (-7.85, -3.95) -- +(3, -0.25);
  \draw[->, >=stealth, >=stealth, line width=1.5, green] 
  (-4.85, -4.2) -- +(2.75, 0.5);
  \draw[->, >=stealth, >=stealth, line width=1.5, green] 
  (-2.1, -3.7) -- +(2.25, 1.75);
  \draw[->, >=stealth, >=stealth, line width=1.5, green] 
  (0.15, -1.95) -- +(1, 4);
  \draw[->, >=stealth, >=stealth, line width=1.5, green] 
  (1.15, 2.05) -- +(0.25, 6);
\end{tikzpicture}
\caption{When the exact trajectories lie on energy level sets with regions of 
high curvature, the numerical trajectories generated by symplectic integrators 
can diverge, rapidly flying off towards the boundaries of phase space.  Unlike 
the slower drift of generic numerical integrators, these divergences, and hence
the failure of the symplectic integrator to provide an accurate solution, are 
straightforward to identify in practice.
}
\label{fig:diverging_integrator}
\end{figure*}

There is, however, one important exception to performance of symplectic 
integrators.  Long time accuracy can be compromised when the exact energy level 
sets feature neighborhoods of high curvature that the finite time discretization 
is not able to resolve.  These neighborhoods induce a \emph{divergence} that almost 
immediately propels the numerical trajectory towards infinite energies (Figure
\ref{fig:diverging_integrator}).  This distinctive behavior proves beneficial in 
practice, however, because it makes the failures of a symplectic integrator 
straightforward to identify and hence diagnose.

Employing symplectic integrators provides the opportunity to translate the 
theoretical performance of the Hamiltonian Monte Carlo method into a practical 
implementation.  There remain, however, two obstructions to realizing this 
translation.  First, even though symplectic integrators are highly accurate, 
the small errors they do introduce will bias the resulting Hamiltonian transitions 
without an exact correction.  Second, we have to be able to select a symplectic 
integrator well-suited to a given target distribution.

\subsection{Correcting for Symplectic Integrator Error}

One particularly natural strategy for correcting the bias introduced by the error 
in a symplectic integrator is to treat the Hamiltonian transition as the proposal 
for a Metropolis-Hastings scheme on phase space.  If we can construct the acceptance 
probability analytically then this correction will yield an exact sample from the 
canonical distribution on phase space which then projects to an exact sample from 
our target distribution.  In order to construct that acceptance probability, 
however, we have to carefully augment the Hamiltonian transition.

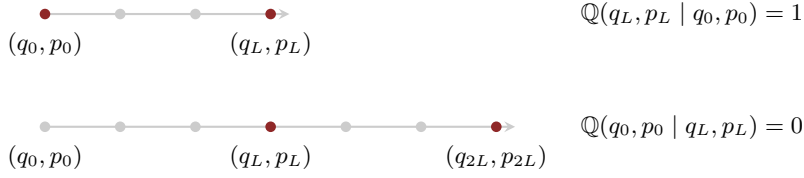
\begin{figure*}
\centering
\begin{tikzpicture}[scale=0.5, thick]

\draw[->, >=stealth, >=stealth, color=gray80] (0, 3) -- +(6.5,0);

\foreach \i in {0, 3} {
 \fill[color=dark] (2 * \i, 3) circle (4pt);
}

\foreach \i in {1, 2} {
 \fill[color=gray80] (2 * \i, 3) circle (4pt);
}

\fill[color=dark] (0, 3 - 0.25) circle (0pt)
node[below, color=black] { $(q_{0}, p_{0})$ };

\fill[color=dark] (6, 3 - 0.25) circle (0pt)
node[below, color=black] { $(q_{L}, p_{L})$ };

\fill[] (14, 3) circle (0pt)
node[right, color=black] { $\mathbb{Q} ( q_{L}, p_{L} \mid q_{0}, p_{0} ) = 1$ };

\draw[->, >=stealth, >=stealth, color=gray80] (0, 0) -- +(12.5,0);

\foreach \i in {0, 1, 2, 4, 5} {
 \fill[color=gray80] (2 * \i, 0) circle (4pt);
}

\foreach \i in {3, 6} {
 \fill[color=dark] (2 * \i, 0) circle (4pt);
}

\fill[color=dark] (0, -0.25) circle (0pt)
node[below, color=black] { $(q_{0}, p_{0})$ };

\fill[color=dark] (6, -0.25) circle (0pt)
node[below, color=black] { $(q_{L}, p_{L})$ };

\fill[color=dark] (12, -0.25) circle (0pt)
node[below, color=black] { $(q_{2L}, p_{2L})$ };

\fill[] (14, 0) circle (0pt)
node[right, color=black] { $\mathbb{Q} ( q_{0}, p_{0} \mid q_{L}, p_{L} ) = 0$ };

\end{tikzpicture}
\caption{Because Hamiltonian trajectories, and their numerical approximations, are 
deterministic and non-reversible, Metropolis-Hastings proposals are always rejected.
In particular, we have positive proposal probabilities going forwards in time but
vanishing proposal probabilities going backwards in time which renders the
Metropolis-Hastings acceptance probability identically zero.
}
\label{fig:nonreversible_trajectories}
\end{figure*}

For example, consider a simple scheme where we integrate the initial state, 
$(q_{0}, p_{0})$, forward for some static time, or equivalently $L$ symplectic 
integrator steps, and then propose the last state of the numerical trajectory,
$(q_{L}, p_{L})$,
\begin{equation*}
\mathbb{Q} ( q', p' \mid q_{0}, p_{0} )
=
\delta (q' - q_{L} ) \, \delta (p' - p_{L} ).
\end{equation*}
Because we can propose only states going forwards and not backwards,
(Figure \ref{fig:nonreversible_trajectories}), the ratio of proposal
densities, and hence the Metropolis-Hastings acceptance probability, 
always vanishes,
\begin{equation*}
\frac{ \mathbb{Q} ( q_{0}, p_{0} \mid q_{L}, p_{L} ) } 
{ \mathbb{Q} ( q_{L}, p_{L} \mid q_{0}, p_{0} ) }
=
\frac{0}{1}.
\end{equation*}

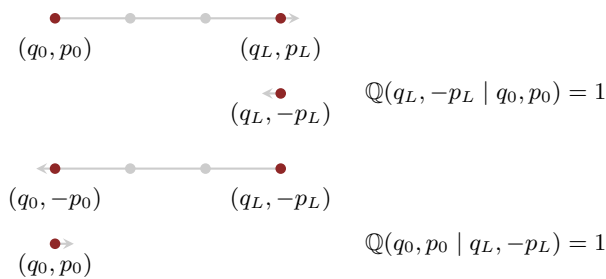
\begin{figure*}
\centering
\begin{tikzpicture}[scale=0.5, thick]

\draw[->, >=stealth, >=stealth, color=gray80] (0, 5) -- +(6.5,0);

\foreach \i in {0, 3} {
 \fill[color=dark] (2 * \i, 5) circle (4pt);
}

\foreach \i in {1, 2} {
 \fill[color=gray80] (2 * \i, 5) circle (4pt);
}

\fill[color=dark] (0, 5 - 0.25) circle (0pt)
node[below, color=black] { $(q_{0}, p_{0})$ };

\fill[color=dark] (6, 5 - 0.25) circle (0pt)
node[below, color=black] { $(q_{L}, p_{L})$ };

\draw[->, >=stealth, >=stealth, color=gray80] (6, 3) -- +(-0.5,0);

\fill[color=dark] (6, 3) circle (4pt)
node[below, color=black] { $(q_{L}, -p_{L})$ };

\fill[] (8, 3) circle (0pt)
node[right, color=black] { $\mathbb{Q} ( q_{L}, -p_{L} \mid q_{0}, p_{0} ) = 1$ };

\draw[->, >=stealth, >=stealth, color=gray80] (6, 1) -- +(-6.5,0);

\foreach \i in {1, 2} {
 \fill[color=gray80] (2 * \i, 1) circle (4pt);
}

\foreach \i in {0, 3} {
 \fill[color=dark] (2 * \i, 1) circle (4pt);
}

\fill[color=dark] (0, 1 - 0.25) circle (0pt)
node[below, color=black] { $(q_{0}, -p_{0})$ };

\fill[color=dark] (6, 1 - 0.25) circle (0pt)
node[below, color=black] { $(q_{L}, -p_{L})$ };

\draw[->, >=stealth, >=stealth, color=gray80] (0, -1) -- +(0.5, 0);

\fill[color=dark] (0, -1) circle (4pt)
node[below, color=black] { $(q_{0}, p_{0})$ };

\fill[] (8, -1) circle (0pt)
node[right, color=black] { $\mathbb{Q} ( q_{0}, p_{0} \mid q_{L}, -p_{L} ) = 1$ };

\end{tikzpicture}
\caption{Augmenting the numerical trajectory with a momentum flip defines a 
reversible Metropolis-Hastings proposal for which the forwards and backwards 
proposal probabilities are both well-behaved and we recover a valid correction 
scheme.
}
\label{fig:augment_reversible}
\end{figure*}

If we modify the Hamiltonian transition to be \emph{reversible}, however, then 
the ratio of proposal densities becomes non-zero and we achieve a useful correction
scheme.  The simplest way of achieving a reversible proposal is to augment the
the numerical integration with a negation step that flips the sign of momentum,
\begin{equation*}
\left(q, p \right) \rightarrow \left(q, -p \right).
\end{equation*}
This yields the reversible proposal (Figure \ref{fig:augment_reversible})
\begin{equation*}
\mathbb{Q} (q', p' \mid q_{0}, p_{0}) 
= 
\delta (q' - q_{L} ) \, \delta (p' + p_{L} ),
\end{equation*}
with the corresponding Metropolis-Hastings acceptance probability 
\begin{align*}
a ( q_{L}, -p_{L} \mid q_{0}, p_{0} )
&=
\min \! \left(1, 
\frac{ \mathbb{Q} ( q_{0}, p_{0} \mid q_{L}, -p_{L} ) \, \pi ( q_{L}, -p_{L} ) }
{ \mathbb{Q} ( q_{L}, -p_{L} \mid q_{0}, p_{0} ) \, \pi ( q_{0}, p_{0} ) } 
\right)
\\
&=
\min \! \left(1, 
\frac{ \delta (q_{L} - q_{L} ) \, \delta (-p_{L} + p_{L} ) \,  \pi ( q_{L}, -p_{L} ) }
{ \delta (q_{0} - q_{0} ) \, \delta (p_{0} - p_{0} ) \, \pi ( q_{0}, p_{0} ) } 
\right)
\\
\\
&=
\min \! \left(1, 
\frac{ \pi ( q_{L}, -p_{L} ) }
{ \pi ( q_{0}, p_{0} ) } 
\right)
\\
&=
\min \! \left(1, 
\frac
{  \exp \left( - H ( q_{L}, -p_{L} ) \right) }
{  \exp \left( - H ( q_{0}, p_{0} ) \right) } 
\right)
\\
&=
\min \! \left(1, 
\exp \left( - H ( q_{L}, -p_{L} ) + H ( q_{0}, p_{0} ) \right)
\right),
\end{align*}
Because we can always evaluate the Hamiltonian, we can immediately construct the 
acceptance probability and then correct for the bias induced by the symplectic 
integrator error.  Moreover, because symplectic integrators oscillate near the 
exact energy level set this acceptance probability will deteriorate only negligibly 
as we consider target distributions of higher and higher dimensions.

Our analysis of optimal integration times in Section 
\ref{sec:optimal_integration_time}, however, motivated taking not the last state 
from a Hamiltonian trajectory but rather sampling points uniformly from the entire 
trajectory to best approximate a sample from the microcanonical distribution.  We
can modify our Metropolis-Hastings approach towards this end by proposing not just
the final point but all points in the numerical trajectory,
\begin{equation*}
\mathbb{Q} (q', p' \mid q_{0}, p_{0}) 
= 
\frac{1}{L} \sum_{l = 0}^{L}
\delta (q' - q_{l} ) \, \delta (p' + p_{l} ).
\end{equation*}
This generalized proposal is still reversible and the acceptance probability
similarly reduces to 
\begin{align*}
a ( q_{l}, -p_{l} \mid q_{0}, p_{0} )
=
\min \! \left(1, 
\exp \left( - H ( q_{l}, -p_{l} ) + H ( q_{0}, p_{0} ) \right)
\right),
\end{align*}
for $0 \le l \le L$ and vanishes otherwise.

Unfortunately, this generalized proposal can suffer from sub-optimal performance
because it proposes states independent of their acceptance probability.  In
particular, we can propose a state with large error and be forced to reject the
proposal even though there are might be other states in the numerical trajectory
with much smaller errors and correspondingly higher acceptance probabilities.
To achieve optimal performance we have to \emph{average} the proposals to all 
states in the trajectory into a single, efficient proposal.

There are numerous schemes for achieving such an average, and they all require
generating numerical trajectories by integrating the initial state not only
forwards in time but also backwards.  Once we start integrating in both directions
it becomes easier to reason about the Hamiltonian Markov transition as a two-stage
process.  First we uniformly sample a trajectory, $\mathfrak{t}$, from all numerical 
trajectories of length $L$ that contain the initial point, and then we sample a 
point from that trajectory with the probabilities
\begin{equation*}
\pi \! \left( q, p \right)
=
\frac{ e^{- H (q, p) } }{ \sum_{(q', p') \in \mathfrak{t}} e^{ - H (q', p') } }.
\end{equation*}
Detailed proofs demonstrating the validity of this scheme, and its extension
to dynamic trajectory lengths, can be found in Appendix \ref{sec:technical_details}.

\subsection{Optimal Choice of Symplectic Integrator}

Now that we know how to implement the Hamiltonian Monte Carlo method with 
symplectic integrators, we are left with the choice of the exact configuration
of the symplectic integrator itself.  These configurations are characterized
with a choice of step size, $\epsilon$, and an order, $K$, that quantifies how
many gradient evaluations are used in each discrete integrator step.  

In general there is a delicate balance between configurations that are more 
accurate but more expensive, and those that are less accurate but cheaper.  For 
example, regardless of the exact order of the integrator, the step size will 
control the number of integrator steps, $L = T / \epsilon$, and hence the overall 
cost of each Hamiltonian transition.  Smaller step sizes yield more accurate 
numerical trajectories and better exploration, but at the expense of many more 
integrator steps.  Similarly, the higher the order of the integrator the more
accurate, but also more expensive, it will be.

\begin{figure}
\centering
\subfigure[]{ \includegraphics[width=2.5in]{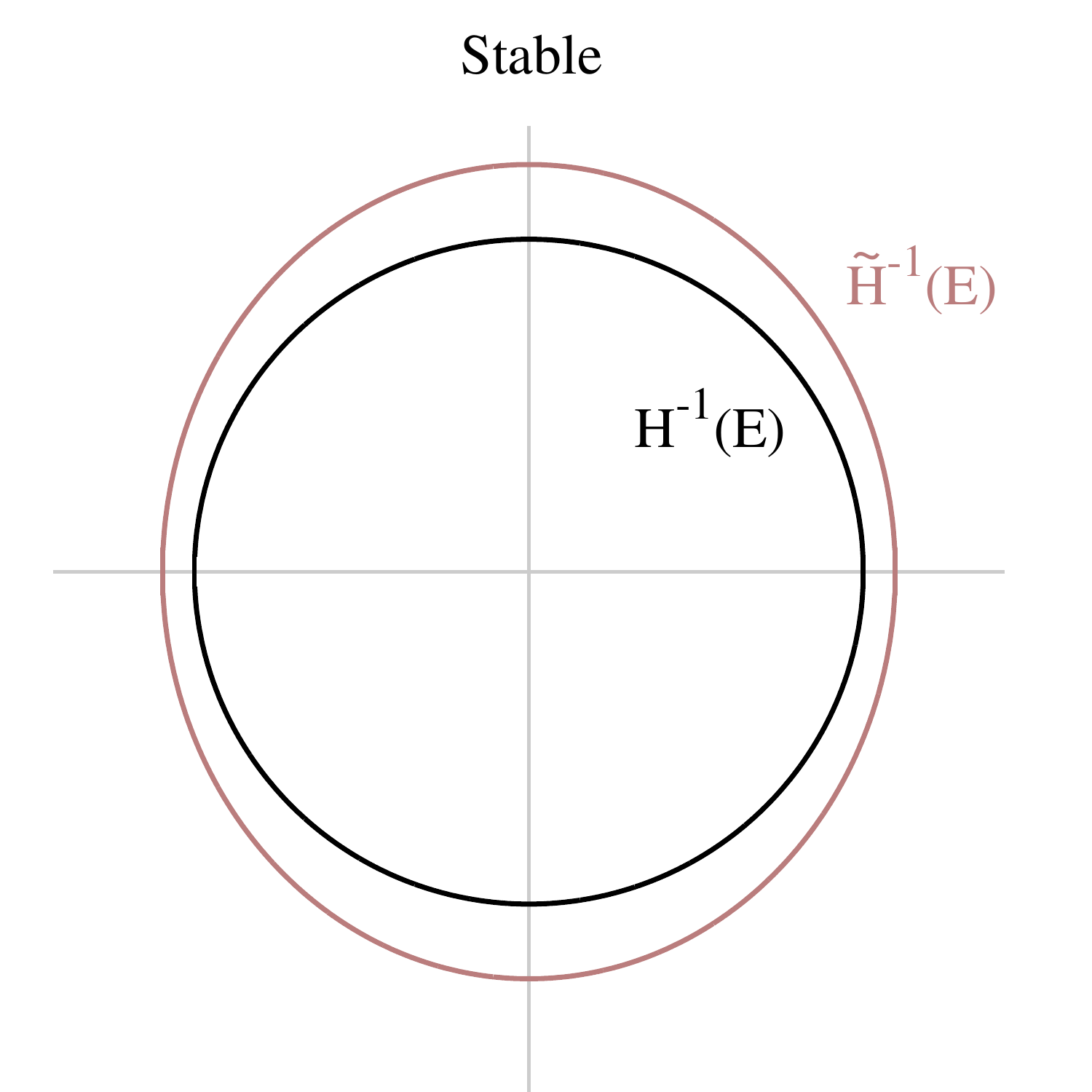} }
\subfigure[]{ \includegraphics[width=2.5in]{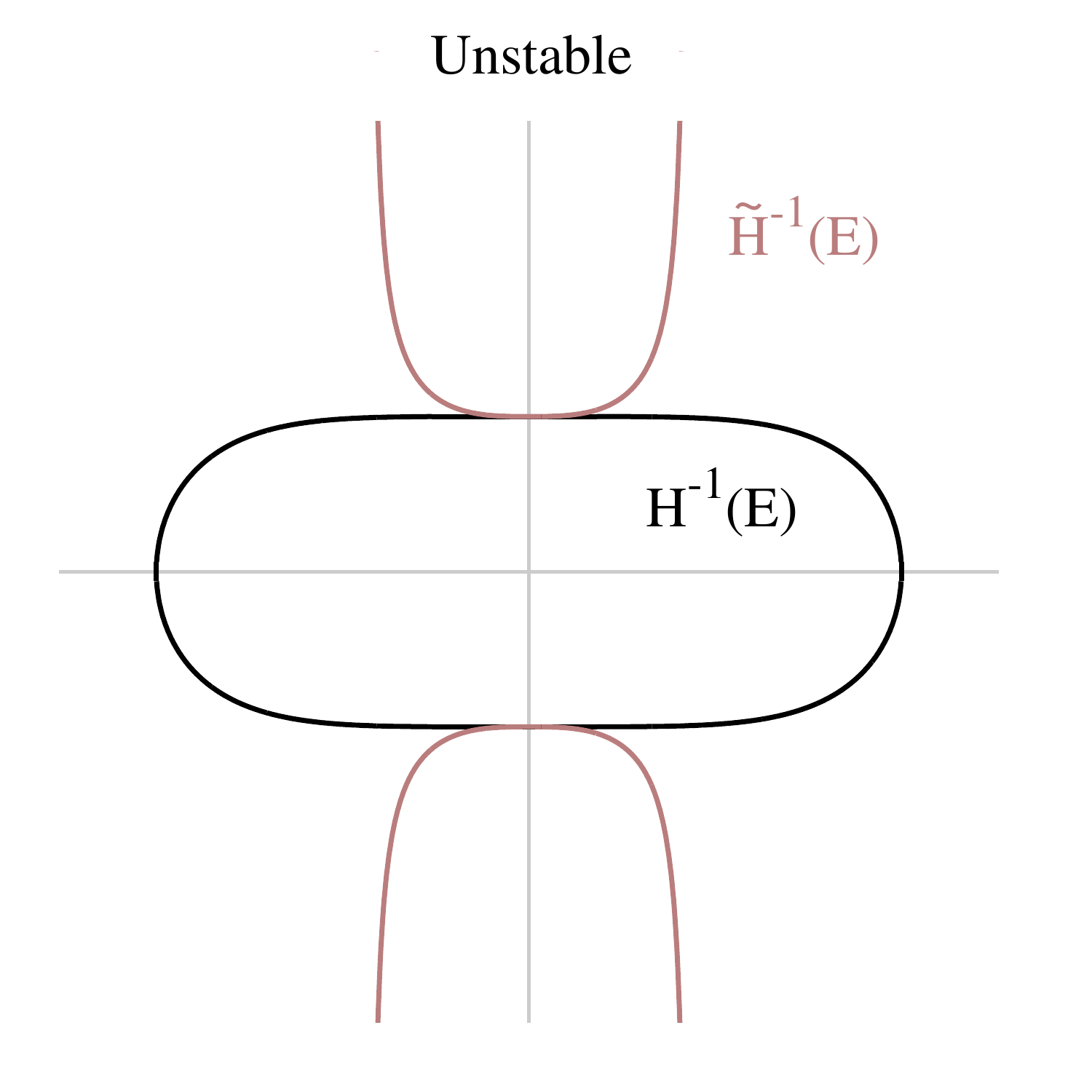} }
\caption{Symplectic integrators generate exact trajectories from a modified Hamiltonian,
$\widetilde{H}$ that resembles the exact Hamiltonian, $H$, and we can quantify the
accuracy of a symplectic integrator by comparing the corresponding energy level sets. 
(a) When the level sets of the modified Hamiltonian share the same topology as the 
level sets of the exact Hamiltonian, the numerical trajectories will be stable and 
highly accurate.  (b) When the modified level sets have a different topology, however, 
the numerical trajectories will become unstable and diverge.}
\label{fig:modified_level_sets}
\end{figure}

Fortunately, this intuition can be formalized and quantified by exploiting the 
geometry inherent to symplectic integrators.  In particular, the volume 
preservation of symplectic integrators guarantees not only that they well-approximate 
trajectories generated by the exact Hamiltonian, but also that they exactly solve 
trajectories generated by some \emph{modified Hamiltonian} or 
\emph{shadow Hamiltonian}.  This implies that numerical trajectories will be 
confined to energy level sets of this modified Hamiltonian, and we can quantify 
the performance of the integrator by studying the differences between the exact and 
modified level sets (Figure \ref{fig:modified_level_sets}a).  For example, divergences 
occur when the modified level sets become non-compact and extend out to the boundaries 
of phase space (Figure \ref{fig:modified_level_sets}b).

\begin{figure}
\centering
\includegraphics[width=3in]{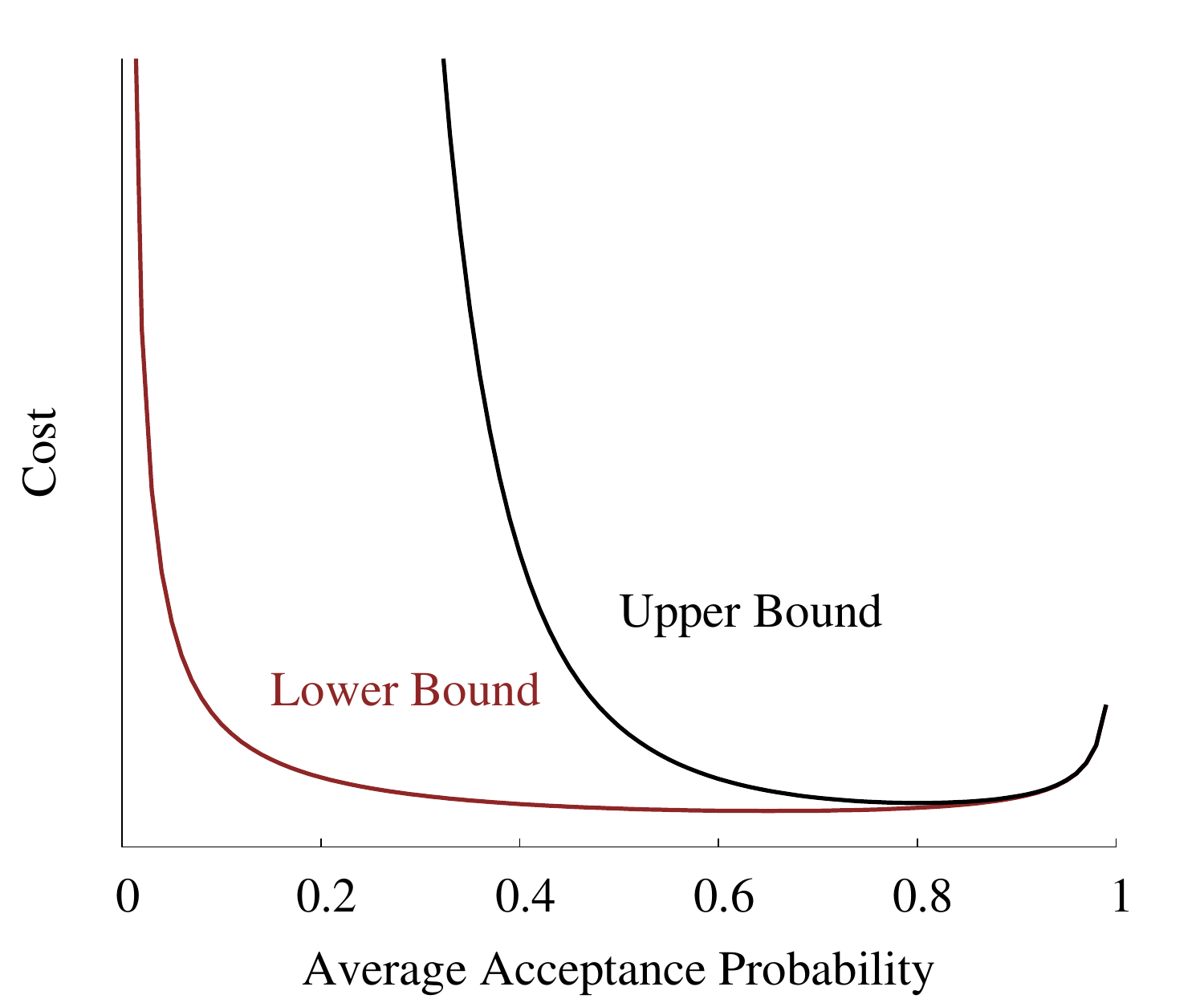}
\caption{By comparing the geometry of the modified level sets induced by a symplectic 
integrator with the exact level sets we can compute bounds between the computational 
cost of a Hamiltonian transition and the average Metropolis acceptance probability, at 
least for simple implementations of the Hamiltonian Monte Carlo method.  These bounds 
then inform adaptation procedures that tune the integrator step size to achieve an 
average acceptance probability that yields minimal cost.}
\label{fig:cost_bounds}
\end{figure}

When the modified level sets are well-behaved and we do not encounter any divergences, 
we can quantify the performance of the symplectic integrator by comparing the shapes 
of the exact and modified level sets.  For the simple implementation of Hamiltonian 
Monte Carlo where we integrate for a static time, $T$, flip the momentum, and apply 
a Metropolis-Hastings correction to the final state, this comparison bounds the 
relationship between the cost of the algorithm and the average Metropolis-Hastings 
acceptance probability, which itself depends on 
the step size~\citep{BetancourtEtAl:2014b} (Figure \ref{fig:cost_bounds}).  These 
bounds provide the basis for an automated tuning algorithm that adapts the step size
during an extended warm-up to achieve an average Metropolis acceptance probability 
between 0.6 and 0.8, similar to the adaptation of the Euclidean metric discussed
in Section \ref{sec:euclidean_gaussian}. Although much of this analysis carries over 
to more sophisticated implementations of Hamiltonian Monte Carlo, more work is needed 
to formalize the corresponding results.

Finally we are left to consider the order of the symplectic integrator.  Most
discussions of Hamiltonian Monte Carlo consider only the first-order leapfrog 
integrator due to its simplicity and empirical success.  Recent work, however, 
has suggested that higher-order integrators may prove more effective, especially 
in higher-dimensional problems~\citep{Blanes:2014aa, Fernandez-PendasEtAl:2015}.
One serious complication introduced by higher-order integrators are additional
degrees of freedom that can significantly affect performance -- poor choices
of these integrator parameters can completely neutralize any gains over the 
leapfrog integrator.  Understanding how these additional integrator degrees of 
freedom interact with a given problem, and then developing automated adaptation 
procedures, is an important next step to the inclusion of higher-order methods.

\section{The Robustness of Hamiltonian Monte Carlo}

Up to this point we have been focused on why Hamiltonian trajectories rapidly 
explore a given target distribution and then how we can configure Hamiltonian 
Markov transitions to maximize the efficiency of this exploration.  We have not 
yet, however, addressed the critical question of how robust that exploration 
will be -- even the fastest algorithm is worthless if it does not quantify all 
of the typical set.  As discussed in Section \ref{sec:pathological_behavior}, 
the formal question we have to consider is for what target distributions will 
an implementation of Hamiltonian Monte Carlo be geometrically ergodic and yield 
well-behaved, unbiased Markov chain Monte Carlo estimators?

Because this question is so general, theoretical analysis is extremely difficult.  
Preliminary results, however, show that even simple implementations of the 
Hamiltonian Monte Carlo method are geometrically ergodic over a large class of 
target distributions~\citep{LivingstoneEtAl:2016}.  In particular, this class is 
significantly larger than the class for non-gradient based algorithms like Random 
Walk Metropolis, consistent with the intuition that gradients are critical to 
robust Markov chain Monte Carlo in high-dimensional problems.

Perhaps more importantly, this theoretical analysis identifies the pathological 
behaviors of a target distribution that obstruct geometric ergodicity.  Moreover, 
the work demonstrates that these pathologies manifest in precise empirical 
behaviors that immediately motivate practical diagnostics unique to Hamiltonian 
Monte Carlo.

\subsection{Diagnosing Poorly-Chosen Kinetic Energies}

For example, one pathological behavior that frustrates geometric ergodicity is 
heavy tails in the target distribution.  Heavy tails cause the typical set to 
push deep into the extremes of parameter space, requiring that any successful 
Markov chain make prolonged sojourns to the tails and back lest the chain miss 
significant regions of the typical set.

The progressively long trajectories enabled by dynamic integration time 
implementations facilitate these sojourns in Hamiltonian Markov chains, but 
sometimes even exponentially growing integration times can be insufficient.  
In particular, if the kinetic energy is poorly-chosen then the marginal energy 
distribution can become heavy-tailed itself in which case the stochastic 
exploration between level sets will become so slow that after any finite 
number of transitions the exploration of the Markov chain will be incomplete.
 
\begin{figure}
\centering
\subfigure[]{ \includegraphics[width=2.5in]{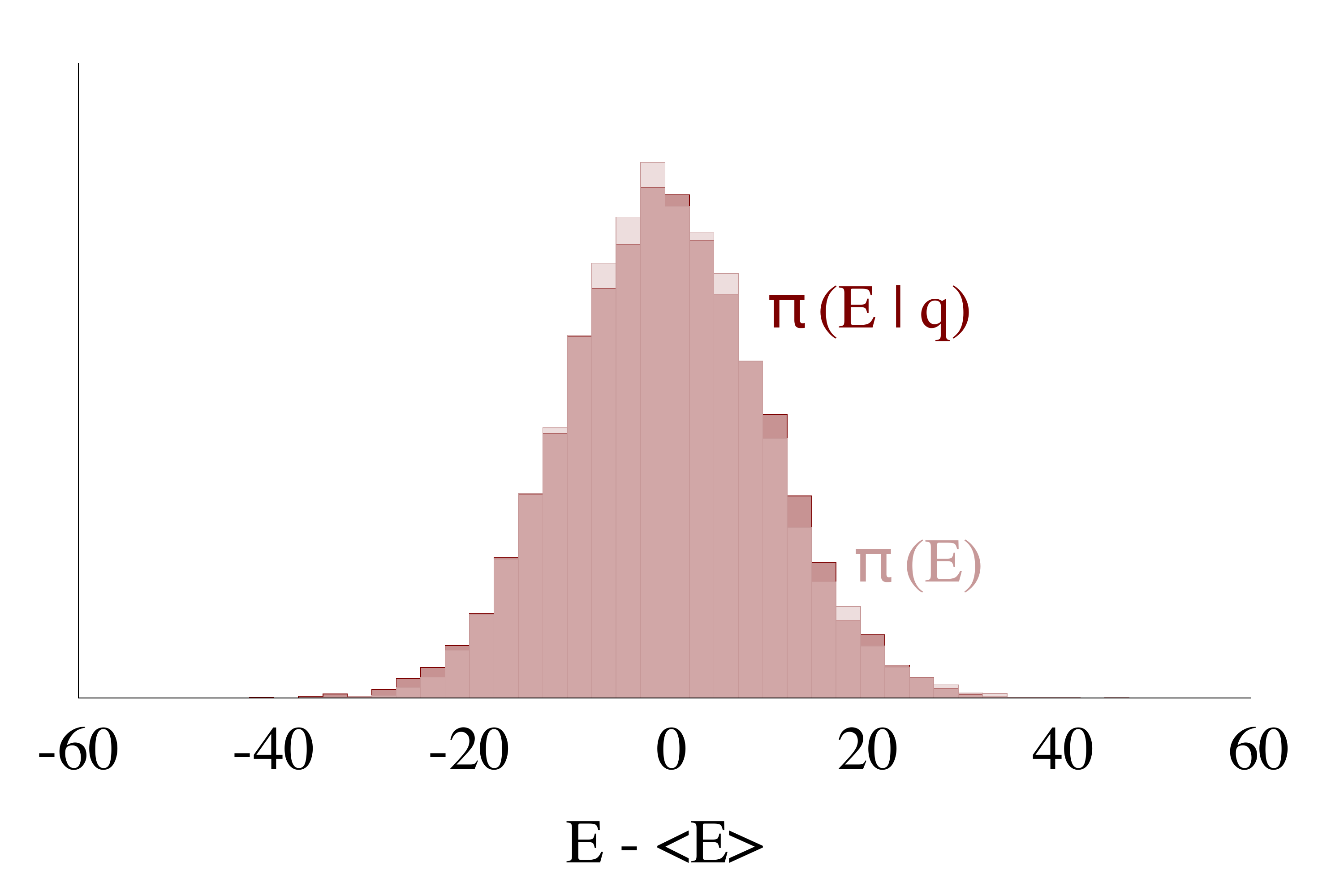} }
\subfigure[]{ \includegraphics[width=2.5in]{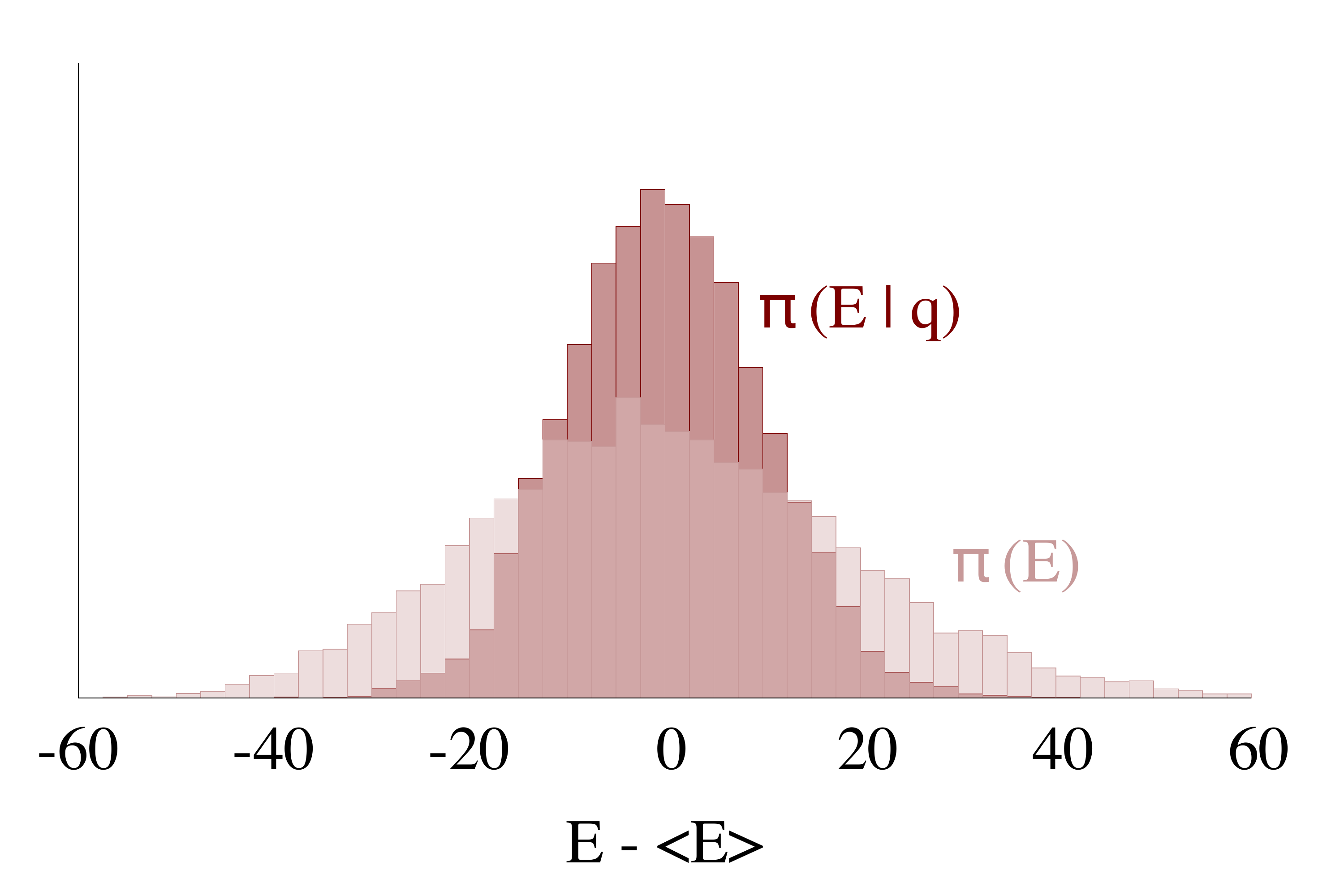} }
\caption{The energies explored by a Hamiltonian Markov chain can be used to visualize 
both the energy transition density, $\pi \! \left( E \mid q \right)$, and the marginal 
energy distribution, $\pi \! \left( E \right)$. (a) When these distributions are 
well-matched the Hamiltonian Markov chain should perform robustly, (b) but if the 
energy transitions density is significantly narrower than the marginal energy 
distribution then the chain may not be able to completely explore the tails of the 
target distribution.}
\label{fig:energy_diagnostic}
\end{figure}

Fortunately, such ill-suited kinetic energies are easy to identify by visualizing 
the marginal energy density and energy transition density using the Markov chain 
itself~\citep{Betancourt:2016b} (Figure \ref{fig:energy_diagnostic}).  This 
visualization quickly identifies when the momentum resampling will inefficiently 
explore the energy level sets, especially when aggregating multiple chains 
initialized from diffuse starting points.  We can also formalize the mismatch 
between these two distributions using the Bayesian fraction of missing information,
\begin{equation*}
\text{E-BFMI} \equiv
\frac{ \mathbb{E}_{\pi} \! 
\left[ \mathrm{Var}_{ \pi_{E \mid q} } \! \left[ E \mid q \right] \right] }
{ \mathrm{Var}_{ \pi_{E} } \! \left[ E \right] }
\approx
\widehat{\text{E-BFMI}} 
\equiv
\frac{ \sum_{n = 1}^{N} \left( E_{n} - E_{n - 1} \right)^{2} }
{ \sum_{n = 0}^{N} \left( E_{n} - \bar{E} \right)^{2} }.
\end{equation*}
Empirically, values of this energy Bayesian fraction of missing information below 
0.3 have proven problematic, although more theoretical work is needed to formalize 
any exact threshold.

\subsection{Diagnosing Regions of High Curvature}

Another common obstruction to geometric ergodicity is neighborhoods in parameter 
space where the target distribution exhibits large curvature.  As we saw in Section \ref{sec:pathological_behavior}, most Markov transitions are not able to resolve 
these narrow neighborhoods, resulting in incomplete exploration and biased Markov 
chain Monte Carlo estimators.  Hamiltonian Markov transitions are no exception.  
Because such neighborhoods are characteristic of many important models, in particular 
hierarchical models~\citep{BetancourtEtAl:2015}, these pathologies cannot be ignored 
in applied practice.

\begin{figure*}
\centering
\begin{tikzpicture}[scale=0.3, thick]
  \draw[white] (-12, -10.5) rectangle (12, 7);

  \foreach \i in {0, 0.05,..., 1} {
    \begin{scope}
      \clip (-0.005, -9.5) rectangle (11, 6.5);
      \draw[line width={30 * \i}, opacity={exp(-8 * \i)}, dark] 
      (0, 5) .. controls (5, 5) and (10, 8) .. (10, 0)
               .. controls (10, -8) and (5, -3) .. (-2, -10);
    \end{scope}
  
    \begin{scope}
      \clip (0.005, -9.5) rectangle (-11, 6.5);
      \draw[line width={30 * \i}, opacity={exp(-8 * \i)}, dark] 
      (0, 5) .. controls (-5, 5) and (-10, 8) .. (-10, 0)
               .. controls (-10, -8) and (-5, -3) .. (2, -10);
    \end{scope}
  }  

  \fill[opacity=0.3, green] (0, -8.5) circle (1);

  \fill[color=green] (-9.8725, 2) circle (5pt);
  
  \begin{scope}
    \clip (-15, -7.8) rectangle (-0.7, 2);
    \draw[->, >=stealth, >=stealth, color=green, thick, line width=1.5] 
        (0, 5) .. controls (-5, 5) and (-10, 8) .. (-10, 0)
                 .. controls (-10, -8) and (-5, -3) .. (2, -10);
  \end{scope}
  
  \draw[->, >=stealth, >=stealth, green, line width=1.5] 
  (-0.7, -7.75) -- (-10, -10);

\end{tikzpicture}
\caption{Neighborhoods of high curvature in the typical set (green) that frustrate 
geometric ergodicity are also pathological to symplectic integrators, causing them 
to diverge.  This confluence of pathologies is advantageous in practice because we 
can use the easily-observed divergences to identify the more subtle statistical 
pathologies.}
\label{fig:divergence}
\end{figure*}
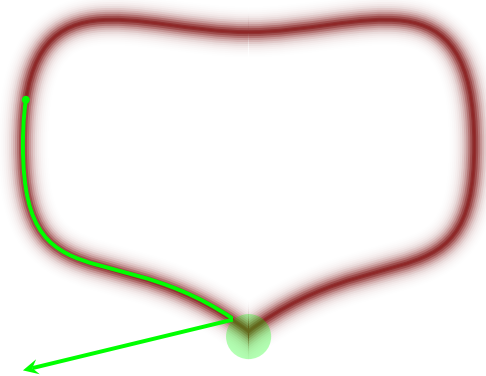

Conveniently, these neighborhoods of high curvature also prove pathological to 
symplectic integrators, which become unstable and diverge once they enter 
(Figure \ref{fig:divergence}).  Importantly, this means that divergent numerical
trajectories are extremely sensitive identifiers of these pathological neighborhoods 
and hence provide a powerful and immediate diagnostic. 

In particular, any divergent transitions encountered in a Hamiltonian Markov chain 
should prompt suspicion of the validity of any Markov chain Monte Carlo estimators. 
When the pathologies are sufficiently mild, the divergences can be eliminated, and 
the validity of the estimators restored, by decreasing the step size of the 
symplectic integrator.  Often, however, the pathologies are too severe and the
divergences will persist regardless of the how much the step size is decreased, 
indicating significant bias in the resulting estimators.  In these cases the target 
distribution itself needs to be regularized, for example with stronger priors or an 
alternative representation.  A prevalent example of the latter is alternating between 
centered and non-centered parameterizations in latent Gaussian 
models~\citep{BetancourtEtAl:2015}.

\subsection{Limitations of Diagnostics}

Although these two diagnostics are powerful means of identifying many pathological 
target distributions, they are only necessary and not sufficient conditions for the 
validity of the Markov chain Monte Carlo estimators.  Without an explicit theoretical 
analysis, there is no guarantee that a target distribution will not manifest more 
subtle pathologies capable of derailing Hamiltonian Monte Carlo, especially as we 
endeavor towards more and more complex problems.

Still, these diagnostics make Hamiltonian Monte Carlo a uniquely robust method, 
especially when combined with generic diagnostics like the split $\hat{R}$
statistic with multiple, parallel chains.

\section{Conclusion}

By exploiting the geometry of the typical set, Hamiltonian Monte Carlo generates 
coherent exploration of smooth target distributions.  This effective exploration 
yields not only better computational efficiency than other Markov chain Monte Carlo 
algorithms, but also stronger guarantees on the validity of the resulting estimators.  
Moreover, careful analysis of this geometry facilitates principled strategies for 
automatically constructing optimal implementations of the method, allowing users 
to focus their expertise into building better models instead of wrestling with the 
frustrations of statistical computation.  Consequently, Hamiltonian Monte Carlo is 
uniquely suited to tackling the most challenging problems at the frontiers of applied 
statistics, as demonstrated by the huge success of tools like Stan~\citep{Stan:2017}.

In hindsight, this success should not be particularly surprising.  The ultimate 
challenge in estimating probabilistic expectations is quantifying the typical set 
of the target distribution, a set which concentrates near a complex surface in 
parameter space.  What more natural way is there to quantify and then exploit 
information about a surface than through its geometry?

Continuing to leverage this geometric perspective will be critical to further 
advancing our understanding of the Hamiltonian Monte Carlo method and its optimal 
implementations.  This includes everything from improving geometric ergodicity 
analysis, refining the existing implementation techniques, and motivating the 
appropriate use of non-Gaussian and non-Euclidean kinetic energies.

Furthermore, techniques from differential geometry may allow us to generalize 
the principles of Hamiltonian Monte Carlo to other, more challenging computational 
problems.  For example, geometric methods have already generalized coherent 
exploration to thermodynamic algorithms in 
\emph{Adiabatic Monte Carlo}~\citep{Betancourt:2014}.  Incorporating more 
sophisticated geometric techniques may allow us to extend these ideas to more 
intricately-structured parameter spaces, such as discrete spaces, tree spaces, 
and Hilbert spaces.

\section{Acknowledgements}

This conceptual description of Hamiltonian Monte Carlo was refined over three
years of frequent speaking engagements coincident with active theoretical research.  
I am hugely indebted to Mark Girolami, Simon Byrne, and Samuel Livingstone with whom 
the theoretical work flourished, and the Stan development team with whom the 
implementations were developed.  I also warmly thank Matthew Hoffman, Gareth Roberts, 
Dan Simpson, Chris Wendl, and everyone who came out to see one of my talks or take 
the time to ask a question.  The manuscript itself benefited from the valuable comments 
of Ellie Sherrard-Smith, Daniel Lee, Maggie Lieu, Chris Jones, Nathan Sanders, 
Aki Vehtari, Cole Monnahan, Bob Carpenter, Jonah Sol Gabry, Andrew Gelman,
Sidharth Kshatriya, Fabian Dablander, Don van den Bergh, and Nathan Evans.

This was work supported by EPSRC grant EP/J016934/1, a Centre for Research in 
Statistical Methodology Fellowship, and a 2014 EPRSC NCSML Award for Postdoctoral 
Research Associate Collaboration.

\appendix
\section{Technical Details of Practical Implementations} \label{sec:technical_details}

As discussed in Section \ref{sec:microcanonical_geometry}, Hamiltonian 
systems naturally decompose into a microcanonical distribution over 
energy level sets and a marginal energy distribution between the energies 
themselves.  Hamiltonian Monte Carlo exploits this structure to generate 
efficient exploration of the canonical distribution, alternating between 
exploring a given energy level set and resampling the momenta to generate 
a random walk over the marginal energy distribution.  

In particular, the exploration over the energy level sets is readily
accomplished by following the trajectories generated by the Hamiltonian.
These trajectories are not only energy preserving, and hence confined
to a single energy level set, they are typically ergodic and will explore
that entire level set as the trajectories are integrated for longer and
longer.  More formally, a uniform sample from an ergodic trajectory 
converges to a sample from the microcanonical distribution.

Unfortunately we cannot implement this idealized method in practice because 
outside of a few exceptional systems we cannot generate Hamiltonian trajectories 
analytically.  Instead we have to work with the approximate trajectories given 
by numerical integrators.  Fortunately, Hamiltonian systems admit very powerful 
symplectic integrators whose numerical errors not only are small but also can 
be exactly rectified with straightforward probabilistic corrections.

In this appendix I will review the technical details of how the numerical 
trajectories from symplectic integrators can be utilized to generate samples 
from the exact microcanonical distribution, first for the case of \emph{static} 
trajectories of a fixed length and then \emph{dynamic} trajectories that grow 
until a given termination criterion is met.  Finally I will comment on how 
these implementations relate to the No-U-Turn sampler and Stan.

\subsection{Notation}

Let $z = (q, p)$ denote a point in phase space.  A symplectic integrator 
generates a numerical trajectory, $\mathfrak{t}$, from any such point by 
integrating either forwards and backwards for an incremental amount of time, 
$\epsilon$, reducing all of phase space to the discrete set of points that 
can be reached by the symplectic integrator.  

Given an initial point, $z_{0}$, let $z_{n}$ denote the point reached by 
integrating forward $n$ times, and $z_{-n}$ the point reached by integrating 
backwards $n$ times.  Example trajectories of length $L = 3$ then include 
$\mathfrak{t} = \left\{ z_{0}, z_{1}, z_{2} \right\}$,
$\mathfrak{t} = \left\{ z_{-1}, z_{0}, z_{1} \right\}$, and
$\mathfrak{t} = \left\{ z_{-2}, z_{-1}, z_{0} \right\}$.

I will write the set of all symplectic integrator trajectories of length 
$L$ that contain the point $z$ as $\mathfrak{T}^{L}_{z}$.  Similarly, I will 
denote the set of numerical trajectories of length $L$ that contain the points 
$z$ and $z'$ as $\mathfrak{T}^{L}_{z, z'}$, i.e.
\begin{equation*}
\mathfrak{T}^{L}_{z, z'} = \mathfrak{T}^{L}_{z} \cap \mathfrak{T}^{L}_{z'}.
\end{equation*} 

\subsection{Static Implementations}

Symplectic integrators are not exactly energy preserving, causing their 
numerical trajectories to deviate from the target energy level set.  In 
particular, sampling a state uniformly from any numerical trajectory will 
not generate a sample from the canonical distribution. This error, however, 
can be exactly corrected by sampling from the trajectory not uniformly but 
rather with weights proportional to the desired canonical density function.  

The easiest way to reason about any such scheme is sequentially, first
by sampling a numerical trajectory that contains the initial point, 
$\mathbb{T} ( \mathfrak{t} \mid z )$, and then sampling a state from that 
trajectory, $\mathbb{T} ( z' \mid \mathfrak{t} )$, yielding the joint
transition
\begin{equation*}
\mathbb{T} ( z' \mid z ) =
\sum_{\mathfrak{t}}
\mathbb{T} ( z' \mid \mathfrak{t} ) \, \mathbb{T} ( \mathfrak{t} \mid z ),
\end{equation*}
where the sum runs over all numerical trajectories.

\subsubsection{Sampling a Trajectory}

In the static case we consider only the trajectories of length $L$ that contain 
the initial point, $z$, in which case $\mathbb{T} ( \mathfrak{t} \mid z )$ is 
supported only on $\mathfrak{T}^{L}_{z}$.  

In order to ensure a valid correction, this transition from states
to trajectories has to exhibit a certain reversibility.  Formally
we require that the probability of transitions to a trajectory is 
the same regardless of from which state in that trajectory we might 
begin,
\begin{equation*}
\mathbb{T} ( \mathfrak{t} \mid z )
=
\mathbb{T} ( \mathfrak{t} \mid z' ), 
\, \mathfrak{t} \in \mathfrak{T}^{L}_{z, z'}.
\end{equation*}
This has the important implication that
\begin{equation*}
\mathbb{I} ( z' \in \mathfrak{t} ) \, \mathbb{T} ( \mathfrak{t} \mid z )
=
\mathbb{I} ( z \in \mathfrak{t} ) \, \mathbb{T} ( \mathfrak{t} \mid z ),
\end{equation*}
where $\mathbb{I}$ is the indicator function.

In practice this condition is most easily satisfied by uniformly
sampling over all of the trajectories that contain the initial point,
\begin{equation*}
\mathbb{T} ( \mathfrak{t} \mid z ) = U \! \left( \mathfrak{T}^{L}_{z} \right),
\end{equation*}
as $\mathbb{T} ( \mathfrak{t} \mid z ) = \mathbb{T} ( \mathfrak{t} \mid z' ) 
= L^{-1}$ if $\mathfrak{T}^{L}_{z, z'} \neq \emptyset$.

\subsubsection{Sampling a State}

Once we have sampled a numerical trajectory, we can generate a sample 
from the exact canonical distribution with any process that yields
transition probabilities proportional to the canonical densities,
\begin{align}
\mathbb{T} ( z' \mid \mathfrak{t} )
&= 
\frac{ \pi \! \left( z' \right) }
{ \sum_{z'' \in \mathfrak{t} } \pi \! \left( z'' \right) }
\mathbb{I} ( z' \in \mathfrak{t} )
\label{eqn:state_sampling}
\\
&=
\frac{ e^{-H( z' )} }{ \sum_{z'' \in \mathfrak{t} } e^{-H(z'')} }
\mathbb{I} ( z' \in \mathfrak{t} ).
\nonumber
\end{align}
If the symplectic integrator were exact, and the numerical trajectories 
confined to the initial energy level set, then these transition 
probabilities would reduce to the uniform distribution over the states 
in the trajectory, as desired.  In practice the error in the numerical 
trajectory induces non-uniform transition probabilities that disfavor 
deviations towards higher energies.

There are various schemes capable of yielding the desired transition 
probabilities.  For example, we could implement a slice sampler by
augmenting the transition with a random variable 
$u \sim U \left(0, \pi \! \left( z \right) \right)$, and then sampling
uniformly from those states in the trajectory satisfying 
$\pi \! \left( z' \right) > u$.

We can also achieve the desired transition probabilities directly,
however, by simply drawing from a multinomial distribution over
the states in the trajectory with probabilities
\begin{equation*}
\mathbb{P} ( z' ) = 
\frac{ \pi \! \left( z' \right) }
{ \sum_{z'' \in \mathfrak{t} } \pi \! \left( z'' \right) }.
\end{equation*}
Not only is this approach straightforward to implement, the Rao-Blackwell 
Theorem establishes that the lack of auxiliary random variables will ensure 
better performance.

\subsubsection{Invariance of the Canonical Distribution}

Symplectic integrators are able to realize such high practical performance 
because they exactly preserve volume in phase space.  Conveniently, this
volume preservation also ensures that we can achieve samples from the
exact canonical distribution with any transition that preserves the
canonical density, 
\begin{equation*}
\pi ( z ) \propto e^{-H(z)}.
\end{equation*}

An explicit calculation shows that the properties laid out above are
sufficient to guarantee this invariance,
\begin{align*}
\sum_{z} \mathbb{T} ( z' | z ) \pi (z)
&=
\sum_{z} \sum_{\mathfrak{t} }
\mathbb{T} (z' \mid \mathfrak{t} ) \, \mathbb{T} ( \mathfrak{t} \mid z )
\pi ( z ) 
\\
&=
\sum_{z} \sum_{\mathfrak{t} }
\frac{ \pi \! \left( z' \right) }
{ \sum_{z'' \in \mathfrak{t} } \pi \! \left( z'' \right) }
\mathbb{I} ( z' \in \mathfrak{t} )
\mathbb{T} ( \mathfrak{t} \mid z )
\pi ( z ) 
\\
&=
\sum_{z} \sum_{\mathfrak{t} }
\frac{ \pi \! \left( z' \right) }
{ \sum_{z'' \in \mathfrak{t} } \pi \! \left( z'' \right) }
\mathbb{I} ( z \in \mathfrak{t} )
\mathbb{T} ( \mathfrak{t} \mid z' )
\pi ( z ) 
\\
&=
\sum_{\mathfrak{t} }
\frac{ \pi \! \left( z' \right) }
{ \sum_{z'' \in \mathfrak{t} } \pi \! \left( z'' \right) }
\mathbb{T} ( \mathfrak{t} \mid z' )
\sum_{z} \mathbb{I} ( z \in \mathfrak{t} ) \pi ( z ) 
\\
&=
\pi \! \left( z' \right) 
\sum_{\mathfrak{t} }
\frac{1}
{ \sum_{z'' \in \mathfrak{t} } \pi \! \left( z'' \right) }
\mathbb{T} ( \mathfrak{t} \mid z' )
\sum_{z \in \mathfrak{t}} \pi ( z ) 
\\
&=
\pi \! \left( z' \right) 
\sum_{\mathfrak{t} }
\mathbb{T} ( \mathfrak{t} \mid z' )
\\
&=
\pi \! \left( z' \right).
\end{align*}

\subsection{Efficient Static Implementations}
\label{sec:efficient_static}

The problem with this initial scheme is that it requires us to keep the 
entire trajectory in memory in order to generate the transition.  For
static trajectories of length $L$, this requires keeping $L$ states in
memory at any given time, which can become burdensome in high-dimensional
problems.  We can construct a much more efficient transition, however, by 
weaving the sampling from $\mathbb{T} ( z' \mid \mathfrak{t} )$ into the 
construction of the trajectory, $\mathfrak{t}$, itself.

\subsubsection{Uniform Progressive Sampling}

Instead of sampling a trajectory containing the initial point at once,
consider generating a random trajectory sequentially.  We begin with
a trajectory containing the initial point, $\mathfrak{t}_{\mathrm{old}}$
and then append a new trajectory, $\mathfrak{t}_{\mathrm{new}}$, by
randomly integrating either forwards or backwards in time.  The union
of these two component trajectories gives the complete trajectory,
$\mathfrak{t} = \mathfrak{t}_{\mathrm{old}} \cup \mathfrak{t}_{\mathrm{new}}$. 

\begin{figure*}
\centering
\subfigure[]{
\begin{tikzpicture}[scale=0.5, thick]

\draw[<->, >=stealth, color=gray80] (-0.5, 0) -- +(7,0);
\draw[<->, >=stealth, color=gray80] ( 7.5, 0) -- +(7,0);

\foreach \i in {0, 1, ..., 7} {
 \fill[color=dark] (2 * \i, 0) circle (4pt);
}

\fill[color=dark] (3, -1) circle (0pt)
node[below, color=black] { $z_{\mathrm{old}} \sim 
                            \mathbb{T} ( z \mid \mathfrak{t}_{\mathrm{old}} )$ };

\fill[color=dark] (11, -1) circle (0pt)
node[below, color=black] { $z_{\mathrm{new}} \sim 
                            \mathbb{T} ( z \mid \mathfrak{t}_{\mathrm{new}} )$ };

\end{tikzpicture}
}
\subfigure[]{
\begin{tikzpicture}[scale=0.5, thick]

\draw[<->, >=stealth, color=gray80] (-0.5, 0) -- +(15,0);
\foreach \i in {0, 1, ..., 7} {
 \fill[color=dark] (2 * \i, 0) circle (4pt);
}

\fill[color=dark] (7, -1) circle (0pt)
node[below, color=black] 
{ 
$
z \sim \mathbb{T} [ z \mid \mathfrak{t}_{\mathrm{old}} \cup \mathfrak{t}_{\mathrm{new}} ]
= \frac{ w_{\mathrm{old}} }{ w_{\mathrm{old}} + w_{\mathrm{new}} }
\mathbb{T} ( z \mid \mathfrak{t}_{\mathrm{old}} )
+
\frac{ w_{\mathrm{new}} }{ w_{\mathrm{old}} + w_{\mathrm{new}} }
\mathbb{T} ( z \mid \mathfrak{t}_{\mathrm{new}} )
$
};

\end{tikzpicture}
}
\caption{(a) Given samples from two adjoining numerical trajectories
(b) we can generate a sample from their union with a Bernoulli
process, effectively mixing the two component transition distributions into
the correct joint transition distribution. This allows us to sample from a 
trajectory progressively during it's construction, transitioning first
between the old and the new components and then a sample therein.}
\label{fig:trajectory_building}
\end{figure*}
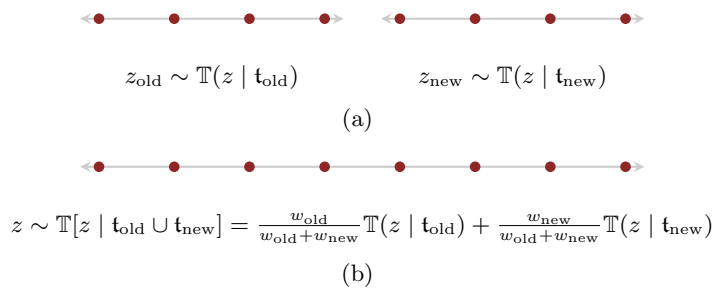

If we had samples from both the old and new trajectories according to
\eqref{eqn:state_sampling},
\begin{equation*}
z_{\mathrm{old}} \sim \mathbb{T} ( z' \mid \mathfrak{t}_{\mathrm{old}} ), \,
z_{\mathrm{new}} \sim \mathbb{T} ( z' \mid \mathfrak{t}_{\mathrm{new}} ), 
\end{equation*}
then we could generate a sample from the entire trajectory with a Bernoulli 
process that takes $z = z_{\mathrm{old}}$ with probability $p_{\mathrm{old}}$ 
and $z = z_{\mathrm{new}}$ otherwise (Figure \ref{fig:trajectory_building}), 
where
\begin{equation*}
p_{\mathrm{old}} = \frac{ w_{\mathrm{old}} }{ w_{\mathrm{old}} + w_{\mathrm{new}} },
\end{equation*}
with $w_{\mathrm{old}} = \sum_{z \in \mathfrak{t}_{\mathrm{old}} } \pi \! \left( z \right)$
and $w_{\mathrm{new}} = \sum_{z \in \mathfrak{t}_{\mathrm{new}} } \pi \! \left( z \right)$.

Formally we have
\begin{align}
\mathbb{T} ( z' \mid \mathfrak{t} )
&=
p_{\mathrm{old}} \,
\mathbb{T} ( z' \mid \mathfrak{t}_{\mathrm{old}} )
+
(1 - p_{\mathrm{old}}) \,
\mathbb{T} ( z' \mid \mathfrak{t}_{\mathrm{new}} )
\nonumber
\\
&=
\frac{ w_{\mathrm{old}} }{ w_{\mathrm{old}} + w_{\mathrm{new}} }
\mathbb{T} ( z' \mid \mathfrak{t}_{\mathrm{old}} )
+
\frac{ w_{\mathrm{new}} }{ w_{\mathrm{old}} + w_{\mathrm{new}} }
\mathbb{T} ( z' \mid \mathfrak{t}_{\mathrm{new}} ),
\label{eqn:uniform_progressive_transition}
\end{align}
which immediately follows from the definition, 
\begin{align*}
\frac{ w_{\mathrm{old}} }{ w_{\mathrm{old}} + w_{\mathrm{new}} }
\mathbb{T} ( z' \mid \mathfrak{t}_{\mathrm{old}} )
+
\frac{ w_{\mathrm{new}} }{ w_{\mathrm{old}} + w_{\mathrm{new}} }
\mathbb{T} ( z' \mid \mathfrak{t}_{\mathrm{new}} )
\\
& \hspace{-30mm} = \quad
\frac{ \sum_{z'' \in \mathfrak{t}_{\mathrm{old}} } \pi \! \left( z'' \right) }
{ \sum_{z'' \in \mathfrak{t} } \pi \! \left( z'' \right) }
\frac{ \pi \! \left( z' \right) }
{ \sum_{z'' \in \mathfrak{t}_{\mathrm{old}} } \pi \! \left( z'' \right) }
\mathbb{I} ( z' \in \mathfrak{t}_{\mathrm{old}} )
\\
& \hspace{-30mm} \quad +
\frac{ \sum_{z'' \in \mathfrak{t}_{\mathrm{new}} } \pi \! \left( z'' \right) }
{ \sum_{z'' \in \mathfrak{t} } \pi \! \left( z'' \right) }
\frac{ \pi \! \left( z' \right) }
{ \sum_{z'' \in \mathfrak{t}_{\mathrm{new}} } \pi \! \left( z'' \right) }
\mathbb{I} ( z' \in \mathfrak{t}_{\mathrm{new}} )
\\
& \hspace{-30mm} =
\frac{ \pi \! \left( z' \right) }
{ \sum_{z'' \in \mathfrak{t} } \pi \! \left( z'' \right) }
\mathbb{I} ( z' \in \mathfrak{t}_{\mathrm{old}} )
+
\frac{ \pi \! \left( z' \right) }
{ \sum_{z'' \in \mathfrak{t} } \pi \! \left( z'' \right) }
\mathbb{I} ( z' \in \mathfrak{t}_{\mathrm{new}} )
\\
& \hspace{-30mm} =
\frac{ \pi \! \left( z' \right) }
{ \sum_{z'' \in \mathfrak{t} } \pi \! \left( z'' \right) }
\left( 
\mathbb{I} ( z' \in \mathfrak{t}_{\mathrm{old}} )
+
\mathbb{I} ( z' \in \mathfrak{t}_{\mathrm{new}} )
\right)
\\
& \hspace{-30mm} =
\frac{ \pi \! \left( z' \right) }
{ \sum_{z'' \in \mathfrak{t} } \pi \! \left( z'' \right) }
\mathbb{I} ( z' \in \mathfrak{t} )
\\
& \hspace{-30mm} =
\mathbb{T} ( z' \mid \mathfrak{t} ).
\end{align*}

This progressive sampling allows us to generate a sample as we build the 
trajectory itself, keeping only a few states in memory at any given time.  
For example, starting with the initial point as the active sample we could 
repeatedly expand the trajectory by integrating one step forward or backwards 
in time and then update the active sample using the transition above, yielding 
at once both $\mathfrak{t} \sim \mathbb{T} ( \mathfrak{t} \mid z )
= U \! \left( \mathfrak{T}^{L + 1}_{z} \right)$ and 
$z' \sim \mathbb{T} ( z \mid \mathfrak{t} )$.  This additive expansion scheme 
also coincides with Neal's windowed sampler when the window length equals the 
full trajectory length~\citep{Neal:1994}.

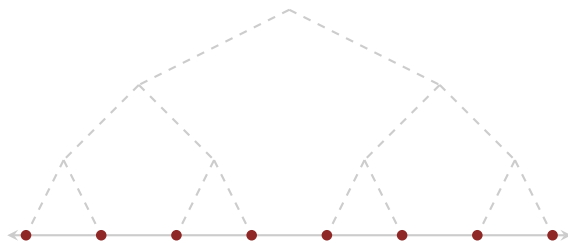
\begin{figure*}
\centering
\begin{tikzpicture}[scale=0.5, thick]
\draw[color=gray80, dashed] (7, 6) -- +(-4, -2);
\draw[color=gray80, dashed] (7, 6) -- +(4, -2);
\foreach \i in {0, 1} {
  \draw[color=gray80, dashed] (8 * \i + 3, 4) -- +(-2, -2);
  \draw[color=gray80, dashed] (8 * \i + 3, 4) -- +(2, -2);
}
\foreach \i in {0, 1, ..., 3} {
  \draw[color=gray80, dashed] (4 * \i + 1, 2) -- +(-1, -2);
  \draw[color=gray80, dashed] (4 * \i + 1, 2) -- +(1, -2);
}
\draw[<->, >=stealth, color=gray80] (-0.5, 0) -- +(15,0);
\foreach \i in {0, 1, ..., 7} { 
  \fill[color=dark] (2 * \i, 0) circle (4pt); 
}
\end{tikzpicture}
\caption{When the length of a numerical trajectory is a power of 2,
$L = 2^{D}$, the trajectory is equivalent to the leaves of a perfect,
ordered binary tree of depth $D$.  
}
\label{fig:trajectory_as_tree}
\end{figure*}

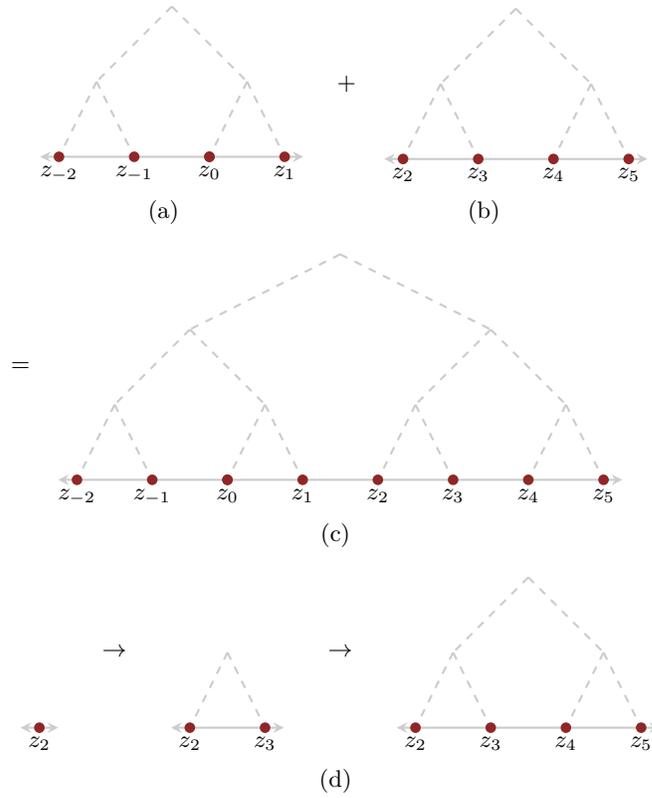
\begin{figure*}
\centering
\subfigure[]{
\begin{tikzpicture}[scale=0.5, thick]

\foreach \i in {0} {
  \draw[color=gray80, dashed] (8 * \i + 3, 4) -- +(-2, -2);
  \draw[color=gray80, dashed] (8 * \i + 3, 4) -- +(2, -2);
}

\foreach \i in {0, 1} {
  \draw[color=gray80, dashed] (4 * \i + 1, 2) -- +(-1, -2);
  \draw[color=gray80, dashed] (4 * \i + 1, 2) -- +(1, -2);
}

\draw[<->, >=stealth, color=gray80] (-0.5, 0) -- +(7,0);
\foreach \i in {0, 1, 2, 3} {
 \fill[color=dark] (2 * \i, 0) circle (4pt);
}

\fill[color=dark] (0, 0) circle (0pt)
node[below, color=black] { $z_{-2}$ };

\fill[color=dark] (2, 0) circle (0pt)
node[below, color=black] { $z_{-1}$ };

\fill[color=dark] (4, 0) circle (4pt)
node[below, color=black] { $z_{0}$ };

\fill[color=dark] (6, 0) circle (0pt)
node[below, color=black] { $z_{1}$ };

\end{tikzpicture}
}
\subfigure[]{
\begin{tikzpicture}[scale=0.5, thick]

\node[color=black] at (-1.5, 2) { $+$ };

\foreach \i in {0} {
  \draw[color=gray80, dashed] (8 * \i + 3, 4) -- +(-2, -2);
  \draw[color=gray80, dashed] (8 * \i + 3, 4) -- +(2, -2);
}

\foreach \i in {0, 1} {
  \draw[color=gray80, dashed] (4 * \i + 1, 2) -- +(-1, -2);
  \draw[color=gray80, dashed] (4 * \i + 1, 2) -- +(1, -2);
}

\draw[<->, >=stealth, color=gray80] (-0.5, 0) -- +(7,0);
\foreach \i in {0, 1, 2, 3} {
 \fill[color=dark] (2 * \i, 0) circle (4pt);
}

\fill[color=dark] (0, 0) circle (0pt)
node[below, color=black] { $z_{2}$ };

\fill[color=dark] (2, 0) circle (0pt)
node[below, color=black] { $z_{3}$ };

\fill[color=dark] (4, 0) circle (0pt)
node[below, color=black] { $z_{4}$ };

\fill[color=dark] (6, 0) circle (0pt)
node[below, color=black] { $z_{5}$ };

\end{tikzpicture}
}
\subfigure[]{
\begin{tikzpicture}[scale=0.5, thick]

\node[color=black] at (-1.5, 3) { $=$ };
\node[color=white] at (15.5, 3) { $=$ }; % Balance

\draw[color=gray80, dashed] (7, 6) -- +(-4, -2);
\draw[color=gray80, dashed] (7, 6) -- +(4, -2);

\foreach \i in {0, 1} {
  \draw[color=gray80, dashed] (8 * \i + 3, 4) -- +(-2, -2);
  \draw[color=gray80, dashed] (8 * \i + 3, 4) -- +(2, -2);
}

\foreach \i in {0, 1, 2, 3} {
  \draw[color=gray80, dashed] (4 * \i + 1, 2) -- +(-1, -2);
  \draw[color=gray80, dashed] (4 * \i + 1, 2) -- +(1, -2);
}

\draw[<->, >=stealth, color=gray80] (-0.5, 0) -- +(15, 0);
\foreach \i in {0, 1, ..., 7} {
 \fill[color=dark] (2 * \i, 0) circle (4pt);
}

\fill[color=dark] (0, 0) circle (0pt)
node[below, color=black] { $z_{-2}$ };

\fill[color=dark] (2, 0) circle (0pt)
node[below, color=black] { $z_{-1}$ };

\fill[color=dark] (4, 0) circle (4pt)
node[below, color=black] { $z_{0}$ };

\fill[color=dark] (6, 0) circle (0pt)
node[below, color=black] { $z_{1}$ };

\fill[color=dark] (8, 0) circle (0pt)
node[below, color=black] { $z_{2}$ };

\fill[color=dark] (10, 0) circle (0pt)
node[below, color=black] { $z_{3}$ };

\fill[color=dark] (12, 0) circle (0pt)
node[below, color=black] { $z_{4}$ };

\fill[color=dark] (14, 0) circle (0pt)
node[below, color=black] { $z_{5}$ };

\end{tikzpicture}
}
\subfigure[]{
\begin{tikzpicture}[scale=0.5, thick]

\draw[<->, >=stealth, color=gray80] (-0.5, 0) -- +(1,0);
\foreach \i in {0} {
 \fill[color=dark] (2 * \i + 0, 0) circle (4pt);
}

\fill[color=dark] (0, 0) circle (0pt)
node[below, color=black] { $z_{2}$ };

\node[color=black] at (2, 2) { $\rightarrow$ };

\foreach \i in {0} {
  \draw[color=gray80, dashed] (4 * \i + 5, 2) -- +(-1, -2);
  \draw[color=gray80, dashed] (4 * \i + 5, 2) -- +(1, -2);
}

\draw[<->, >=stealth, color=gray80] (3.5, 0) -- +(3,0);
\foreach \i in {0, 1} {
 \fill[color=dark] (2 * \i + 4, 0) circle (4pt);
}

\fill[color=dark] (4, 0) circle (0pt)
node[below, color=black] { $z_{2}$ };

\fill[color=dark] (6, 0) circle (0pt)
node[below, color=black] { $z_{3}$ };

\node[color=black] at (8, 2) { $\rightarrow$ };

\foreach \i in {0} {
  \draw[color=gray80, dashed] (8 * \i + 13, 4) -- +(-2, -2);
  \draw[color=gray80, dashed] (8 * \i + 13, 4) -- +(2, -2);
}

\foreach \i in {0, 1} {
  \draw[color=gray80, dashed] (4 * \i + 11, 2) -- +(-1, -2);
  \draw[color=gray80, dashed] (4 * \i + 11, 2) -- +(1, -2);
}

\draw[<->, >=stealth, color=gray80] (9.5, 0) -- +(7,0);
\foreach \i in {0, 1, 2, 3} {
 \fill[color=dark] (2 * \i + 10, 0) circle (4pt);
}

\fill[color=dark] (10, 0) circle (0pt)
node[below, color=black] { $z_{2}$ };

\fill[color=dark] (12, 0) circle (0pt)
node[below, color=black] { $z_{3}$ };

\fill[color=dark] (14, 0) circle (0pt)
node[below, color=black] { $z_{4}$ };

\fill[color=dark] (16, 0) circle (0pt)
node[below, color=black] { $z_{5}$ };

\end{tikzpicture}
}
\caption{
In multiplicative trajectory expansion we start with (a) an old trajectory 
component of length $L = 2^{D}$ that contains the initial point, $z_{0}$, 
and append (b) a new trajectory component of the same length by randomly 
integrating either forwards or backwards in time.  (c) This yields a trajectory 
twice as long as the initial trajectory sampled from 
$\mathfrak{t} \sim \mathbb{T} ( \mathfrak{t} \mid z_{0} )
= U \! \left( \mathfrak{T}^{2L}_{z_{0}} \right)$,
with a sampled state $z' \sim \mathbb{T} ( z' \mid \mathfrak{t} )$
generated concurrently from \eqref{eqn:uniform_progressive_transition}.
(d) The same process can be also be used to recursively build up the new
trajectory component, here initialized from $z_{2}$, and a corresponding
sample $z_{\mathrm{new}} \sim \mathbb{T} ( z' \mid \mathfrak{t}_{\mathrm{new}} )$.}
\label{fig:tree_building}
\end{figure*}

We can also expand faster by doubling the length of the trajectory 
at every iteration, yielding a sampled trajectory
$\mathfrak{t} \sim \mathbb{T} ( \mathfrak{t} \mid z )
= U \! \left( \mathfrak{T}^{2L}_{z} \right)$ with
the corresponding sampled state $z' \sim \mathbb{T} ( z' \mid \mathfrak{t} )$.
In this case both the old and new trajectory components at every 
iteration are equivalent to the leaves of perfect, ordered binary trees 
(Figure \ref{fig:trajectory_as_tree}).  This allows us to build the
new trajectory components recursively, propagating a sample at each step
in the recursion using \eqref{eqn:uniform_progressive_transition} until 
we have both a new trajectory and a new sample (Figure \ref{fig:tree_building}).

\subsubsection{Biased Progressive Sampling}

This progressive approach to sampling a state from a random trajectory also 
provides an opportunity to introduce anti-correlations into the transition
and potentially improve performance.  Specifically, we can use an alternative
transition that satisfies
\begin{equation*}
\mathbb{T} ( z' \mid \mathfrak{t} )
=
\min \! \left( 1, \frac{ w_{\mathrm{new}} }{ w_{\mathrm{old}} } \right) \,
\mathbb{T} ( z' \mid \mathfrak{t}_{\mathrm{new}} )
+
\left( 1 - \min \! \left( 1, \frac{ w_{\mathrm{new}} }{ w_{\mathrm{old}} } \right) \right) \,
\mathbb{T} ( z' \mid \mathfrak{t}_{\mathrm{old}} ),
\end{equation*}
where $\mathbb{T} ( z' \mid \mathfrak{t}_{\mathrm{old}} )$ and
$\mathbb{T} ( z' \mid \mathfrak{t}_{\mathrm{new}} )$ are given by
\eqref{eqn:state_sampling} as in the uniform progressive sampling approach.

Any such transition favors samples from the new trajectory component and
hence biases the transition away from the initial state.  When we average 
over all possible initial states in a given trajectory, however, these 
repulsions cancel to ensure that we preserve the canonical distribution
restricted to the sampled trajectory,
\begin{align*}
\sum_{z} \mathbb{T} ( z' \mid z )
\pi \! \left( z \right) \mathbb{I} ( z \in \mathfrak{t} )
&=
\sum_{z} 
\mathbb{T} ( z' \mid z )
\pi \! \left( z \right) \mathbb{I} ( z \in \mathfrak{t}_{\mathrm{old}} )
+
\sum_{z} 
\mathbb{T} ( z' \mid z )
\pi \! \left( z \right) \mathbb{I} ( z \in \mathfrak{t}_{\mathrm{new}} )
\\
&=
\sum_{z} 
\mathbb{T} ( z' \mid \mathfrak{t}_{\mathrm{old}} )
\pi \! \left( z \right) \mathbb{I} ( z \in \mathfrak{t}_{\mathrm{old}} )
+
\sum_{z} 
\mathbb{T} ( z' \mid \mathfrak{t}_{\mathrm{new}} )
\pi \! \left( z \right) \mathbb{I} ( z \in \mathfrak{t}_{\mathrm{new}} )
\\
&=
\mathbb{T} ( z' \mid \mathfrak{t}_{\mathrm{old}} )
\sum_{z}  \pi \! \left( z \right) \mathbb{I} ( z \in \mathfrak{t}_{\mathrm{old}} )
+
\mathbb{T} [ z' \mid \mathfrak{t}_{\mathrm{new}} ] 
\sum_{z}  \pi \! \left( z \right) \mathbb{I} ( z \in \mathfrak{t}_{\mathrm{new}} )
\\
&=
\mathbb{T} ( z' \mid \mathfrak{t}_{\mathrm{old}} )
w_{\mathrm{old}}
+
\mathbb{T} ( z' \mid \mathfrak{t}_{\mathrm{new}} )
w_{\mathrm{new}}
\\
&= \quad
\min \! \left( 1, \frac{ w_{\mathrm{new}} }{ w_{\mathrm{old}} } \right)
\frac{ \pi (z') }{ w_{\mathrm{new}} }  
\mathbb{I} ( z' \in \mathfrak{t}_{\mathrm{new}} )
w_{\mathrm{old}}
\\
& \quad +
\left( 1 - \min \! \left( 1, \frac{ w_{\mathrm{new}} }{ w_{\mathrm{old}} } \right) \right)
\frac{ \pi (z') }{ w_{\mathrm{old}} }  
\mathbb{I} ( z' \in \mathfrak{t}_{\mathrm{old}} )
w_{\mathrm{old}}
\\
& \quad +
\min \! \left( 1, \frac{ w_{\mathrm{old}} }{ w_{\mathrm{new}} } \right)
\frac{ \pi (z') }{ w_{\mathrm{old}} }  
\mathbb{I} ( z' \in \mathfrak{t}_{\mathrm{old}} )
w_{\mathrm{new}}
\\
& \quad +
\left( 1 - \min \! \left( 1, \frac{ w_{\mathrm{old}} }{ w_{\mathrm{new}} } \right) \right)
\frac{ \pi (z') }{ w_{\mathrm{new}} }  
\mathbb{I} ( z' \in \mathfrak{t}_{\mathrm{new}} )
w_{\mathrm{new}}
\\
&= \quad
\min \! \left( \frac{ w_{\mathrm{old}} }{ w_{\mathrm{new}} }, 1 \right)
\pi (z') \mathbb{I} ( z' \in \mathfrak{t}_{\mathrm{new}} )
\\
& \quad +
\left( 1 - \min \! \left( 1, \frac{ w_{\mathrm{new}} }{ w_{\mathrm{old}} } \right) \right)
\pi (z') \mathbb{I} ( z' \in \mathfrak{t}_{\mathrm{old}} )
\\
& \quad +
\min \! \left( \frac{ w_{\mathrm{new}} }{ w_{\mathrm{old}} }, 1\right)
\pi (z') \mathbb{I} ( z' \in \mathfrak{t}_{\mathrm{old}} )
\\
& \quad +
\left( 1 - \min \! \left( 1, \frac{ w_{\mathrm{old}} }{ w_{\mathrm{new}} } \right) \right)
\pi (z') \mathbb{I} ( z' \in \mathfrak{t}_{\mathrm{new}} )
\\
&=
\pi (z') \, \mathbb{I} ( z' \in \mathfrak{t}_{\mathrm{old}} )
+ \pi (z') \, \mathbb{I} ( z' \in \mathfrak{t}_{\mathrm{new}} )
\\
&=
\pi (z') \, \mathbb{I} ( z' \in \mathfrak{t} ).
\end{align*}

\subsection{Dynamic Implementations}

No matter how efficient we can make a static scheme, it will unfortunately be 
fundamentally limited.  Aside from a few exceptionally simple problems, each 
energy level set will require a different integration time to achieve the same 
effective exploration.  Some level sets can be surveyed with only short 
integration times, while others require very long integration times.  In order 
to ensure optimal performance of the Hamiltonian Monte Carlo method we need 
dynamic implementations that can automatically tune the integration time to 
approximate uniform exploration across all energy level sets.

Fortunately, the efficient static schemes discussed in Section
\ref{sec:efficient_static} can be iterated to achieve a dynamic implementation 
once we have chosen a criterion for determining when a trajectory has grown 
long enough to satisfactory explore the corresponding energy level set.  We 
first building an initial trajectory of a given length and then check the 
termination criterion.  If the criterion is not met, we expand the trajectory 
further and repeat, but if the criterion is met then we return the trajectory
and its corresponding sample.  

Care must be taken, however, to ensure that such a dynamic scheme does 
not obstruct the reversibility of the trajectory sampling and hence 
the desired invariance of the canonical distribution.  In this section I 
will first demonstrate how naive dynamic schemes can frustrate reversibility 
and what steps we have to take to avoid these problems.  With a valid dynamic 
implementation understood I will then review potential termination criteria.

\subsubsection{Maintaining Reversibility}

\begin{figure*}
\centering
\begin{tikzpicture}[scale=0.5, thick]

\draw[color=gray80, dashed] (7, 6) -- +(-4, -2);
\draw[color=gray80, dashed] (7, 6) -- +(4, -2);

\foreach \i in {0, 1} {
  \draw[color=gray80, dashed] (8 * \i + 3, 4) -- +(-2, -2);
  \draw[color=gray80, dashed] (8 * \i + 3, 4) -- +(2, -2);
}

\foreach \i in {0, 1, ..., 3} {
  \draw[color=gray80, dashed] (4 * \i + 1, 2) -- +(-1, -2);
  \draw[color=gray80, dashed] (4 * \i + 1, 2) -- +(1, -2);
}

\draw[color=black] (9, 2) -- +(-1, -2);
\draw[color=black] (9, 2) -- +(1, -2);

\draw[<->, >=stealth, color=gray80] (-0.5, 0) -- +(15,0);
\foreach \i in {0, 1, ..., 7} {
 \fill[color=dark] (2 * \i, 0) circle (4pt);
}

\fill[color=dark] (0, 0) circle (4pt)
node[below, color=black] { $z_{-2}$ };

\fill[color=dark] (2, 0) circle (4pt)
node[below, color=black] { $z_{-1}$ };

\fill[color=dark] (4, 0) circle (4pt)
node[below, color=black] { $z_{0}$ };

\fill[color=dark] (6, 0) circle (4pt)
node[below, color=black] { $z_{1}$ };

\fill[color=dark] (8, 0) circle (4pt)
node[below, color=black] { $z_{2}$ };

\fill[color=dark] (10, 0) circle (4pt)
node[below, color=black] { $z_{3}$ };

\fill[color=dark] (12, 0) circle (4pt)
node[below, color=black] { $z_{4}$ };

\fill[color=dark] (14, 0) circle (4pt)
node[below, color=black] { $z_{5}$ };

\end{tikzpicture}
\caption{Naively expanding a trajectory until a dynamic termination
criterion is satisfied is not sufficient to ensure reversibility and
hence the validity of the resulting Markov chain.  When expanding
trajectories multiplicatively, for example, if the termination criterion 
is satisfied by any sub-tree, for example the trajectory beginning at 
$z_{2}$ and ending at $z_{3}$, then both 
$\mathbb{T} \! \left( \mathfrak{t} \mid z_{2} \right) = 0$ and
$\mathbb{T} \! \left( \mathfrak{t} \mid z_{3} \right) = 0$
despite $\mathbb{T} \! \left( \mathfrak{t} \mid z_{0} \right) \neq 0$.
To avoid this pathology we have to reject any trajectory expansions
that contain such pathological sub-trajectories.
}
\label{fig:premature_termination}
\end{figure*}
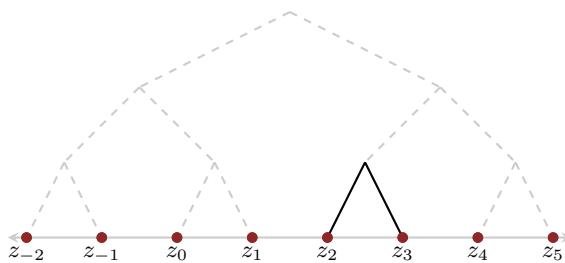

The problem with the naive dynamic scheme discussed above is that
it can violate the reversibility condition,
\begin{equation*}
\mathbb{T} ( \mathfrak{t} \mid z )
=
\mathbb{T} ( \mathfrak{t} \mid z' ), 
\, \mathfrak{t} \in \mathfrak{T}^{L}_{z, z'},
\end{equation*}
for the expanded trajectory, and hence the bias the stationary 
distribution of the resulting Markov chain.  In particular, if
we initialize from a different point in the final trajectory then
the dynamic trajectory expansion might terminate prematurely
before reaching the final trajectory, causing 
$\mathbb{T} ( \mathfrak{t} \mid z )$ to vanish for some initial 
states but not others (Figure \ref{fig:premature_termination}).

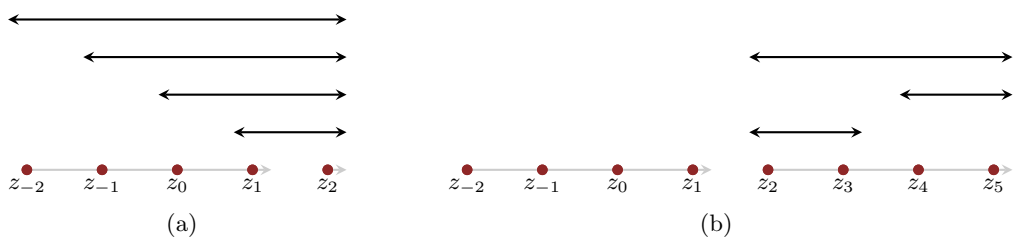
\begin{figure*}
\centering
\subfigure[]{
\begin{tikzpicture}[scale=0.5, thick]

\draw[color=white] (-1, 0) -- (9.5, 0);

\foreach \i in {0, 1, ..., 3} {
  \draw[<->, >=stealth, color=black] (2 * \i - 0.5, 4 - \i) -- (8.5, 4 - \i);
}

\draw[->, >=stealth, color=gray80] (0, 0) -- +(6.5, 0);
\foreach \i in {0, 1, ..., 4} {
 \fill[color=dark] (2 * \i, 0) circle (4pt);
}

\fill[color=dark] (0, 0) circle (4pt)
node[below, color=black] { $z_{-2}$ };

\fill[color=dark] (2, 0) circle (4pt)
node[below, color=black] { $z_{-1}$ };

\fill[color=dark] (4, 0) circle (4pt)
node[below, color=black] { $z_{0}$ };

\fill[color=dark] (6, 0) circle (4pt)
node[below, color=black] { $z_{1}$ };

\draw[->, >=stealth, color=gray80] (8, 0) -- +(0.5, 0);

\fill[color=dark] (8, 0) circle (4pt)
node[below, color=black] { $z_{2}$ };

\end{tikzpicture}
}
\subfigure[]{
\begin{tikzpicture}[scale=0.5, thick]

\draw[color=white] (-1.5, 0) -- (15, 0);

\draw[<->, >=stealth, color=black] (7.5, 1) -- +(3, 0);
\draw[<->, >=stealth, color=black] (11.5, 2) -- +(3, 0);
\draw[<->, >=stealth, color=black] (7.5, 3) -- +(7, 0);

\draw[->, >=stealth, color=gray80] (0, 0) -- +(6.5, 0);
\foreach \i in {0, 1, ..., 3} {
 \fill[color=dark] (2 * \i, 0) circle (4pt);
}

\fill[color=dark] (0, 0) circle (4pt)
node[below, color=black] { $z_{-2}$ };

\fill[color=dark] (2, 0) circle (4pt)
node[below, color=black] { $z_{-1}$ };

\fill[color=dark] (4, 0) circle (4pt)
node[below, color=black] { $z_{0}$ };

\fill[color=dark] (6, 0) circle (4pt)
node[below, color=black] { $z_{1}$ };

\draw[->, >=stealth, color=gray80] (8, 0) -- +(6.5, 0);
\foreach \i in {0, 1, ..., 3} {
 \fill[color=dark] (2 * \i + 8, 0) circle (4pt);
}

\fill[color=dark] (8, 0) circle (4pt)
node[below, color=black] { $z_{2}$ };

\fill[color=dark] (10, 0) circle (4pt)
node[below, color=black] { $z_{3}$ };

\fill[color=dark] (12, 0) circle (4pt)
node[below, color=black] { $z_{4}$ };

\fill[color=dark] (14, 0) circle (4pt)
node[below, color=black] { $z_{5}$ };

\end{tikzpicture}
}
\caption{(a) In an additive expansion scheme we have to check the 
termination criterion for all sub-trajectories between each point 
in the old trajectory and the single point in the new trajectory.  
(b) In a multiplicative expansion scheme, however, we need to check 
the termination criterion across only the binary sub-trees of the 
new trajectory.
}
\label{fig:termination_checks}
\end{figure*}

In order to equalize the transition probabilities we have to reject
any trajectory expansions that contains certain sub-trajectories 
satisfying the termination criterion and hence states that would
have prematurely terminated if we had instead begun there. For example, 
if we expand the trajectories additively then we need to check the 
termination criterion for the sub-trajectories beginning at every state 
in the old trajectory and ending at the single state in the new trajectory, 
inducing a cost linear in the size of the old trajectory (Figure 
\ref{fig:termination_checks}a).  Doing this for each expansion results 
in a cost quadratic in the final trajectory length.

If we expand the trajectories multiplicatively, however, then we need 
check only the sub-trajectories equivalent to binary sub-trees of the new
trajectory, at a cost just logarithmic in the size of the old trajectory 
(Figure \ref{fig:termination_checks}b) and a total cost quadratic only 
in the log of final trajectory length.  Moreover, because the checks 
occur across sub-trees we can compute them concurrently as we build the
new trajectory.  If any sub-tree satisfies the termination criterion
prematurely then we can immediately reject the new trajectory without
having to build it to completion.

\subsubsection{Dynamic Termination Criteria}
\label{sec:term_criteria}

Now we can finally consider explicit termination criteria that can,
to varying degrees of success, identify when a trajectory is long
enough to yield sufficient exploration of the neighborhood around
the current energy level set.  In particular, we want to avoid both
integrating too short, in which case we would not take full advantage 
of the Hamiltonian trajectories, and integrating too long, in which
case we waste precious computational resources on exploration that 
yields only diminishing returns.

\begin{figure*}
\centering
\begin{tikzpicture}[scale=0.5, thick]

\draw[<->, >=stealth, color=gray80] (-0.5, 0) -- +(15,0);
\foreach \i in {-2, -1, ..., 5} {
 \fill[color=dark] (2 * \i + 4, 0) circle (4pt)
 node[below, color=black] { $z_{\i}$ };
}

\fill[color=dark] (0, -1) circle (0pt)
node[below, color=black] { $z_{-} ( \mathfrak{t} ) = z_{-2}$ };

\fill[color=dark] (14, -1) circle (0pt)
node[below, color=black] { $z_{+} ( \mathfrak{t} ) = z_{5}$ };

\end{tikzpicture}
\caption{The No-U-Turn termination criterion considers only the
boundaries of a trajectory.  The negative boundary, $z_{-} ( \mathfrak{t} )$,
refers to the boundary that can be reached by integrating backwards
in time, while the positive boundary, $z_{+} ( \mathfrak{t} ) $ refers 
to the boundary that can be reached by integrating forwards in time.}
\label{fig:trajectory_boundaries}
\end{figure*}
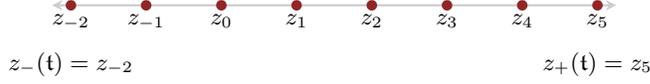

The first criterion of this form was a heuristic introduced in the 
No-U-Turn sampler~\citep{HoffmanEtAl:2014} that considered only the 
boundaries of a trajectory.  Let $z_{-} ( \mathfrak{t} )$ and 
$z_{+} ( \mathfrak{t} )$ be the boundaries of the trajectory $\mathfrak{t}$ 
that can be reached by integrating backwards and forwards in time, 
respectively (Figure \ref{fig:trajectory_boundaries}). Similarly, denote 
$q_{\pm} ( \mathfrak{t} )$ and $p_{\pm} ( \mathfrak{t} )$ as the position 
and momentum at each boundary.  On Euclidean space we can then write the 
No-U-Turn termination criterion as
\begin{align}
p_{+} ( \mathfrak{t} )^{T} 
\cdot \left( q_{+} ( \mathfrak{t} ) - q_{-} ( \mathfrak{t} ) \right) 
&< 0
\nonumber
\\
\mathrm{AND} \;
p_{-} ( \mathfrak{t} )^{T} 
\cdot \left( q_{-} ( \mathfrak{t} ) - q_{+} ( \mathfrak{t} ) \right) 
&< 0.
\label{eqn:original_nuts_criterion}
\end{align}

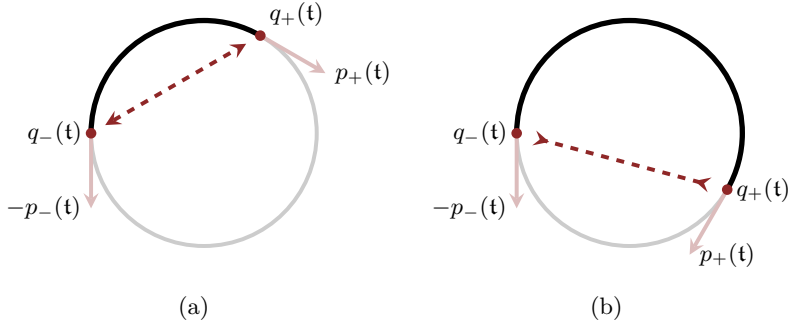
\begin{figure*}
\centering
\subfigure[]{
\begin{tikzpicture}[scale=0.5, thick]

\draw[color=white] (0, -4) -- (0, 4); % Alignment

\draw[color=gray80, line width=1.5] (3, 0) circle (3);

\draw[line width=2] (0, 0) arc (180:60:3);

\draw[->, >=stealth, color=light, line width=1.5] (0, 0) -- +(0, -2)
node[left, color=black] { $-p_{-} ( \mathfrak{t} )$ };

\draw[<->, >=stealth, color=dark, dashed, line width=1.5] ({0 + 0.25 * (1 + cos(60))}, 
                                {0 + 0.25 * sin(60)}) 
                               -- ({3 * (1 + cos(60)) - 0.25 * (1 + cos(60))}, 
                                   {3 * sin(60) - 0.25 * sin(60)});

\fill[color=dark] (0, 0) circle (4pt)
node[left, color=black] { $q_{-} ( \mathfrak{t} )$ };

\draw[->, >=stealth, color=light, line width=1.5] ({3 + 3 * cos(60)}, {3 * sin(60)}) 
                                   -- +({2 * sin(60)}, -{2 * cos(60)})
node[right, color=black] { $p_{+} ( \mathfrak{t} )$ };

\fill[color=dark] ({3 + 3 * cos(60)}, {3 * sin(60)}) circle (4pt)
node[above right, color=black] { $q_{+} ( \mathfrak{t} ) $ };

\end{tikzpicture}
}
\subfigure[]{
\begin{tikzpicture}[scale=0.5, thick]

\draw[color=white] (0, -4) -- (0, 4); % Alignment

\draw[color=gray80, line width=1.5] (3, 0) circle (3);

\draw[line width=2] (0, 0) arc (180:-30:3);

\draw[->, >=stealth, color=light, line width=1.5] (0, 0) -- +(0, -2)
node[left, color=black] { $-p_{-} ( \mathfrak{t} )$ };

\draw[>-<, >=stealth, color=dark, dashed, line width=1.5] ({0.25 * (1 + cos(-30))}, 
                                {0.25 * sin(-30)}) 
                                -- ({3 * (1 + cos(-30)) - 0.25 * (1 + cos(-30))}, 
                                    {3 * sin(-30) - 0.25 * sin(-30)});

\fill[color=dark] (0, 0) circle (4pt)
node[left, color=black] { $q_{-} ( \mathfrak{t} )$ };

\draw[->, >=stealth, color=light, line width=1.5] ({3 + 3 * cos(-30)}, {3 * sin(-30)}) 
                         -- +({2 * sin(-30)}, -{2 * cos(-30)})
node[right, color=black] { $p_{+} ( \mathfrak{t} )$ };

\fill[color=dark] ({3 + 3 * cos(-30)}, {3 * sin(-30)}) circle (4pt)
node[right, color=black] { $q_{+} ( \mathfrak{t} ) $ };

\end{tikzpicture}
}
\caption{(a) When the No-U-Turn termination criterion is not satisfied, 
expanding the trajectory in either direction typically extends the 
trajectory further across the energy level set (grey) towards unexplored
neighborhoods.  (b) When the No-U-Turn termination criterion is satisfied, 
however, further expansion typically contracts the boundaries of the 
trajectory towards each other and neighborhoods that have already been
explored.  Stopping when the No-U-Turn termination criterion is just 
satisfied typically ensures that each trajectory efficiently explores 
about half of the energy level set.  Although this can be too aggressive,
it works well empirically and is far superior to any scheme with only a 
static integration time.}
\label{fig:no_u_turn_cartoon}
\end{figure*}

This termination criterion is satisfied only when the momentum at 
both ends of the trajectory are anti-aligned along the line connecting 
the positions at each boundary.  In this case we would expect further 
integration in either direction to bring the ends of the trajectory 
closer together, which typically happens only when the trajectory has
already expanded across the width of the energy level set (Figure
\ref{fig:no_u_turn_cartoon}).

Unfortunately, the original No-U-Turn termination criterion is well-defined 
only for Euclidean manifolds where we can rigorously define subtraction 
between points, and hence is inapplicable for non-Euclidean kinetic energies.  
Fortunately, however, the criterion can be readily generalized.  First note 
that on a Euclidean manifold
\begin{equation*}
q_{+} ( \mathfrak{t} ) - q_{-} (\mathfrak{t} )
= \int_{t = 0}^{t = T ( \mathfrak{t} )} 
\mathrm{d}t \, M^{-1} \cdot p(t),
\end{equation*}
where $T ( \mathfrak{t} )$ is the integration time needed to reach
the positive boundary of the trajectory, $z_{+} (\mathfrak{t})$, from 
the negative boundary, $z_{+} (\mathfrak{t})$.  Substituting this into
\eqref{eqn:original_nuts_criterion}, for example, allows us to write 
the first term of the No-U-Turn termination criterion as
\begin{equation*}
p_{+} (\mathfrak{t} )^{T} 
\cdot \left( q_{+} (\mathfrak{t} ) - q_{-} (\mathfrak{t} ) \right)
= 
p_{+} (\mathfrak{t} )^{T} \cdot M^{-1} \cdot 
\int_{t = 0}^{t = T ( \mathfrak{t} )} 
\mathrm{d}t \, p(t)
= p^{\sharp}_{+} (\mathfrak{t} )^{T} \cdot \rho (\mathfrak{t} )
\end{equation*}
where
\begin{equation*}
p^{\sharp} (\mathfrak{t} ) \equiv M^{-1} \cdot p (\mathfrak{t} )
\end{equation*}
and
\begin{equation*}
\rho (\mathfrak{t} ) \equiv 
\int_{t = 0 ( \mathfrak{t} )}^{t = T ( \mathfrak{t} )} 
\mathrm{d}t \, p(t).
\end{equation*}

Continuing the substituting into the second term, and being careful
with the signs, then gives the equivalent termination criterion,
\begin{equation*}
p^{\sharp}_{+} (\mathfrak{t} )^{T} \cdot \rho (\mathfrak{t} )< 0
\quad \mathrm{AND} \quad
p^{\sharp}_{-} (\mathfrak{t} )^{T} \cdot \rho (\mathfrak{t} ) < 0.
\end{equation*}
This new form is well-defined for any Riemannian manifold and hence
defines a proper generalization of the No-U-Turn termination criterion 
to Riemannian manifolds~\citep{Betancourt:2013a}.

Although we can't compute $\rho$ analytically in practice, we can 
readily approximate it as a discrete sum over the momenta in the 
numerical trajectory generated by the symplectic integrator,
\begin{equation*}
\rho (\mathfrak{t} ) \approx \sum_{z \in \mathfrak{t} } p (z).
\end{equation*}
Because of the high accuracy of symplectic integrators, this approximation 
yields almost identical results as at the original No-U-Turn criterion. 

The geometry of the Hamiltonian systems behind Hamiltonian Monte Carlo
also motivates another termination criterion that doesn't require any
Riemannian structure at all.  \emph{Exhaustions} track a quantity inherent
to Hamiltonian systems called the virial to determine when an energy level set 
has been sufficiently explored~\citep{Betancourt:2016}.  Unfortunately exhaustions 
introduce an additional tuning parameter that strongly effects the performance 
of the resulting implementation. Although careful hand-tuning can yield better 
performance than implementations that use the No-U-Turn termination criterion, 
we need to understand how to automatically tune exhalations before they can 
become a practical competitor.

\subsection{The No-U-Turn Sampler and the Current State of Stan}

With the hindsight of this analysis, we see that the original No-U-Turn 
sampler~\citep{HoffmanEtAl:2014} is a dynamic implementation of Hamiltonian 
Monte Carlo.  In addition to the employing the No-U-Turn termination criterion 
discussed in Section \ref{sec:term_criteria}, the No-U-Turn sampler uses a 
multiplicative expansion of each trajectory and a slice sampler to sample 
states from those trajectories.  Each new trajectory component is sampled 
using uniform progressive sampling, but the sample update when appending 
the new component is generated with biased progressive sampling.

Although Stan~\citep{Stan:2017} was first developed around this original 
No-U-Turn sampler, more recent releases have adopted some modifications.  
In addition to the generalized No-U-Turn termination criterion, the 
Hamiltonian Monte Carlo implementation in Stan uses multinomial sampling 
from each trajectory instead of slice sampling, which provides a significant 
improvement in overall performance.

\bibliography{hmc_intro}
\bibliographystyle{imsart-nameyear}

\end{document}